\documentclass[a4paper,12pt]{article}
\pdfoutput=1
\usepackage{epsfig}
\usepackage{amssymb}
\usepackage{cancel}
\usepackage{setspace}
\usepackage[usenames,dvipsnames,svgnames,table]{xcolor}
\usepackage{amsmath}
\usepackage{graphicx}
\usepackage{subfigure}
\usepackage{url}
\usepackage{slashed}
\usepackage{booktabs}
\usepackage{cite}
\newcommand\ddfrac[2]{\frac{\displaystyle #1}{\displaystyle #2}}

\newcommand{\GeV}{\,{\rm GeV}}
\newcommand{\TeV}{\,{\rm TeV}}

\usepackage{siunitx}
\newcolumntype{L}{>{$}l<{$}}
\newcolumntype{C}{>{$}c<{$}}
\newcolumntype{R}{>{$}r<{$}}
\newcolumntype{P}{>{$}p<{$}}

\newcommand{\be}{\begin{equation}}
\newcommand{\ee}{\end{equation}}
\newcommand{\br}{\begin{eqnarray}}
\newcommand{\bea}{\begin{eqnarray}}
\newcommand{\eea}{\end{eqnarray}}
\newcommand{\er}{\end{eqnarray}}
\newcommand{\ba}{\begin{array}}
\newcommand{\ea}{\end{array}}
\newcommand{\bi}{\begin{itemize}}
\newcommand{\ei}{\end{itemize}}
\newcommand{\bn}{\begin{enumerate}}
\newcommand{\en}{\end{enumerate}}
\newcommand{\bc}{\begin{center}}
\newcommand{\ec}{\end{center}}

\newcommand{\eq}[1]{eq.~(\ref{#1})}

\newcommand{\beq}{\begin{equation}}
\newcommand{\eeq}{\end{equation}}

\newcommand{\U}{\scriptscriptstyle U}
\newcommand{\D}{\scriptscriptstyle D}

\newcommand{\gsim}{\lower1.0ex\hbox{$\;\stackrel{\textstyle>}{\sim}\;$}}
\newcommand{\lsim}{\lower1.0ex\hbox{$\;\stackrel{\textstyle<}{\sim}\;$}}

\newcommand{\bs}{\begin{small}}
\newcommand{\es}{\end{small}}

\renewcommand{\Re}[1]{\text{Re}\!\left[#1 \right]}

\newcommand{\pB}{p_{\scriptscriptstyle B}}
\newcommand{\pK}{p_{\scriptscriptstyle K}}

\newcommand{\mKs}{m_{K^{*}}}
\newcommand{\mB}{m_{B}}
 
\newcommand{\hmKs}{\hat{m}_{K^{*}}}
\newcommand{\hmK}{\hat{m}_{K}}

\newcommand{\hmb}{\hat{m}_{b}}

\newcommand{\hs}{\hat{s}}
\newcommand{\hu}{\hat{u}}
\newcommand{\bu}{\bar{u}(\hs)}
\newcommand{\rl}{r_{\ell}}
\newcommand{\rKs}{r_{K^{*}}}
\newcommand{\Ks}{\scriptscriptstyle{K^*}}
\newcommand{\K}{\scriptscriptstyle{K}}
\newcommand{\LO}{\scriptscriptstyle{{\rm LO}}}
\newcommand{\M}{\scriptscriptstyle{{\rm M}}}
\newcommand{\Som}{\scriptscriptstyle{{\rm S}}}
\newcommand{\BR}{{\rm BR}}
\newcommand{\eff}{{\rm eff}}

\def\di{\mbox{d}}

\begin{document}
\begin{center}
  {\Large {\bf Magnetic-dipole corrections to $R_{K}$ and $R_{K^*}$
      \\ \vspace{0.5cm}
      in the Standard Model and Dark Photon scenarios}}
\\
\vspace*{1.5cm}
{

 {\bf Emidio Gabrielli$^{a,b,c}$} and   {\bf Marco Palmiotto$^{a}$}}

\vspace{0.5cm}
{\it
(a)  Dipartimento di Fisica, Theoretical section, Universit\`a di 
Trieste, \\ Strada Costiera 11, I-34151 Trieste, Italy and\\
INFN, Sezione di Trieste, Via Valerio 2, I-34127 Trieste, Italy
\\
(b) NICPB, R\"avala 10, Tallinn 10143, Estonia
\\
(c) IFPU, Via Beirut 2, 34151 Trieste, Italy
\\[1mm] }

\vspace*{2cm}{\bf ABSTRACT}
\end{center} 
\vspace{0.3cm}

\noindent
In this work we evaluate the long-distance QED contributions, induced by the magnetic-dipole corrections to the final charged leptons, on the $B$ meson decay widths $B\to (K,K^*) \ell^+\ell^-$ and ratios $R_{K,K^*}=\Gamma(B\to (K,K^*) \mu^+\mu^-)/\Gamma(B\to (K,K^*) e^+e^-)$, as well as on $R^{\tau}_{K,K^*}$ (with $\mu$ replaced by the $\tau$ lepton). QED long-distance contributions induced by the Coulomb potential corrections (Fermi-Sommerfeld factors) were also included. Corresponding corrections to the inclusive decay widths of $B\to X_s \ell^+\ell^-$, with $\ell=e,\mu,\tau$, are also analyzed for completeness. The magnetic-dipole corrections, which are manifestly Lepton Flavor Universality violating and gauge-invariant, are expected to be particularly enhanced in  $R_{K^*}$ for the dilepton mass region close to the threshold. However, we find that the largest contribution of all these corrections to the $R_{K,K^*}$ observables do not exceed a few per mille effect, thus reinforcing the validity of previous estimates about the leading QED corrections to $R_{K,K^*}$.
Finally, viable new physics contributions to $R_{K,K^{*}}$ induced by the exchange of a massless dark-photon via magnetic-dipole interactions, which provide the leading contribution to the corresponding $B\to (K,K^*) \ell^+\ell^-$  amplitudes in this scenario, are analyzed in light of the present $R_{K,K^{*}}$ anomalies.
\newpage

\section{Introduction} 
\label{sec:Introduction}
The semi-leptonic flavor-changing neutral currents (FCNC) $B$ meson decays
$B\to (X_s,X_d) \ell^+\ell^-$ with $\ell=e,\mu,\tau$, with $X_s$ representing any hadron with overall strangeness $S=\pm 1$, are very powerful tools to test the Standard Model (SM) predictions \cite{Misiak:1992bc, Buras:1994dj,Buchalla:1995vs,Chetyrkin:1996vx,Bobeth:1999mk,Bobeth:2001jm,Gambino:2003zm,Gorbahn:2004my,Gorbahn:2005sa,Altmannshofer:2008dz} and also sensitive probes to any New Physics (NP) beyond it \cite{Cho:1996we,Ali:1999mm}. Indeed, due to the fact that the FCNC processes are forbidden at the tree-level, the sensitivity to any potential NP contribution turns out to be strongly enhanced in the $b\to s \ell^+\ell^-$ decays.

Concerning the exclusive decays $B\to (K,K^*) \ell^+\ell^-$, great efforts have been devoted to achieve accurate SM predictions for the corresponding branching ratios and their distributions
\cite{Khodjamirian:2010vf,Beylich:2011aq,Bobeth:2011nj,DescotesGenon:2012zf,Khodjamirian:2012rm,Jager:2012uw,Descotes-Genon:2014uoa,Jager:2014rwa,Straub:2015ica,Ciuchini:2015qxb,Capdevila:2017ert,Chobanova:2017ghn}. However, these observables are affected by large theoretical uncertainties, mainly due to the evaluation of the form factors and estimate of the non-factorizable hadronic corrections.

Recently, the exclusive $B$ meson decays $B\to K \ell^+\ell^-$ and $B\to K^* \ell^+\ell^-$ with $\ell=e,\mu$ have been measured and in particular, the Lepton Flavor Universality (LFU) ratios $R_{K}$\cite{Aaij:2014ora,Aaij:2019wad} and $R_{K^*}$ \cite{Aaij:2017vbb,BelleMoriond,Abdesselam:2019wac} defined as 
\bea 
R_{K,K^*}&=&\ddfrac{\int_{q^2_{\rm min}}^{q^2_{\rm max}}
  \ddfrac{d\Gamma(B\to (K,K^*)\mu^+\mu^-)}{dq^2}dq^2}
{\int_{q^2_{\rm min}}^{q^2_{\rm max}}
  \ddfrac{d\Gamma(B\to (K,K^*)e^+e^-)}{dq^2}dq^2}\, ,
\label{RK} 
\eea
where $q^2$ represents the invariant mass of the dilepton $\ell^+\ell^-$ system. As suggested in \cite{Hiller:2003js}, the $R_{K,K^*}$ ratios can provide a clean test of the LFU of weak interactions predicted by the SM and are also sensitive probes of any new interactions that could potentially couple to electron and muons in a non-universal way \cite{Hiller:2014yaa}. Indeed, due to the LFU of gauge interactions, the SM prediction for these ratios is almost 1, for $q^2\gg (4m_{\mu})^2$ \cite{Bouchard:2013mia,Bordone:2016gaq}. Moreover, the hadronic matrix elements mainly factorize in $R_{K,K^*}$, thus reducing the main theoretical uncertainties and enhancing the sensitivity to any potential NP that might induce LFU violations.

The LHCb experimental result ~\cite{Aaij:2017vbb} reported for the following two $q^2$ bins is
\beq\label{eq:RK*LHCB}
R_{K^*}  = \left\{\begin{array}{ll}
  0.660^{+0.110}_{-0.070} \pm 0.024 & 0.045 \GeV^2 < q^2 < 1.1 \GeV^2  \\[2mm]
  0.685^{+0.113}_{-0.069} \pm 0.047& 1.1  \GeV^2 < q^2 < 6   \GeV^2 \, .
\end{array}
\right.
\eeq
This should be compared with a recent SM prediction \cite{Bordone:2016gaq}
\beq
R^{SM}_{K^*}  = \left\{\begin{array}{ll}
  0.906 \pm 0.028 & 0.045 \GeV^2 < q^2 < 1.1 \GeV^2  \\[2mm]
  1.00 \pm 0.01 & 1.1  \GeV^2 < q^2 < 6   \GeV^2 \, ,
\end{array} 
\right.
\eeq   
where the effect of soft and collinear LFU-violating radiative QED corrections, has been taken into account.
As we can see, the experimental measurements in the above $q^2$ bins show a substantial deviation from the SM expectations, although it is still within $2.6\sigma$ significance level.

Recently, new measurements of $R_{K^*}$ come also from the Belle Collaboration \cite{BelleMoriond,Abdesselam:2019wac}. Combining charged and neutral channels, the corresponding values in the $ q^2 < 6 {\rm GeV}^2$ region are \cite{BelleMoriond,Abdesselam:2019wac}
\beq
R_{K^*}  = \left\{\begin{array}{ll}
  0.52^{+0.36}_{-0.26}\pm 0.05 &  0.045 \GeV^2  < q^2 < 1.1 \GeV^2  \\[2mm]
  0.96^{+0.45}_{-0.29} \pm 0.11 & 1.1  \GeV^2 < q^2 < 6   \GeV^2 \, .
\end{array}
\right.
\label{RKSM}
\eeq
These results also show some substantial deviation from the SM central value, especially at  low bins close to the dimuon mass threshold, although the combined error is large enough to leave the deviation on the statistical significance within $\sim 1 \sigma$.

Concerning the $R_K$, the most recent measurement by LHCb in the low $q^2$ region recently appeared  \cite{Aaij:2014ora,Aaij:2019wad}
\beq
R_{K}  =  0.846^{+0.060+0.016}_{-0.054-0.014} ~~ 1.1 \GeV^2 < q^2 < 6 \GeV^2  \, ,
\eeq
where the first and second uncertainties correspond to the systematic and statistic errors respectively.
There, the full $5{\rm fb}^{-1}$ of data have been analyzed, including only the charged channel $B^+\to K^+ \mu^+\mu^-$.
Compared to the previous LHCb measurements, we can see that the new experimental central value is more close to the  SM one, although the significance tension is slightly reduced from previous $2.6\sigma$ (at $3{\rm fb}^{-1}$ of data)  to $2.5\sigma$.

In recent years, a large number of papers have suggested the possibility that all these deviations could be interpreted as signal of NP interactions, that potentially couple in non-universal way to muon and electrons\cite{Descotes-Genon:2015uva,Altmannshofer:2017fio,Capdevila:2017bsm,DAmico:2017mtc,Altmannshofer:2017yso,Geng:2017svp,Ciuchini:2017mik,Hiller:2017bzc,Alok:2017sui,Hurth:2017hxg,Alguero:2019aa,Alok:2019ufo,Ciuchini:2019usw,Aebischer:2019mlg,Kowalska:2019ley,Alguero:2018nvb,Alguero:2019pjc,Datta:2019zca,Greljo:2015mma,Barbieri:2015yvd,Barbieri:2016las,Bordone:2017anc,Bordone:2017lsy,Buttazzo:2017ixm,Barbieri:2017tuq}.
However, in order to quantify if the observed SM deviations in $R_{K,K^*}$ could be addressed to a genuine NP contributions, a precise estimation of the SM uncertainties is required.
In this respect, in \cite{Bordone:2016gaq} the LFU-violating contributions in the SM coming from the real photon emissions and its virtual effects, have been analyzed and the results are summarized in Eq.(\ref{RKSM}). This task consists in the computation and re-summation of the large $(\alpha/\pi)\log^2{(m_B/m_{\ell})}$ terms originating from the one-loop QED radiative contributions. These corrections depend by the choice of the $m^{\rm rec}_B$ mass for the reconstructed $K^*\ell^+\ell^-$ system, and are safe from infrared and collinear divergencies. Using the same $m^{\rm rec}_B$ in the range adopted for instance at the LHCb \cite{Aaij:2017vbb}, it has been shown that the largest QED effect does not exceed a few percent in $R_{K,K^*}$ \cite{Bordone:2016gaq}.

Larger SM uncertainties in $R_{K^*}$ are expected in  the $q^2$ region closer to the mass threshold $q^2\sim 4 m_{\ell}^2$ of the final lepton states. As shown in \cite{Bordone:2016gaq}, contributions coming from the direct photon emission amplitudes induced by the light-hadron mediated amplitudes should be also taken into account. These are of the type $B\to K^* P^0\to K^* \ell^+\ell^-\gamma$, where $P^0$ stands for an on-shell $\eta$ or $\pi^0$ meson state. In \cite{Bordone:2016gaq} it has been estimated that these contributions give an effect on
  $R_{K^*}$  of the order of $\Delta R_{K^*}\sim -0.017$ for $0.045 \GeV^2 < q^2 < 1.1 \GeV^2 $,
while they become negligible for $q^2 > 0.1\GeV^2$, where the meson-mediated amplitudes becomes lepton universal.

In this paper we focus on a new class of QED radiative corrections that have not been considered so far in the literature and that can add a new source of LFU-violating contributions to $R_{K,K^*}$ in the low $q^2$ region. In particular, we consider the effect induced by the QED magnetic-dipole corrections to the final lepton pair, on the $B\to K \ell^+\ell^-$ and $B\to K^* \ell^+\ell^-$ decay rates and in the ratios $R_{K,K^*}$. These corrections represent an independent set of standalone QED radiative corrections, being gauge invariant and free from any infrared and collinear divergencies. 
Indeed, due to the intrinsic long-distance nature of magnetic-dipole interactions, mediated by the one-photon exchange amplitude, their contribution is particularly enhanced at low $q^2$. 

We can see that, by means of chirality arguments, the interference between the amplitude containing the magnetic-dipole correction and the leading order SM amplitude turns out to be chiral suppressed, naturally providing a LFU-violating corrections to $R_{K,K^*}$. This automatically guarantees that the magnetic-dipole contribution to the width in the electron channel turns out to be much smaller than the corresponding muon one.
On the other hand, the magnetic form factor $F_2(q^2)$ can partially compensate for this chiral suppression in the dimuon final state, since
${\rm Re}[F_2(q^2)]\to 1/m_{\mu}$  in the region close to the mass threshold $q^2\to 4m_{\mu}^2$.

Due to the presence of infrared double  poles $m_{\ell}^2/(q^2)^2$ proportional to the magnetic-form factor $F_2$ in the width distribution $\frac{d\Gamma}{dq^2}$, sizeable LFU-violating contributions to $R_{K,K^*}$ are possible, if the dilepton mass threshold is included in the $q^2$ integration region.
Indeed, for the muon channel, by integrating the double poles from $q^2> 4 m_{\mu}^2$ the chiral suppression is removed, while the contributions to the $e^+e^-$ channel results suppressed by terms of order $m_e^2/m_{\mu}^2$.
Moreover, larger corrections in $R_{K^*}$, with respect to $R_K$, are also expected, since the magnetic-dipole contributions are more enhanced in the $B\to K^*$  transitions than in $B\to K$, due to the effects of the longitudinal polarization of $K^*$. By a n\"aive estimation these effects could be of the order of a few percent on $R_{K^*}$, if no cancellation among the leading contributions of the double-poles terms, that is the ones proportional to $(m_B/m_{K^*})^2\simeq 35$ enhancement factor, takes place. More details about this issue are reported in section 3.1. Therefore, it would be mandatory to provide an exact evaluation of the magnetic-dipole contributions in order to establish the hierarchy of the QED corrections in this context.
  
We will provide analytical results for the QED vertex corrections to the corresponding differential $B$ decay widths, induced by the magnetic-dipole corrections to the final lepton pair.  Then, we evaluate the impact of these corrections on the $R_{K,K^*}$ observables and compare these predictions with the corresponding LFU-violating results induced by collinear and infrared QED corrections \cite{Bordone:2016gaq}. To complete our study, we will include another set of corrections induced by the long-distance contributions to $R_{K,K^*}$. In particular, the ones related to the Coulomb potential corrections, that can be eventually absorbed in the so-called Sommerfeld-Fermi factor \cite{Sommerfeld,Fermi,Isidori:2007zt}.

Finally, concerning the LFU-violating soft-photon emissions by the magnetic-dipole interactions in the final-state leptons, we estimate these radiative corrections to be very small and negligible in both the rates and the $R_{K,K^*}$ observables, being  of order ${\cal O}(\alpha^2)$ and also chiral suppressed. Indeed, these contributions arise from the interference between the amplitude with soft-photon emissions by the magnetic-dipole interactions and the corresponding leading-order amplitude with tree-level soft-photon emission to final-state leptons.

Due to the gauge-invariant structure of the magnetic-dipole operators, 
these results could be easily generalized to include potential contributions from NP scenarios that are mediated by long-distance interactions.  So far, the majority of the beyond SM interpretations of the B-anomalies rely on NP contributions affecting the Wilson coefficients of the short-distance 4-fermion operators ${\cal O}_{L}$ and ${\cal O}_{9}$ \cite{Descotes-Genon:2015uva,Altmannshofer:2017fio,Capdevila:2017bsm,DAmico:2017mtc,Altmannshofer:2017yso,Geng:2017svp,Ciuchini:2017mik,Hiller:2017bzc,Alok:2017sui,Hurth:2017hxg,Alguero:2019aa,Alok:2019ufo,Ciuchini:2019usw,Aebischer:2019mlg,Kowalska:2019ley,Alguero:2018nvb,Alguero:2019pjc,Datta:2019zca}. In this respect, we explore the possibility to explain the $R_{K,K^*}$ anomalies by means of a new physics mediating long-distance interactions. In particular, we analyze the contribution of a massless dark-photon exchange in the $b\to s \ell^+\ell^-$ transitions, which has the feature to couple to both quarks and leptons via the leading magnetic-dipole interactions \cite{Fabbrichesi:2020wbt,Holdom:1985ag}, and estimate its impact on the $R_{K,K^*}$ ratios.

The paper is organized as follows: in section 2 and 3 we provide the analytical results for the magnetic-dipole corrections to the widths of $b\to s\, \ell^+\ell^-$ and $B\to (K,K*^*) \ell^+\ell^-$  respectively. Section 4 is devoted to the implementation of the Fermi-Sommerfeld corrections to the widths.
Numerical predictions  for the corresponding branching ratios of these processes and for the ratios $R_{K,K^*}$ are provided in section 5. In 
section 6 we analyze the impact of a NP scenario on the $R_{K,K^*}$ observables, given by the exchange of a massless dark-photon via magnetic-dipole interactions. Our conclusions are provided in section 7.

\section{Magnetic-dipole corrections to $b\to s \ell^+\ell^-$} 
We start this section by providing the notation used in the Effective Hamiltonian relevant for the semileptonic quark decay
\bea
b(p_b)&\to& s(p_s)\, \ell^+(p^+)\, \ell^-(p^-)
\label{bproc}
\eea
where corresponding momenta are associated in parenthesis. After integrating out the $W^{\pm}$ and top-quark, the effective Hamiltonian relevant for the $\Delta B=1$ transitions in Eq.(\ref{bproc}), is given by
\bea
H_{\rm eff}=-4\frac{G_F}{\sqrt{2}}V^*_{ts} V_{tb} \sum_{i=1}^{10}
C_i(\mu) Q_i(\mu)\, . 
\label{Heff} 
\eea
Here we adopt the definitions of operators $Q_i(\mu)$ as provided in \cite{Misiak:1992bc,Buras:1994dj,Buchalla:1995vs,Chetyrkin:1996vx} and the results for the corresponding Wilson coefficients $C_i(\mu)$ evaluated at the next-to-next-to-leading order (NNLO) \cite{Bobeth:1999mk,Bobeth:2001jm,Gambino:2003zm,Gorbahn:2004my,Gorbahn:2005sa,Altmannshofer:2008dz}, with the renormalization scale $\mu$ chosen at the b-quark pole mass $\mu=m_b$.

\begin{figure}
\begin{center}
\hspace{0.cm}
\includegraphics[width=1\textwidth]{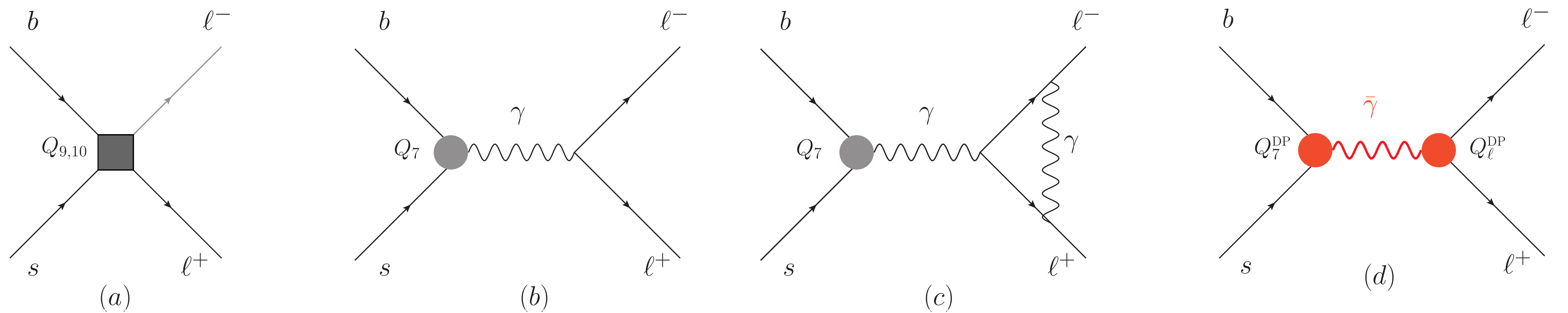}
\caption{Feynman diagrams for the b quark decay $b\to s \ell^+\ell^-$: (a)-(c) represent the SM contribution including vertex radiative corrections in the final leptons (c), while diagram (d) corresponds to the exchange of a massless dark-photon, mediated by magnetic-dipole interactions. The gray square and circular vertices represent the insertion of the 4-fermions $Q_{9,10}$ and magnetic-dipole $Q_7$ operators respectively, while the red circular vertex represent the insertion of the magnetic-dipole operators with a dark photon, namely $Q^{\rm DP}_{7}=[\bar{s} \sigma^{\mu\nu}b ]F_{\mu\nu}^D$ and $Q^{\rm DP}_{\ell}=[\bar{\ell}\sigma^{\mu\nu}\ell]F_{\mu\nu}^D$ with $F_{\mu\nu}^D$ the dark-photon field strength.}
\label{diagrams}
\end{center}
\end{figure}

Starting from the effective Hamiltonian in Eq.(\ref{Heff}), the SM Feynman diagrams contributions to the $b\to s \ell^+\ell^-$ amplitude are given in Fig.\ref{diagrams}(a)-(c). The diagrams (c) represents the contribution of the QED radiative corrections to the final lepton states, proportional to the magnetic-dipole form factor, that will be discussed in the following. The gray square vertex stands for the insertion of the local 4-fermion operators $Q_{9,{10}}$, while the gray circular vertex corresponds to the contribution of the magnetic-dipole operator $Q_7$. The diagrams (b) and (c) describe the long-distance contributions to the amplitude mediated by the virtual photon. The NP contribution, characterized by the diagram (d), will be discussed in section 6.

The amplitude for $b\to s \ell^+\ell^-$ decay can be simply described by introducing the effective Wilson coefficients $C_9^{\rm eff}(\hs)$, $C_{7}^{\rm eff}$, 
$C_{10}^{\rm eff}$ (for their definition in terms of $C_i$ in Eq.(\ref{Heff}) see \cite{Altmannshofer:2008dz}), in particular we have
\bea
    {\cal M}(b\to s \ell^+\ell^-)&=&\frac{G_F\alpha}{\sqrt{2}\pi}V^*_{ts} V_{tb}
    \Big\{C_9^{\rm eff}(\hs) \left[\bar{s}_L\gamma_{\mu} b_L\right]\left[\bar{\ell}\gamma^{\mu}\ell\right]\nonumber +  C^{\eff}_{10} \left[\bar{s}_L\gamma_{\mu} b_L\right]\left[\bar{\ell}\gamma^{\mu}\gamma_5\ell\right]\nonumber\\
    &-&2i \frac{C_7^{\rm eff}}{s}\left\{
    m_b\left[\bar{s}_L\sigma_{\mu\nu}q^{\nu}b_R\right]
    +
    m_s\left[\bar{s}_R\sigma_{\mu\nu}q^{\nu}b_L\right]\right\}
    \left[\bar{\ell}\gamma^{\mu}\ell\right]
    \Big\}
    \label{Mb} 
    \eea 
    where $s=q^2$, $q=p^++p^-$, $\sigma_{\mu\nu}=i/2\left[\gamma_{\mu},\gamma_{\nu}\right]$, and $\psi_{L/R}\equiv (1\mp \gamma_5)/2 \psi$, with $\psi$ the corresponding Dirac spinor in momentum space. Terms in parenthesis $[\cdots ]$ in Eq.(\ref{Mb}) stand for the usual bi-spinorial matrix elements in momentum space, where sum over spin and color indices (for the quark spinors)  is understood. The last term in the amplitude in Eq.(\ref{Mb}) comes from the photon exchange between the contribution of the $\Delta B=1$ matrix element of magnetic-dipole operator $Q_7$ and the tree-level EM current $\bar{\ell}\gamma_{\mu} \ell$. There, we have also retained the contribution of the Flavor Changing (FC) magnetic-dipole operator proportional to the strange-quark mass $m_s$.
All along the paper we will use the results for the effective 
Wilson coefficients $C_{7,9,10}^{\eff}(\mu)$ computed at the NNLO as provided in \cite{Altmannshofer:2008dz}, and evaluated at the renormalization $\mu=m_b$ scale, with $m_b$ corresponding to the b-quark pole mass. For their numerical values see table \ref{tabinput1}. Notice that only the  $C_{9}^{\eff}(\hs)$ has a $q^2$ dependence (or analogously $\hs$), due to the inclusion of the matrix elements of operators in its definition \cite{Altmannshofer:2008dz}. We removed the $\mu$ dependence from all $C_{i}^{\eff}$,  while  retained the $\hs$ dependence only in  $C_{9}^{\eff}$.

Now, we analyze the QED radiative corrections. Since we are mainly interested in analyzing the effect of a specific class of virtual corrections which are manifestly gauge-invariant as well as LF non-universal, we restrict our choice to the selected contribution of the magnetic-dipole corrections into the final lepton pair $\ell^+\ell^-$. A complete treatment of the full EM radiative corrections on the decay widths of $b\to s e^+e^-$, that would require the computation of all virtual corrections and real photon emissions at 1-loop, goes beyond the purpose of the present paper.

We start by substituting the tree-level vertex $\gamma^{\mu}$ appearing in the matrix element of the leptonic current $[\bar{l}\gamma^{\mu} l]$, with the full vertex $\Gamma^{\mu}(q^2)$ as follows
\bea
\Gamma^{\mu}(q^2)=\gamma^{\mu} F_1(q^2) + iF_2(q^2) \sigma^{\mu\nu}\hat{q}_{\nu}\,  ,
\label{Gamma}
\eea
where $\hat{q}=q/m_b$ and so $F_2(q^2)$ turns out to be a dimensionless form factor. In order to isolate the magnetic-dipole contribution, we retain only the $F_2$ term in Eq.(\ref{Gamma}) and set $F_1\to 1$.
The form factor $F_1(q^2)$ is almost LF universal. Its main LFU-violating contributions are contained in the terms proportional to $\alpha\log{(m_B/m_l)}$. The contribution of these large log terms, from virtual corrections and real emission have been consistently included and resummed in the analysis of \cite{Bordone:2016gaq}.

The contribution from the magnetic-dipole form factor $F_2(q^2)$ is a gauge invariant and IR-safe observable, and it is manifestly flavor non-universal being proportional to the lepton mass. Moreover, it does not vanish in the $q^2\to 0$. Indeed, the $q^2\to 0$ limit of $F_2(q^2)$ is related to the well-known contribution to the anomalous magnetic moment $(g-2)_l$. Moreover, as we will show in section 3, the $F_2$ correction does not factorize in the $R_K$ and $R_{K^*}$ observables and it is dominant in the $q^2$ regions close to the dilepton mass thresholds. More details about this issue can be found in section 3.

In the following, we will change the argument dependence $F_2(q^2) \to F_2(\hs)$ where we have defined the symbol $\hat{s}\equiv q^2/m_b^2$.
In QED, at one-loop the $F_2(\hs)$ expression, for positive $q^2>0$ is given by
\bea
F_2(\hs)&=&\frac{\alpha }{2\pi}\frac{\sqrt{r_{\ell}}} {\sqrt{\hat{s}\left(\hat{s}-4r_{\ell}\right)}}
  \log{\left(\frac{2 r_{\ell}-\hat{s}+
      \sqrt{\hat{s}\left(\hat{s}-4r_{\ell}\right)}}{2r_{\ell}}\right)}
  \label{F2}
\eea
where $r_\ell\equiv m_{\ell}^2/m_b^2$. In the limit $q^2\to 0$, or analogously
$\hs\to 0$, the $F_2(\hs)$ reproduces the well-known result for the anomalous-magnetic moment correction to $g-2$, in particular at one-loop we have
  \bea
\frac{1}{m_b}\lim_{\hs\to 0} F_2(\hs)&=&\frac{\alpha}{4 \pi m_{\ell}}\, .
\eea

The Feynman diagrams for the SM amplitude of $b\to s\ell^+\ell^-$, including the magnetic-dipole contributions, are shown in Fig.[1].
The corresponding total decay width can be decomposed as
\bea
\Gamma(b\to s\ell^+\ell^-) &=& \Gamma^b_0+\Gamma^b_M\, ,
\label{Gammabll}
\eea
where  $\hat{u}=u/m_b^2$, with $u=2((p_b\cdot p^+)-(p_b\cdot p^-))$, and  $\Gamma^b_0$ include the differential contribution to the width of the square amplitude of the SM without the ${\cal O}(\alpha)$ magnetic-dipole corrections.
The $\Gamma^b_M$ absorbs the contributions of both the interference of the ${\cal O}(\alpha)$ magnetic-dipole amplitude with the zero order in the SM and its square term. Although the latter is of order ${\cal O}(\alpha^2)$, for completeness we included it in our analysis. The reason is because this contribution has a higher infrared singularity at small $q^2$ and it could in principle give a potential $m_B^2/m_{\ell}^2$ enhancement in the rate, although suppressed by a higher power of $\alpha$. Eventually, we will see that its effect is tiny and can be fully neglected in the analysis.

After computing the square amplitude and summing over polarizations, 
the corresponding  expressions for the differential width are given by
\bea
\frac{d^2{\Gamma}^b_0}{d\hat{s}\,d\hat{u}} &=& \hat{\Gamma}\Big[\left|C_9^{\rm eff}(\hs)\right|^2K^{9,9}+
  \left(C^{\eff}_{10}\right)^2K^{10,10}+\left(C^{\rm eff}_{7}\right)^2K^{7,7}\nonumber\\
  \nonumber\\
  &+&  C^{\rm eff}_{7}{\rm Re}\left[C_9^{\rm eff}(\hs)\right]K^{7,9}
 + C^{\eff}_{10}\Big(C^{\rm eff}_{7}K^{7,10}+{\rm Re}\left[C_9^{\rm eff}(\hs)\right]K^{9,10}\Big)
\Big]\, ,
\\
\frac{d^2{\Gamma}^b_M}{d\hat{s}\,d\hat{u}} &=&\hat{\Gamma}\Big[C^{\rm eff}_{7}\Big(
  {\rm Re}\left[C_9^{\rm eff}(\hs)F_2^{*}(\hs)\right]
  M^{7,9}+C_{10}^{\rm eff}{\rm Re}\left[F_2(\hs)\right]M^{7,10}\Big)\nonumber\\
  &+&\left(C^{\rm eff}_{7}\right)^2
  \left({\rm Re}\left[F_2(\hs)\right]M_1^{7,7}+
  \left|F_2(\hs)\right|^2M_2^{7,7} 
  \right)
  \Big]
\label{ddwidthb}
\eea
where $\hat{\Gamma}=\frac{G_F^2}{512\pi^5}m_b^5|V_{ts}^*V_{tb}|^2 \alpha^2$.
Above, we used the property that only $C_9^{\rm eff}(\hs)$ is complex.

Then, by retaining all mass corrections, the coefficients $K^{i,j}$ are given by
\bea
K^{7,7}&=&
\frac{2}{\hat{s}^2} \Big(4 r_{\ell}
\left(1 - \hs - r_s (1 + r_s - r_s^2 + (6 + r_s) \hs)\right)\nonumber\\
&+&\hat{s} \left(
1 - r_s - r_s^2 + r_s^3 - 8 r_s \hs - \hs^2 - r_s \hs^2 + (1 + r_s) \hu^2\right)
\Big)
\nonumber\\
K^{9,9}&=&\frac{1}{2} \left(4 r_{\ell}
   (r_s-\hat{s}+1)+r_s^2-2
   r_s-\hat{s}^2-\hat{u}^2+1\right)
   \nonumber\\
K^{10,10}&=&   \frac{1}{2} \left(-4 r_{\ell}
   (r_s-\hat{s}+1)+r_s^2-2
r_s-\hat{s}^2-\hat{u}^2+1\right)
\nonumber\\
K^{7,9}&=&  \frac{4}{\hat{s}} (2 r_{\ell}+\hat{s})
(1-\hs  - r_s (2 + \hs)  +r_s^2)
\nonumber\\
K^{7,10}&=& 4 \hat{u}(1+r_s)
\nonumber\\
K^{9,10}&=&2 \hat{s} \hat{u}
\label{Kij}
\eea
and for the $M^{ij}$ coefficients we have
\bea
M^{7,7}_1&=&
\frac{8 \sqrt{r_{\ell}}}{\hat{s}}
\Big(2 + 2 r_s^3 - \hs - \hs^2 - r_s^2 (2 + \hs) - r_s (2 + 14 \hs + \hs^2)\Big)
\nonumber\\
M^{7,7}_2&=&\frac{8 r_{\ell}}{\hat{s}} \Big(
1 - r_s^2 + r_s^3 - \hs^2 - r_s (1 + 8 \hs + \hs^2)\Big)\nonumber\\
&+&2
\Big(1 + r_s^3 - \hs - r_s^2 (1 + \hs) - \hu^2 - r_s (1 + 6 \hs + \hu^2)\Big)
\nonumber\\
M^{7,9}&=&12 \sqrt{r_{\ell}}\, (1 + r_s^2 - \hs - r_s (2 + \hs))
\nonumber\\
M^{7,10}&=& 8 \sqrt{r_{\ell}}\, \hat{u}(1+r_s)\, ,
\label{Mij}
\eea 
where $r_s=m_s^2/m_b^2$.  

The kinematic region for the variables $\hu$ and $\hs$ is \cite{Ali:1999mm}
\bea
4\rl \le &\hs&\le(1-\sqrt{r_s})^2\\
-\bu\le &\hu& \le \bu\, .
\eea
with $\bu=\sqrt{\lambda\left(1-\frac{4r_{\ell}}{\hat{s}}\right)}$
and $\lambda \equiv 1 + r_s^2 + \hat{s}^2 - 2\hat{s} - 2r_s\left(1 + \hat{s}\right)$.
The variable $\hu$ corresponds to the $\theta$ angle between the momentum of the $b$ quark and the antilepton $\ell^+$ in the dilepton center of mass system frame, expressed in this frame by the relation $\hu=-\bu\cos{\theta}$.

After integrating in $\hat{u}$ on its whole kinematic region, the corresponding distributions $d\Gamma^b/d\hat{s}$ can be obtained from Eq.(\ref{ddwidthb}) by replacing the $K^{i,j}\to \hat{K}^{i,j}$ and $M^{i,j}\to \hat{M}^{i,j}$, where
$\hat{K}^{i,j},~\hat{M}^{i,j}$ are
\bea
\hat{K}^{7,7}&=&\frac{8\bu}{3
  \hat{s}^2} (\hat{s}+2\rl)
\Big(2 (1 - r_s)^2 (1 + r_s) - \hs - r_s (14 + r_s) \hs - (1 + r_s) \hs^2\Big)\, ,
\nonumber\\
\hat{K}^{9,9}&=&\frac{2 \bu}{3 \hat{s}} (\hat{s}+2\rl)
   \left(r_s^2+r_s (\hat{s}-2)-2
   \hat{s}^2+\hat{s}+1\right)\, ,
   \nonumber\\
\hat{K}^{10,10}&=&\frac{2 \bu }{3 \hat{s}}\Big(2 r_{\ell} \left(4 \hat{s}^2-5 (r_s+1)   \hat{s}+(r_s-1)^2\right)+\hat{s}
   \left(r_s^2+r_s (\hat{s}-2)-2
   \hat{s}^2+\hat{s}+1\right)\Big)\, ,
\nonumber\\
\hat{K}^{7,9}&=&\frac{8 \bu}{\hat{s}}
(\hat{s}+2\rl)
(1 + r_s^2 - \hs - r_s (2 + \hs))\, ,
\label{Ksij}
\\
\nonumber
\\
\hat{M}^{7,7}_1&=&\frac{16 \sqrt{r_{\ell}}\,\bu }{\hat{s}}
\Big(
2 (1 - r_s)^2 (1 + r_s) - \hs - r_s (14 + r_s) \hs - (1 + r_s) \hs^2\Big)
\nonumber\\
\hat{M}^{7,7}_2&=&\frac{4 \bu}{3
  \hat{s}} (8 r_{\ell}+\hat{s})\Big(
2 (1 - r_s)^2 (1 + r_s) - \hs - r_s (14 + r_s) \hs - (1 + r_s) \hs^2\Big)
\nonumber\\
\hat{M}^{7,9}&=&24 \sqrt{r_{\ell}}\,\bu
\Big((1 - r_s)^2 - (1 + r_s) \hs\Big)
\label{Msij}
\eea
while $\hat{K}^{7,10}=\hat{K}^{9,10}=\hat{M}^{7,10}=0$.
The results in Eqs.(\ref{Kij}),(\ref{Ksij}) for the SM contribution without magnetic-dipole corrections, agree with the corresponding ones in \cite{Gabrielli:1998sw} in the $r_s\to 0$ limit.

Now we compute the effect induced by this correction on the total integrated branching ratios. Results will be obtained by using the values of masses and other  SM inputs reported in table \ref{tabinput1}.
\begin{table}\centering
\begin{tabular}{LS[table-format = +1.11e+1]|LS[table-format = +1.7e+1]}
\toprule
m_{B^0}  &5.27964  & \tau_{B^0} & 1.519\hspace{-2cm} $\times10^{-12}$   \\
m_{B^+}  &5.27933  &  \tau_{B^+} & 1.638\hspace{-2cm} $\times10^{-12}$ \\
m_{K^*}  &0.89176  &  C_7(M_W) & 0.139\\
m_{K^0}  &0.497611 &  C_7^{\eff}(\mu)&  -0.304\\
m_{K^+}  &0.493677 &   C^{\eff}_9(\mu) &  2.211\\
m_b     &4.8      &  C^{\rm eff}_{10}(\mu)    &  -4.103\\
m_c     &1.7      &  \alpha(\mu) &  \multicolumn{1}{C}{1/129}\hspace{2cm}\\
m_s     &0.095    &  |V_{ts}^\ast V_{tb}| &  0.0401\\
\bottomrule
\end{tabular}
\caption[]{
  Numerical inputs for the relevant parameters used in the analysis.
Central values of meson and quark masses $m_i$ and mean lifetimes $\tau_i$ are expressed in GeV and seconds units respectively \cite{Tanabashi:2018oca}. The quark masses $m_{b}$ and  $m_{c}$ correspond to the pole masses of bottom and charm quarks respectively. Effective Wilson coefficients at the NNLO $C^{\eff}_{7,9,10}(\mu)$ (from \cite{Altmannshofer:2008dz}) and EM fine structure constant $\alpha(\mu)$ are intended to be evaluated at the scale $\mu=m_b$. The SM $C_7(M_W)$ corresponds to $m_t=170\GeV$ \cite{Buchalla:1995vs}.}
\label{tabinput1}
\end{table}
The total branching ratio for a particular lepton state $\ell$ is obtained by integrating $\frac{d^2{\Gamma}^b}{d\hat{s}\,d\hat{u}}$ over the entire kinematically-allowed range for that $\ell$. In particular, neglecting the strange quark mass, the $\hat{s}$ range of integration  is $4r_{\ell} <\hat{s}<1$.

To avoid intermediate charmonium resonances and non-perturbative phenomena near the end point, we integrate over a particular range of $\hat{s}$. Following for instance the prescription in \cite{Cho:1996we}, for $\ell=e,\mu$ case we have 
\bea
\left(m_b^2\hat{s}\right) &\in&
\left\{4m_{\ell}^2, (2.9\, {\rm GeV})^2\right\}
\cup
\left\{(3.3{\rm GeV})^2 , (3.6\, {\rm GeV})^2\right\}\cup
\left\{(3.8{\rm GeV})^2 , (4.6\, {\rm GeV})^2\right\}
\label{range-emu}
\eea
while for $\ell=\tau$ we get
\bea
\left(m_b^2\hat{s}\right) &\in&
\left\{4m_{\tau}^2, (3.6\, {\rm GeV})^2\right\}
\cup
\left\{(3.8{\rm GeV})^2 , (4.6\, {\rm GeV})^2\right\}\, .
\label{range-tau}
\eea
\noindent
In order to reduce the uncertainties in the $b\to s \ell^+\ell^-$ partial width it is customary to normalizing the  width to the inclusive semileptonic B decay
$B\to X_ce^+\nu$, that is given by
\bea
\Gamma(b\to c e^+\nu)&=&\frac{G_F^2 m_b^2 |V_{cb}|^2}{192 \pi^3}
g\left(\frac{m_c}{m_b}\right)\left\{1-\frac{2\alpha_s(m_b)}{3\pi}\left[\left(\pi^2 -\frac{31}{4}\right)\left(1-\frac{m_c}{m_b}\right)^2+\frac{3}{2}\right]\right\}\, 
\eea
where $g(z)=1-8z^2+8z^6-z^8-24z^4\log{z}$, $\alpha_S(m_b)$ is the strong coupling evaluated at the $m_b$ mass. The bottom and charm masses entering above are understood as pole masses. Then the branching ratio is obtained as
\bea
\BR(b\to s \ell^+\ell^-)&=& \BR^{\rm exp}(B\to X_ce^+\nu) \frac{\Gamma(b\to s \ell^+\ell^-)}{\Gamma(b\to c e^+\nu)}\, ,
\eea
with the measured $\BR^{\rm exp}(B\to X_ce^+\nu)=(10.64\pm 0.17\pm 0.06)\%$ \cite{Aubert:2009qda}.

Now, it is useful to decompose the total BR as follows
\bea
{\BR}^{\ell}=&{\BR}^{\ell}_0\left(1+\Delta^{\ell}\right)\, ,
\eea
where $\Delta^{\ell}$ absorbs the ${\cal O}(\alpha)$ effect of the magnetic-dipole corrections, while $\BR^{\ell}_0$ is the leading contribution, without these corrections.
In table \ref{tab1} we report the corresponding results for the
total BR integrated over the various bin regions of $m_b^2\hat{s}$, and on the total range as provided in Eq.(\ref{range-emu},\ref{range-tau}), where regions $R_i$ stand for $R_1\cup R_2\cup R_3$ in the case of $\ell=e,\mu$ and $R_1\cup R_2$ for $\ell=\tau$. 
The $\BR_0^l$ values correspond to the central values of $\BR^{\rm exp}(B\to X_ce^+\nu)$ \cite{Aubert:2009qda} and $|V_{cb}|=4.22\times 10^{-2}$ \cite{Tanabashi:2018oca}.

\begin{table} \begin{center}    
\begin{tabular}{|c||c|c|c|c|c|c|}
\hline 
$\hat{s}$ bins
& $10^7\cdot \BR^{e}_0$
& $\Delta^{e}$ 
& $10^7\cdot \BR^{\mu}_0$
& $\Delta^{\mu}$ 
& $10^7\cdot \BR^{\tau}_0$
& $\Delta^{\tau}$ 
\\ \hline
${\rm R}_1$
& $60.0$
& $-1.4 \times10^{-4}$
& $35.8$
& $\,\,-1.8 \times10^{-4}$
& $~0.071$
& $2.8\times10^{-4}$
\\ \hline
${\rm R}_2$
& $4.10$
& $~~~\,3.5\times 10^{-10}$
& $4.09$
& $~~~\,5.1\times 10^{-6}$
& $\,2.03$
& $2.5\times10^{-4}$
\\ \hline
${\rm R}_3$
& $4.12$
& $~~~\,2.2\times 10^{-10}$
& $\,4.12$
& $~~~\,3.3\times 10^{-6}$
& --
& --
\\ \hline
$\sum_i {\rm R}_i$
& $68.2$
& $-1.2 \times10^{-4}$
& $44.0$
& $\,\,-1.5 \times10^{-4}$
& $\,2.10$
& $\,2.6\times 10^{-4}$
\\ \hline 
\end{tabular} 
\caption[]{
  Results for the total branching ratio $\BR^{\ell}$ of
  $B\to X_s \ell^+\ell^-$, integrated on the various $R_i$ bins of $\hat{s}$ as defined in the text, where
  $\BR^{\ell}=\BR_0^{\ell}(1+\Delta^{\ell})$ for $\ell=e,\mu,\tau$, and $\Delta^{\ell}$ includes the magnetic-dipole corrections.
}
\label{tab1}
\end{center} \end{table}

In analogy with the strategy adopted in the $B\to (K,K^*) \ell^+\ell^-$ decays (see next section) for analyzing the LFU, we consider the ratios $R_b^{\mu,\tau}$ for the $b\to s \ell^+\ell^-$ decays defined as
\bea
R^{\ell}_{b}&=&\ddfrac{\int_{\hat{s}_{\rm min}}^{\hat{s}_{\rm max}}
  \ddfrac{d\Gamma(b\to s\ell^+\ell^-)}{d\hat{s}}d\hat{s}}
{\int_{\hat{s}_{\rm min}}^{\hat{s}_{\rm max}}
  \ddfrac{d\Gamma(b\to s e^+e^-)}{d\hat{s}}d\hat{s}}\, ,
\eea
where $\ell=\mu,\tau$.
At this purpose is convenient to define the deviation $\Delta_R^{\ell}$ as
\bea
R_b^{\ell}&=&\bar{R}_{b}^{\ell}\left(1+\Delta_R^{\ell}\right)
\eea
where $\Delta_R^{\ell}$ absorbs here the contribution of the magnetic-dipole correction.
The results are reported below for two representative integrated bin regions of $\hat{s}$ close to the lepton mass thresholds, in particular for the muon case
\bea
\bar{R}_b^{\mu}&=& 0.849 \,\, , ~~~\Delta_R^{\mu}\,=\,-1.1\times 10^{-3} \, ,~~~~~ (m_b^2 \hat{s})\, \in \left\{4m_{\mu}^2, (0.5\, {\rm GeV})^2\right\}\nonumber \\
\bar{R}_b^{\mu}&=& 0.928 \,\, , ~~~\Delta_R^{\mu}\,=\,-6.5\times 10^{-4} \, ,~~~~~ (m_b^2 \hat{s})\, \in \left\{4m_{\mu}^2, (1\, {\rm GeV})^2\right\}\nonumber \\
\bar{R}_b^{\mu}&=& 0.979 \,\, , ~~~\Delta_R^{\mu}\,=\,-1.7\times 10^{-4} \, ,~~~~~ (m_b^2 \hat{s})\, \in \left\{4m_{\mu}^2, (2.9\, {\rm GeV})^2\right\}\, , 
\eea
while for the $\tau$ lepton we get 
\bea
\bar{R}_b^{\tau}&=& 0.117 \,\, , ~~~\Delta_R^{\tau}\,=\,3.3\times 10^{-5} \, ,~~~~~ (m_b^2 \hat{s})\, \in \left\{4m_{\tau}^2, (3.6\, {\rm GeV})^2\right\}\, .
\eea
As we can see from these results, the corrections induced by the magnetic-dipole contributions to the $R^{\mu,e}_b$ ratios, do not exceed the 1 per mille effect in the first bin, while it is two order of magnitude in the tau lepton case. The smallness of the contribution is somehow expected, since the contribution of the flavor-changing magnetic dipole operator $Q_7$ to the amplitude is not dominant in the $b\to s \ell^+\ell^-$ process. Moreover, the interference between the magnetic-dipole correction term and the rest of the amplitude is chiral suppressed. In conclusion, these results show that the expected chiral enhancement induced by the magnetic form factor $F_2$ in Eq.(\ref{F2}) near the mass threshold ($\hs\to 4r_{\ell}$), can only partially compensate the chiral suppression induced by the interference, when integrated over all the kinematic region in Eqs.(\ref{range-emu}),(\ref{range-tau}).

From these results, we can see that UV new physics contributions to the $R^{\ell}_b$ ratios, providing new short-distance corrections to the magnetic-dipole form factor $F_2$, like for instance the large effects expected in technicolor models \cite{Appelquist:2004mn}, turn out to be chiral suppressed in all range of $q^2$, and also negligible with respect to the leading QED corrections to $F_2$. The reason is because the short-distance contributions to $F_2$ are independent of $q^2$ and so they do not provide any infrared enhancement at low $q^2$ to compensate for the associated chiral suppression of the magnetic-dipole corrections to the $b\to s \ell^+\ell^-$ rates. The same conclusions hold for the analogous contributions to the $R_{K,K^*}$ observables.
\section{Magnetic-dipole corrections to $B\to (K^*,K) \ell^+\ell^-$ } 
Here, we analyze the contributions induced by the magnetic-dipole corrections to the final lepton pair, in the exclusive $B$ meson decays $B\to (K^*,K) \ell^+\ell^-$. We parametrize the momenta of the generic decay as
\bea
B(p_B)\to M (p_K) \, \ell^+(p^+)\, \ell^-(p^-)
\eea
where $M$ stands for $K$ or $K^*$ and in parenthesis are reported the corresponding momenta.
As mentioned in the introduction, the $B\to K^* \ell^+\ell^-$ decay is characterized by an enhancement of the long-distance contributions induced by the photon-pole $1/s$ coming from the $b\to s \gamma^*$ transitions (with $\gamma^*$ standing for a virtual photon), proportional to the effective Wilson coefficient  $C_7^{\rm eff}$. This enhancement is absent in the b-quark decay $b\to s \ell^+\ell^-$ as well as in the exclusive $B\to K \ell^+\ell^-$ decay.
In particular, for $s< 1$ GeV, the photon-pole gives the dominant contribution to the rate and it still contributes about $30\%$ around $s\approx 3 {\rm GeV}^2$. This is mainly due to the longitudinal polarizations of the vector meson $K^*$, which can contribute to the rate with enhancement factors proportional to $\mB^2/\mKs^2 \sim 25$. For this reason, we expect the magnetic-dipole corrections to $B\to K^* \ell^+\ell^-$ to be larger than in the $B\to K \ell^+\ell^-$ channel. Here we will evaluate the impact of these corrections in the corresponding decay widths and on the $R_{K^*,K}$ ratios. 

\subsection{Decay width for $B\to K^*\ell^+\ell^-$}

We start by fixing the notation for the following two kinematic variables
\bea
s&=&q^2\,=\,(p^++p^-)^2\,
\nonumber\\
u&=&(p_B-p^-)^2-(p_B-p^+)^2\, ,
\label{variables}
\eea
together with the corresponding dimensionless ones $\hs=s/m_B^2$ and $\hu=u/m_B^2$.
Following the notations of Ref. \cite{Ali:1999mm}, the corresponding amplitude can be simply obtained from the one in Eq.(\ref{Mb}), by replacing the bi-spinorial quark products appearing in Eq.(\ref{Mb}) with the corresponding B meson matrix elements \cite{Ali:1999mm}
\bea
2\langle K^*(\pK) |\left[\bar{s}_L\gamma_{\mu}b_R\right] |B(\pB)\rangle
&=& -i\varepsilon^{\dag}_{\mu}\left(\mB+\mKs\right)A_1(s)
\,+\,i\left(\pB+\pK\right)_{\mu}\left(\varepsilon^{\dag}\cdot \pB\right)\frac{A_2(s)}{m_B+m_K^*}\nonumber\\
&+&iq_{\mu}\left(\varepsilon^{\dag}\cdot \pB\right)\frac{2m_K^*}{s}\left(A_3(s)-A_0(s)\right)\nonumber\\
&+&\epsilon_{\mu\nu\alpha\beta}\,\varepsilon^{\dag \nu}\pB^{\alpha}\pK^{\beta}
\frac{2V(s)}{\mB+\mKs}\, ,
\eea
and 
\bea
2\langle K^*(\pK) |\left[\bar{s}_L\sigma_{\mu\nu} q^{\nu}b_L\right] |B(\pB)\rangle &=& i2\epsilon_{\mu\nu\alpha\beta}\, \varepsilon^{\dag \nu} \pB^{\alpha}\pK^{\beta} T_1(s)\nonumber\\
&+&T_2(s)\left[\varepsilon^{\dag}_{\mu}\left(\mB^2-\mKs^2\right)-
  \left(\varepsilon^{\dag}\cdot \pB\right)\left(\pB+\pK\right)_{\mu}\right]
\nonumber\\
&+&T_3(s)\left(\varepsilon^{\dag}\cdot \pB\right)\left[q_{\mu}-
  \frac{s}{\mB^2-\mKs^2}\left(\pB+\pK\right)_{\mu}\right]\, ,
\label{KstarME}
\eea
where $A_{0,1,23}(s)$, $V(s)$ and $T_{1,2,3}(s)$ are form factors which depend on $s$, with the convention $\epsilon_{0123}=+1$ for the Levi-Civita tensor. The following exact relations hold for the form factors
\bea
A_3(s)&=&\frac{\mB+\mKs}{2\mKs}A_1(s)-\frac{\mB-\mKs}{2\mKs}A_2(s)\\
\nonumber
A_0(0)&=&A_3(0)\, ,\, T_1(0)=T_2(0)\, .
\eea
We have used the updated light-cone sum rules (LCSR) approach of Ref. \cite{Straub:2015ica} to evaluate the form factors.  In particular, for a generic transition 
$B\to V f\bar{f}$, with $V$ a vector meson, the generic hadronic form factor $F_i$ can be decomposed as \cite{Straub:2015ica}
\bea
F_i(s)&=&P_i(s)\sum_k \alpha_k^i\left[z(s)-z(0)\right]^k
\eea
where the variable $z(t)$ is defined as
\bea
z(t)&=&\frac{\sqrt{t_+-t_-}-\sqrt{t_+-t_0}}{\sqrt{t_+-t_-}+\sqrt{t_+-t_0}}
\eea
where $t_{\pm}=(\mB\pm m_{\scriptscriptstyle{V}}$, and $t_0\equiv t_{+}(1-\sqrt{1-t_-/t_+})$, and $P_i(s)=1/(1-s/m_R^i)$ is a simple pole corresponding to the first resonance in the spectrum. Corresponding values for the parameters $\alpha_k^i$ for the present process can be found in \cite{Straub:2015ica} and in table \ref{tabinput2}.

Following the notation of Ref.\cite{Ali:1999mm}, we rewrite the amplitude for the process  $B\to K^* \ell^+\ell^-$ in a compact way as
\bea
    {\cal M}&=& \frac{G_F\alpha}{2\sqrt{2}\pi}\Big\{M_{\mu}^1
    \left[\bar{\ell}\gamma^{\mu}\ell\right]+M_{\mu}^2 \left[\bar{\ell}\gamma^{\mu}\gamma_5\ell\right]+M_{\mu}^3\left[\bar{\ell}\sigma^{\mu\nu}\hat{q}_{\nu}\ell\right]
    \Big\}
\label{MBKS}
    \eea
    where the last term includes the magnetic-dipole corrections. The terms $M^i_{\mu}$ are given by
\bea
M^1_{\mu}&=&A(\hs) \epsilon_{\mu\rho\alpha\beta}\, \varepsilon^{\dag \rho}
\hat{p}_B^{\alpha}\hat{p}_K^{\beta} -iB(\hs)\varepsilon^{\dag}_{\mu}+iC(\hs)
\left(\varepsilon^{\dag}\cdot \hat{p}_B\right)\left(\hat{p}_B+\hat{p}_K\right)
+iD(\hs)\left(\varepsilon^{\dag}\cdot \hat{p}_B\right)\hat{q}_{\mu}
\nonumber\\
M^2_{\mu}&=&E(\hs) \epsilon_{\mu\rho\alpha\beta}\, \varepsilon^{\dag \rho}
\hat{p}_B^{\alpha}\hat{p}_K^{\beta} -iF(\hs)\varepsilon^{\dag}_{\mu}+iG(\hs)
\left(\varepsilon^{\dag}\cdot \hat{p}_B\right)\left(\hat{p}_B+\hat{p}_K\right)
+iH(\hs)\left(\varepsilon^{\dag}\cdot \hat{p}_B\right)\hat{q}_{\mu}
\nonumber\\
M^3_{\mu}&=&\bar{A}(\hs) \epsilon_{\mu\rho\alpha\beta}\, \varepsilon^{\dag \rho}
\hat{p}_B^{\alpha}\hat{p}_K^{\beta} -i\bar{B}(\hs)\varepsilon^{\dag}_{\mu}+
i\bar{C}(\hs)
\left(\varepsilon^{\dag}\cdot \hat{p}_B\right)\left(\hat{p}_B+\hat{p}_K\right)
+i\bar{D}(\hs)\left(\varepsilon^{\dag}\cdot \hat{p}_B\right)\hat{q}_{\mu}
\eea
where $\hs=s/m_B^2$, $\hat{p}_{K,B}=p_{K,B}/m_B$, $\hat{q}=q/m_B$, $\hat{m}_b=m_b/m_B$ and 
\bea
A(\hs)&=&\frac{2}{1+\hmKs}C_9^{\eff}(\hs)V(s)+\frac{4\hmb}{\hs} C_7^{\eff} T_1(s)
\nonumber\\
  B(\hs)&=&(1+\hmKs)\left[C_9^{\eff}(\hs)A_1(s)+
    \frac{2\hmb}{\hs}\left(1-\hmKs\right)C_7^{\eff}T_2(s)\right]
\nonumber\\ 
C(\hs)&=&\frac{1}{1-\hmKs^2}\left[ (1-\hmKs)C_9^{\eff}(\hs)A_2(s)+
  2\hmb C_7^{\eff} \left(T_3(s)+\frac{1-\hmKs^2}{\hat{s}}T_2(s)\right)\right]
\nonumber\\
D(\hat{s})&=&\frac{1}{\hat{s}}\left[C_9^{\eff}(\hs)\Big((1+\hmKs)A_1(s)-
  (1-\hmKs)A_2(s)-2\hmKs A_0(s)\Big)\right.\nonumber\\
  &-&\left. 2\hmb C_7^{\eff}T_3(s)\right]
\nonumber\\
E(\hat{s})&=&\frac{2}{1+\hmKs}C^{\eff}_{10}V(s)
\nonumber\\
F(\hat{s})&=&(1+\hmKs)C^{\eff}_{10}A_1(s)
\nonumber\\
G(\hat{s})&=&\frac{1}{1+\hmKs}C^{\eff}_{10}A_2(s)
\nonumber\\
H(\hat{s})&=&\frac{1}{\hs}C^{\eff}_{10}\left[(1+\mKs)A_1(s)-(1-\hmKs)A_2(s)-2\hmKs A_0(s)\right]\, ,
\label{formfactors}
\eea 
where $\hmKs\equiv \mKs/\mB$.
The quantities with bar $\bar{I}(\hat{s})\equiv \lim_{C_9^{\eff}\to 0}\left\{I(\hat{s})\right\}$, with
$I=A,B,C,D$.

In analogy with the notation adopted in Eq.(\ref{Gammabll}), we decompose the expression for the corresponding decay width $\Gamma^{K^*}$ as
\bea
\Gamma^{K^*}&=&\Gamma^{K^*}_0+\Gamma^{K^*}_M
\eea
where as usual $\Gamma^{K^*}_0$ includes the SM results at the zero order in the magnetic-dipole corrections, while $\Gamma^{K^*}_M$ absorbs the terms containing the interference and square of the amplitude containing the magnetic-dipole corrections with the rest of the zero order amplitude.

The analytical expressions for the differential distributions $\frac{d^2{\Gamma}^{K^*}_0}{d\hat{s}\,d\hat{u}}$ and
$\frac{d{\Gamma}^{K^*}_0}{d\hat{s}}$ can be found in \cite{Ali:1999mm}, where we agree with the corresponding results. Below we provide the new expressions for the new contributions induced by the magnetic dipole corrections, in particular
\bea
\frac{d^2{\Gamma}^{K^*}_M}{d\hat{s}\,d\hat{u}}&=&
\frac{\hat{\Gamma}_B}{\rKs}
\Big\{\sqrt{\rl}\Big[\Re{F_2\bar{A}A^*}
2\rKs\lambda\hs
+\Re{F_2\bar{B}B^*}\left(\lambda+12 \rKs \hs\right)
+\Re{F_2\bar{C}C^*}\lambda^2
   \nonumber\\
   &-&\Re{F_2\left(\bar{A}F^*+\bar{B}E^*\right)}
   4\rKs\hs\hu
+ \Re{F_2\left(\bar{B}C^*+\bar{C}B^*\right)}
\left(\rKs +\hs -1\right)\lambda
\Big]
\nonumber\\
&+&\frac{|F_2|^2}{4}\Big[
|\bar{A}|^2 \rKs \hs \left(\lambda (4 \rl + s) - \hs \hu^2\right)
+|\bar{C}|^2\lambda \left(4 \lambda \rl + \hs \hu^2\right)
\nonumber\\
&+&
|\bar{B}|^2
\left(4 \rl (\rKs^2 + (\hs -1 )^2 + \rKs (6 \hs -2)) + \hs (4 \rKs \hs + \hu^2)\right)
\nonumber\\
&+&2\Re{\bar{B}\bar{C}^*}\left(
  \left( \hs -1 - \rKs^2 + \rKs (2 + 3 \hs)\right) \hu^2 + \lambda (4 \rl ( \hs -1 + \rKs ) + \hu^2\right)
\Big]
\Big\}
\label{dGammaKsu}
\eea
where 
$\hat{\Gamma}_B=G_F^2\alpha^2m_B^5 |V_{tb}^* V_{ts}|^2/(2^{11} \pi^5)$, with $\alpha$ evaluated at the $B$ meson scale, and
\bea
\lambda&\equiv& 1+\rKs^2+\hs^2-2\hs-2\rKs(1+\hs)\, .
\eea
with $\rKs\equiv \mKs^2/\mB^2$. For practical purposes, we omitted the $\hs$ dependence inside the
form factors of Eq.(\ref{formfactors}).
As expected by chirality arguments, the interference terms (proportional to  the $F_2$ form factor) in Eq.(\ref{dGammaKsu}), are all chiral suppressed, being proportional to $\sqrt{\rl} =m_{\ell}/m_B$.

The integration regions of the kinematic variables $\hu$ and $\hs$ are given by \cite{Ali:1999mm}
\bea
4\rl \le &\hs&\le(1-\sqrt{\rKs})^2\\
-\bu\le &\hu& \le \bu \\ \nonumber
\bu&\equiv&\sqrt{\lambda\left(1-4\frac{\rl}{\hs}\right)}\, .
\eea
To avoid confusion, we have used the same symbol for $\bu$ as in $b\to s \ell^+\ell^-$ decay, although  the definition is different due to the hadronic mass of $B$ and $K^*$ involved. As above, the variable $\hu$ corresponds to the $\theta$ angle between the momentum of the $B$ meson and the antilepton $\ell^+$ in the dilepton center of mass system frame, expressed in this frame by the relation $\hu=-\bu\cos{\theta}$. After integrating over the $\hu$ variable, we get
\bea
\frac{d{\Gamma}^{K^*}_M}{d\hat{s}}&=&
\hat{\Gamma}_B\frac{\bu}{\rKs}\Big\{2\sqrt{\rl}\Big[
\Re{F_2\bar{A}A^*}2\lambda \rKs \hs+
\Re{F_2\bar{B}B^*}(\lambda + 12 \rKs \hs)
\nonumber\\
&+&\Re{F_2\bar{C}C^*}\lambda^2 +
\Re{F_2\left(\bar{C}B^*+\bar{B}C^*\right)}\lambda (\rKs + \hs -1)
\Big]
\nonumber\\
&+&\frac{|F_2|^2}{6}\Big[\left(|\bar{C}|^2\lambda
+2\left(|\bar{A}|^2\rKs \hs + \Re{\bar{C}\bar{B}^*}(\rKs + \hs-1)\right)\right)
(8 \rl + \hs)\lambda
\nonumber\\
&+&|\bar{B}|^2(8 \rl + \hs)(\lambda + 12 \rKs \hs)
\Big]
\Big\}
\label{dGammaKs}
\eea

As we can see from these results, the magnetic-dipole correction to the above distribution turns out to be chiral suppressed as expected, and proportional to $r_{\ell}$. The origin of the overall $r_{\ell}$ factor  in the terms of order ${\cal O}(\alpha)$ comes from the $\sqrt{r_{\ell}}$ in the interference of SM amplitude with magnetic dipole-operator, times the $\sqrt{r_{\ell}}$ factor which is  contained in the $F_2$ form factor. On the other hand, in the ${\cal O}(\alpha^2)$ terms the $r_{\ell}$ suppression factor directly arise from $F^2_2$.
However, by a more careful inspection of the infrared behaviour of the above expression for $\hs \to 0$,  one can see that potential contributions of order $r_{\ell}/\hat{s}^2$ and even   $r_{\ell}/\hat{s}^3$ could appear in some of the terms in Eq.(\ref{dGammaKs}).
While the  triple-poles $r_{\ell}/\hat{s}^3$ terms have to vanish due to the absence of infrared power singularity (for $m_l\to 0$) in the total width, for the former terms $r_{\ell}/\hat{s}^2$ there is no guarantee a priori that they would cancel out at any order in the $r_{K^*}$ expansion.

When integrated in $d\hat{s}$, in the region including the dilepton mass threshold $\hat{s}> 4 r_{\ell}$, the double-pole terms $r_{\ell}/\hat{s}^2$ generate finite contributions of order $\alpha$ to the total width that are not chiral suppressed. In contrast, these corrections to $B\to K^* e^+e^-$ turns out to be chiral suppressed by terms of order $m_e^2/m_{\mu}^2$ if the integration region starts from
$s> 4m_{\mu}^2$. Therefore, non-universal and potentially large ${\cal O}(\alpha)$ corrections to $R_{K^*}$ are expected, especially if these are enhanced by the $1/r_{K^*}$ contribution, induced by the $K^*$ longitudinal polarizations.

However, if we expand the differential width in powers of $\hs$ we can see that
all the triple poles  $r_{\ell}/\hat{s}^3$ terms cancel out. On the other hand, non vanishing contributions from the double-pole terms survive, leaving to potentially large contributions to the $R_{K^*}$ as explained above. 
In particular, by using the definition of $\bar{u}(\hat{s})$ and $F_2$, and the results in Eq.(\ref{formfactors}), we get
\bea
\frac{1}{\hat{\Gamma}_B} \frac{d \Gamma^{K^*}_M}{d\hat{s}} \sim \left(\frac{\alpha}{\pi}\right) 32 \hat{m}^2_b (C_7^{\rm eff})^2(T_2^2(s)+T_1^2(s))
\left(\frac{r_{\ell} {\rm Re}[L(\hat{s})]}{\hat{s}^2}\right)\, ,
\eea
where ${\rm Re}[L(\hat{s})]=\log{\left(\frac{\hat{s}-2r_{\ell}-\sqrt{\hat{s}\left(\hat{s}-4r_{\ell}\right)}}{2r_{\ell}}\right)}$ for $\hat{s}>4r_{\ell}$ and
$T_1$ and $T_2$ are the form factors defined before.
Remarkably, the $1/\hat{s}^2$ power singularity removes the chiral suppression $r_{\ell}$ in the numerator, when the differential width is integrated from $\hat{s}> 4r_{\ell}$, namely
    \bea
    \int^{\hat{s}^{\rm max}}_{4 r_{\ell}} \left(\frac{r_{\ell}{\rm Re}[L(\hat{s})]}{\hat{s}^2}\right) \, d\hat{s} = -\frac{1}{2} + {\cal O}(r_{\ell})\, ,
\label{integral}
      \eea
      provided $\hat{s}^{\rm max}\gg 4r_{\ell}$. This is a genuine lepton-mass discontinuity,  already present in the leading-order contribution to the decay rate $d\Gamma/d\hat{s}$ in the quark decay $b\to s \ell^+ \ell^-$, as  pointed out in \cite{Cho:1996we} (see double-pole terms in $\hat{K}^{77}$ expression of Eq.(\ref{Ksij})), that is associated to the photon emissions in chirality-flip transitions.
      
Finally, assuming the values of the form factors $T_{1,2}(s)$ almost constant in the integration region, in particular setting them at $s\sim 0$ (which is a good approximation, since the integral gets its largest value at the threshold), we get
    \bea
    \frac{1}{\hat{\Gamma}_B} \int^{\hat{s}^{\rm max}}_{4 r_{\ell}} \left| \frac{d \Gamma^{K^*}_M}{d\hat{s}}\right| d\hat{s} \,\sim\,\left(\frac{\alpha}{\pi}\right) \hat{m}^2_b 16 (C_7^{\rm eff})^2\left(T^2_2(0)+ T_1^2(0)\right)\, \simeq \, 0.05\%\, .
    \label{NLOdoublepoles}
    \eea
   corresponding to the input values $C_7^{eff}\simeq -0.3$, ${\rm Re}[C_9^{eff}]\simeq 4.8$,  $T_2(\hat{s}\sim 0)\simeq 0.28$ $A_1(\hat{s}\sim 0)\simeq 0.27$.
   This is a contribution of order ${\cal O}(\alpha)$ which is not chiral suppressed, but it is quite small being not enhanced by $1/r_{K^*}$.
When integrated in the range $4 m_{\mu}^2 < s < 1.1 {\rm GeV}^2$, the corresponding correction to $R_{K^*}$ would give a contribution of order $0.1\%$, in agreement (within the order of magnitude) with what will be found by the exact computation in section 5.

As we can see from these results, the leading contributions of the double poles enhanced by $1/r_{K^*}$ exactly cancel out. This is a very crucial result, since if this cancellation would not have occurred, the effect would have been approximately one order of magnitude larger. In order to show how the impact of these potential correction would have affected the  $R_{K^*}$ we report as an example, the results corresponding to  the integrated contribution of double-pole enhanced by  $1/r_{K^*}$  proportional to the $\bar{B} B F_2$ term. By using the analogue ansatz as in Eq.(\ref{NLOdoublepoles}) we get for this term
    \bea
    \frac{1}{\hat{\Gamma}_B} \int^{\hat{s}^{\rm max}}_{4 r_{\ell}} \left| \frac{d \Gamma^{K^*}_M}{d\hat{s}}\right| d\hat{s} \,\sim\, \left(\frac{\alpha}{\pi r_{K^*}}\right)\hat{m}_b C_7^{\rm eff}{\rm Re}[C^{eff}_9] T_2(0) A_1(0)\, \simeq \, 0.9\%\, ,
    \eea
where we retained only the contribution proportional to ${\rm Re}[C_9^{\rm eff}]$ since ${\rm Re}[C_9^{\rm eff}]\gg {\rm Im}[C_9^{\rm eff}]$.
Then, if there would not been any cancellation among the leading double-pole terms $r_{\ell}/(r_{K^*}\hs^2)$, a coherent effect of all of them would have pushed up to an estimated 6\% correction to $R_{K^*}$, in the above integrated range. This  would have been a non-negligible contribution, comparable and even larger than the leading log-enhanced QED contributions of collinear photon emissions.

    \subsection{Decay width for $B\to K\ell^+\ell^-$}
We write the amplitude of the process $B\to K\ell^+\ell^-$ using the same notation as in \cite{Ali:1999mm}. Concerning the $B\to K$ form factors $f_{+,0,T}$, these are usually defined as
\bea
    \langle K(\pK) |\left[\bar{s}\gamma_{\mu}b\right] |B(\pB)\rangle
    &=& f_+(s)\Big( (p_B+p_K)_{\mu}-\frac{m_B^2-m_K^2}{s}q_{\mu}\Big) \nonumber \\
 &+&\frac{m_B^2-m_K^2}{s}f_0(s)q_{\mu}
\\
\langle K(\pK) |\left[\bar{s}\sigma_{\mu\nu} q^{\nu}b\right] |B(\pB)\rangle &=& i\Big((p_B+p_K)_{\mu}s-q_{\mu}(m_B^2-m_K^2)\Big)\frac{f_T(s)}{m_B+m_K}\, ,
\label{KFF}
\eea    
while the other matrix elements involving a $\gamma_5$ inside the operator are vanishing by parity. Regarding the form factors
$f_i=\left\{f_+(s), f_0(s), f_T(s)\right\}$, we will use the parametrization adopted in \cite{Bobeth:2011nj}, where they have been computed in the framework of LCSR
\bea
f_i(s)&=&\frac{f_i(0)}{1-c_i s/m_{\rm res,i}^2}\left\{
1+b_1^i\left(z(s)-z(0)+\frac{1}{2}\left(z(s)^2-z(0)^2\right)\right)\right\}\, ,
\label{K_formF}
\eea
with $s=q^2$ and
\bea
z(s)=\frac{\sqrt{\tau_+-s}-\sqrt{\tau_+-\tau_0}}{\sqrt{\tau_{+}-s}+\sqrt{\tau_{+}-\tau_0}}, ~~\tau_0=\sqrt{\tau_{+}}\left(\sqrt{\tau_{+}}-
\sqrt{\tau_{+}-\tau_{-}}\right), ~~ \tau_{\pm}=\left(m_B\pm m_K\right)^2\, .
\eea
The numerical values for the $b_1^i$ and  $f_i(0)$ coefficients and resonance masses $m_{\rm res,+,T}$ can be found in \cite{Bobeth:2011nj}, with $c_i=1$ for $i=+,T$, and $c_0=0$ due to the absence of a pole for $f_0(s)$. 

Then, the total amplitude, including the magnetic-dipole corrections, can be formally expressed as in Eq.(\ref{MBKS}) with $M^{1,2,3}_{\mu}$ given by
\bea
M^1_{\mu}&=&A^{\prime}(\hs)(\hat{p}_B+\hat{p}_K)+B^{\prime}(\hs)q_{\mu}^{\prime}
\nonumber\\
M^2_{\mu}&=&C^{\prime}(\hs)(\hat{p}_B+\hat{p}_K)+D^{\prime}(\hs)q_{\mu}^{\prime}
\nonumber\\
M^3_{\mu}&=&\bar{A}^{\prime}(\hs)(\hat{p}_B+\hat{p}_K)+\bar{B}^{\prime}(\hs)q_{\mu}^{\prime}\, ,
\eea
where
\bea
A^{\prime}(\hs)&=&C_9^{\eff}(\hs) f_{+}(\hs)+\frac{2\hmb}{1+\hmK} C_7^{\eff} f_T(\hs)\, \nonumber\\
B^{\prime}(\hs)&=&C_9^{\eff}(\hs) f_{-}(\hs)-\frac{2\hmb}{\hs} C_7^{\eff} f_T(\hs)\, , \nonumber\\
C^{\prime}(\hs)&=&C^{\eff}_{10}f_{+}(\hs)\, , \nonumber\\
D^{\prime}(\hs)&=&C^{\eff}_{10}f_{-}(\hs)\, , \nonumber\\
\bar{A}^{\prime}(\hs)&=&\frac{2\hmb}{1+\hmK} C_7^{\eff} f_T(\hs)\, , \nonumber\\
\bar{B}^{\prime}(\hs)&=&-\frac{2\hmb}{\hs} C_7^{\eff} f_T(\hs)\, ,
\label{CoeffBK}
\eea
where $f_{-}(\hs)=(1-\hat{m}_K^2)(f_0(\hs)-f_{+}(\hs))/\hs$.

As in the $B\to K^*$ transition above, we decompose the differential decay width as follows
\bea
\Gamma^{K}&=&\Gamma^{K}_0+\Gamma^{K}_M
\eea
with the term $\Gamma^{K}_M$ containing the magnetic-dipole corrections.  We report below only the results for the differential decay width $\Gamma^{K}_M$, which is given by
\bea
\frac{d^2{\Gamma}^{K}_M}{d\hat{s}\,d\hat{u}}&=&
\hat{\Gamma}_B
\Big\{4\sqrt{\rl}\, \Re{F_2 \bar{A}^{\prime}A^{\prime ~*}}\lambda+|F_2|^2|\bar{A}^{\prime}|^2\left(\hs \hu^2+4\lambda\rl\right)\Big\}\, .
\eea
 After integrating over $\hu$ the results is
\bea
\frac{d{\Gamma}^{K}_M}{d\hat{s}}&=&
\frac{2}{3}\hat{\Gamma}_B\bu\lambda
\Big\{12\sqrt{\rl}\, \Re{F_2 \bar{A}^{\prime}A^{\prime ~*}}+|F_2|^2|\bar{A}^{\prime}|^2\left(8\rl+\hs\right)\Big\}\, .
\eea
The kinematic variables used above are the same as in the $K^*$ case in Eq.(\ref{variables}), but with the replacement of $m_{K^*}\to m_K$. 
Regarding the corresponding expressions for the $d\Gamma^{K}_0$ differential distributions as a function of the parametrization in Eq.(\ref{KFF}), these can be found in \cite{Ali:1999mm} and we fully agree on their results.

As we can see from a simple inspection of the above results, there are not any triple ($r_{\ell}/\hs^3$) or double-poles ($r_{\ell}/\hs^2$) contributions to the distribution $\frac{d{\Gamma}^{K}_M}{d\hat{s}}$, due to the fact that it is proportional to the $A^{\prime}$ and $\bar{A}^{\prime}$ functions.

\section{The Sommerfeld-Fermi factor}
We consider here  the QED long-distance contributions induced by the soft photon exchange \cite{Sommerfeld,Fermi}. In particular, the re-summation of the leading log terms induced by the soft photon corrections is equivalent to the inclusion of the Coulomb interaction in the wave functions of initial and final charged states. These contributions could become relevant in the kinematic regime where the final charged particles are non-relativistic (in the rest frame of the decaying particle). In our case, this would correspond to the $q^2$ bin regions close to the dilepton mass threshold. Therefore,  it is expected to contribute to the $R_{K^*,K}$ mainly in the $q^2$ bin region close to the dimuon threshold $q^2\sim 4 m_{\mu}^2$.
Since the magnetic-dipole corrections are also expected to mainly contribute to the same $q^2$ regions, by completeness we will include these corrections in our analysis.

In general, the decay width
$d\Gamma^0(s_{ij})$ for a generic $N$-body decay is modified by a universal factor~\cite{Weinberg:1965nx}  that takes into account these soft-photon emission corrections. Following the notation of \cite{Isidori:2007zt}), we have
\be
\di \Gamma(s_{ij}, E)  = \Omega (s_{ij}, E)\,  \di \Gamma^0  (s_{ij})
\label{corr} \, ,
\ee
where the kinematic variables $s_{ij}$ are defined as
\be
s_{ij} = \begin{cases}
   (p_i +p_j)^2 &  i \neq 0, \, j \neq 0\\
    (p_0 - p_j)^2     &  i = 0, \, j \neq 0
\end{cases} 
\ee
with $p_i$ the momenta of the final states and $p_0$ that of the decaying particle. The corresponding  variables
\be
\beta_{ij} =\sqrt{1 - \frac{ 4 m_i^2 m_j^2}{(s_{ij} - m_i^2 - m_j^2)^2}}
\ee
can also be  defined. The energy $E$ is the maximum  energy that goes undetected in the process because of the physical limitations of the detector.

Here we retain only the soft-photon corrections
that  become important when the final states are produced near  threshold (in the regime where $\beta_{ij}\rightarrow 0$) and so \eq{corr} becomes
\be
\Omega (s_{ij}, E) = \Omega_C (\beta_{ij} ) 
\ee
where
\be 
\Omega_C (\beta_{ij} ) = \prod_{0<i<j} \frac{2 \pi \alpha q_i q_j}{\beta_{ij}} \frac{1}{\exp \left[{\frac{2 \pi \alpha q_i q_j}{\beta_{ij}}}\right] -1}
\label{Omega}
\ee
 is the (re-summed) correction due to the  Coulomb interaction~\cite{Sommerfeld,Fermi} between  pairs of fermions with charges $q_i$ and $q_j$. 
We  neglect all other ($E$ and non $E$-depending) soft-photon corrections that could become important only in the limit $\beta_{ij}\rightarrow 1$.
 
Since in our numerical analysis we do not include the ${\cal O}(\alpha^2)$ corrections, by consistency we retain in the corresponding widths only the interference terms of magnetic-dipole corrections with the leading order amplitude, and switch off the contribution of the Sommerfeld factor ($\Omega(\hs)\to 1$), in the $\Gamma^M$ contributions.

\section{Numerical results for $R_{K^*}$ and $R_{K}$ ratios}

We provide here the numerical results for the magnetic dipole corrections on the branching ratios of $B\to (K,K^*)$ and the $R_{K^*,K}$.
The values for relevant masses and other SM inputs used to evaluate the BR can be found in table \ref{tabinput1}.

Concerning the evaluation of the form factors, provided by the LCSR method, this is one of the main sources of theoretical uncertainties in the predictions of the BRs. However, since the perturbative and non-perturbative QCD contributions mainly cancel out in the $R_{K^*,K}$, these hadronic uncertainties are expected to be strongly reduced on these observables.
This is not the case of the QED corrections, where the QED collinear singularities, inducing corrections of the order
$(\alpha/\pi) \log^2(m_B/m_{\ell})$, could largely affect the $R_{K^*,K}$ \cite{Bordone:2016gaq}. The same is expected for the QED magnetic dipole corrections, which are manifestly non-universal. 
For this reason,  we present here our numerical results only for a specific values of the form factors, corresponding to the central values of the free parameters entering in the LCSR parametrization.
By consistency, in our analysis we retain only the interference terms proportional to $F_2$ and set to zero the contributions induced by the $|F_2|^2$ terms, since the latter are of order ${\cal O}(\alpha^2)$. A consistent embedding of these contributions should require a full NNLO order analysis in $\alpha$, that goes beyond the aims of the present work.

\subsection{\bf The $B\to K^*$ transition}
We start by analyzing the $B\to K^* \ell^+\ell^-$ decay width.
The numerical results corresponds to the central values of the $\alpha_l^i$ parameters and resonance masses entering in the $P_i(s)$ terms, as provided in \cite{Straub:2015ica}.
The corresponding numerical values are reported in table \ref{tabinput2}.
\begin{table}\centering
\begin{tabular}{C|S[table-format = +1.6]S[table-format = +1.6]S[table-format = +1.6]S[table-format = +1.3]}
\toprule
 {B\to K^\ast \text{ form factors}} &   ${\alpha_0}$ &   ${\alpha_1}$ &   ${\alpha_2}$ & ${m_\textrm{pole} }[\GeV]$ \\
\midrule
       A_0 &   0.355851 &  -1.04363 &    1.12403 &      5.366 \\

       A_1 &   0.269264 &   0.304578 &   -0.10662 &      5.829 \\

    A_{12} &   0.255783 &   0.601902 &   0.117626 &      5.829 \\

         V &   0.341428 &   -1.04834 &    2.37143 &      5.415 \\

       T_1 &    0.28235 &    -0.888396 &    1.94823 &      5.415 \\

       T_2 &    0.28235 &   0.398974 &    0.36137 &      5.829 \\

    T_{23} &   0.667768 &    1.47676 &    1.92352 &      5.829 \\
\bottomrule
\end{tabular}
\caption[]{Central values of the $\alpha_{0,1,2}$ parameters and resonance masses $m_{\rm pole}$, entering in the evaluation of the $B\to K^*$ form factors $A_{0,1,12}$, $V$, and $T_{1,2,23}$, as provided in \cite{Straub:2015ica}.}
\label{tabinput2}
\end{table}

\begin{figure}
\begin{center}
\hspace{0.cm}
\includegraphics{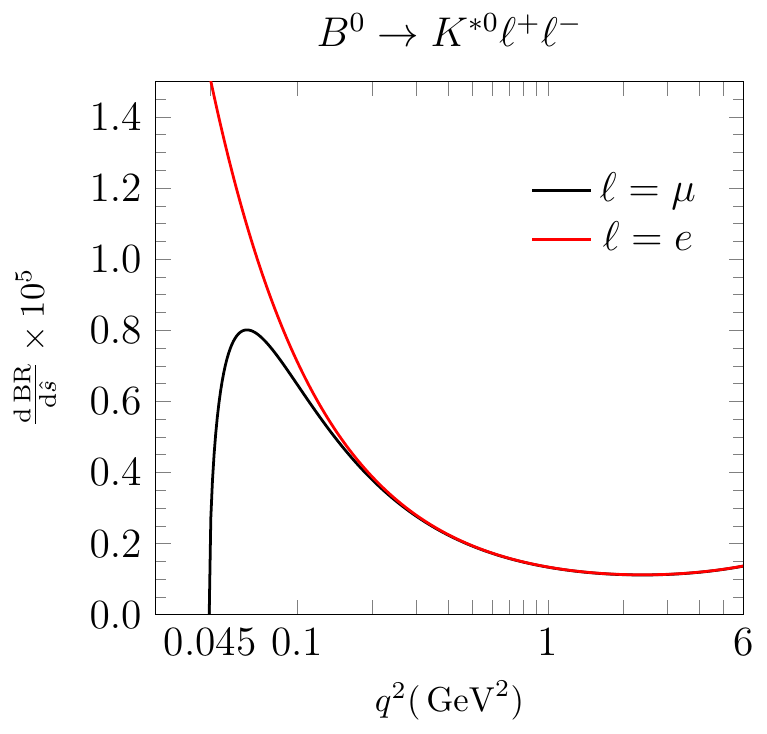}
\hspace{0.3cm}
\includegraphics{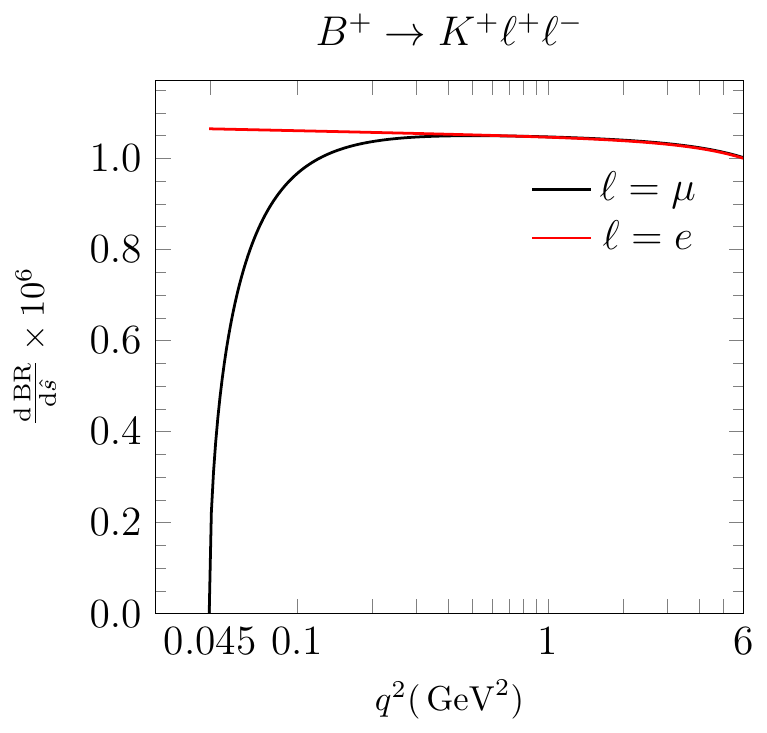}
\caption{Distributions for the differential branching ratio $\frac{d {\rm BR}}{d{\hs}}$ in the SM at the leading order, as a function of $q^2$ (the invariant mass square of the dilepton final state), for the $l=\mu$ (black)  and $l=e$ (red) cases. Left and right plots correspond to the $B\to K^*\ell^+\ell^-$ and $B\to K\ell^+\ell^-$ decays respectively. These results correspond to the parametrization of form factors and input values as reported in the text.}
\label{fig1}
\end{center}
\end{figure}

\begin{figure}
\begin{center}
\hspace{0.cm}
\includegraphics{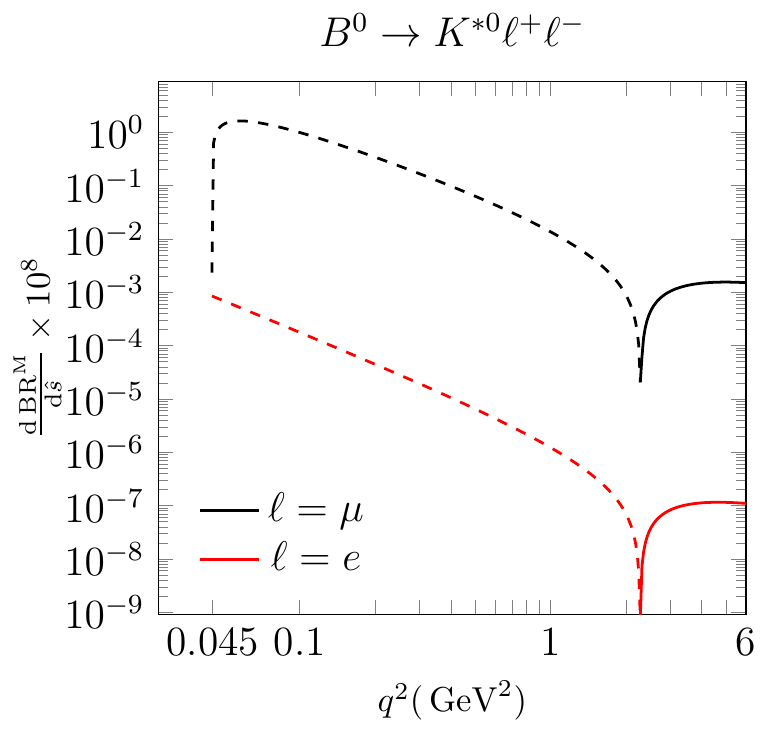}
\hspace{0.3cm}
\includegraphics{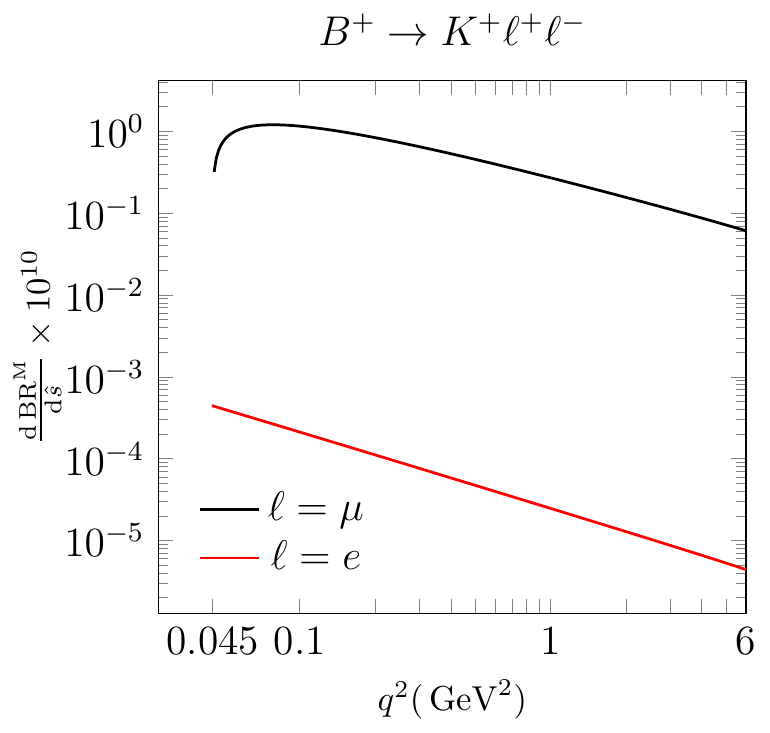}
\caption{Differential distributions of branching ratios as in Fig\ref{fig1}  for the pure magnetic-dipole corrections, as a function of $q^2$.  Left and right plots correspond to the $B\to K^*\ell^+\ell^-$ and $B\to K\ell^+\ell^-$ decays respectively. In the left plot, where the absolute value of the distribution is plotted, values of the curves for $q^2$ less (greater) than the dip point (at $q^2\sim 2 {\rm GeV}^2$) are negative (positive) respectively. Distributions at the dip point are understood to vanish. }
\label{fig2}
\end{center}
\end{figure}

\begin{figure}
\begin{center}
\hspace{0.cm}
\includegraphics{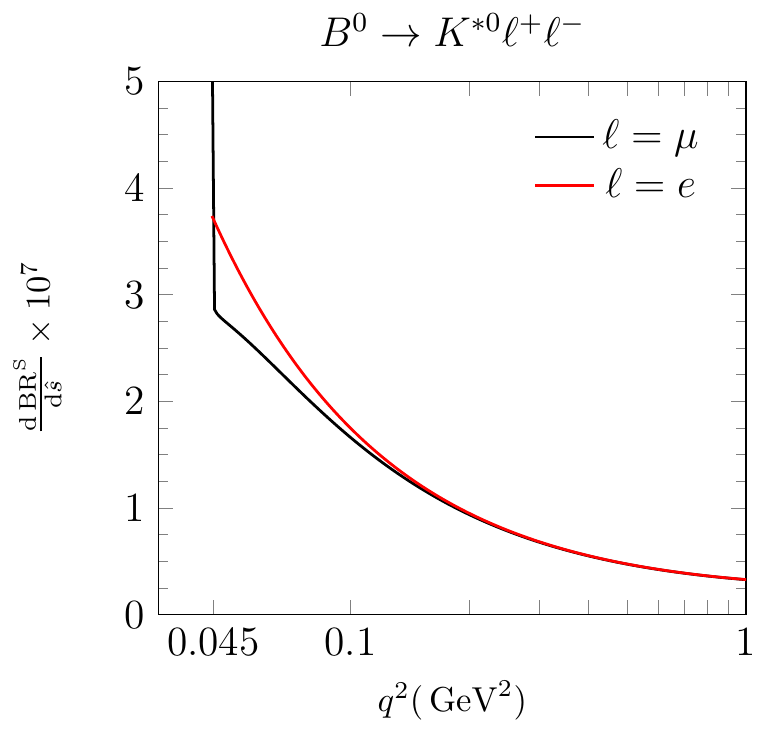}
\caption{Differential distributions of the branching ratio for the pure Sommerfeld correction, as a function of $q^2$ for the $B\to K^*\ell^+\ell^-$ decay. }
\label{fig2S}
\end{center}
\end{figure}

In the left plot of Fig.\ref{fig1} we show the curves for the LO corresponding BR distributions $d\BR/d\hat{s}$ for both final muon and electron pairs, in the relevant $q^2$ range $4 m_{\mu}^2 < q^2< 6 {\rm GeV}^2$. In the left plot of Fig.\ref{fig2}, we plot the absolute value of the differential BR for the pure magnetic-dipole corrections. This correction is defined as
\bea
\frac{d\BR}{d\hat{s}}&=& \frac{d\BR^{\LO}}{d\hat{s}} +\frac{d\BR^{\M}}{d\hat{s}}\, ,
\eea
where $\frac{d\BR}{d\hat{s}}$ represents the total contribution and $\BR^{\LO}$ the leading SM one, without magnetic-dipole corrections.
The curves for $q^2$ smaller (larger) than the dip point (at $q^2\approx 2 {\rm GeV}$) are negative (positive) respectively. The dip point in this plot, where curves are understood to vanish, is due to a change of sign of the correction.

As expected by chirality arguments, the magnetic-dipole correction to the final electron channel is very suppressed in the relevant $q^2$ region of $q^2> 4m_{\mu}^2$, due to the corresponding chiral suppression in the magnetic form factor $F_2$.
As we can see from these results the most relevant effect of these corrections for the muon case is achieved in regions of $q^2$ close to the its threshold. Then, due to the manifest lepton non-universality of these corrections, an impact on the $R_{K^*}$ observable is expected at low $q^2$.

Finally we stress that the ${\cal O}(\alpha)$ distribution for the magnetic-dipole contribution does not vanish at the threshold. This is due to the interference between the LO SM amplitude with the magnetic-dipole one proportional to ${\rm Im}[F_2]$. Indeed ${\rm Im}[F_2]$ scales as ${\rm Im}[F_2]\sim 1/\sqrt{q^2 -4 m^2_{\ell}}$ for $q^2\to 4 m^2_{\ell}$, that can compensate the usual phase space suppression term $\bar{u}\sim \sqrt{q^2 -4 m^2_{\ell}}$. Then, the value at the threshold for the interference term is
$\lim_{q^2\to 4m_{\mu}^2} \frac{d\BR^{\M}}{d\hat{s}}(B\to K \mu^+\mu^-)=-2.2 \times 10^{-11}\,$ .
Notice that, at the order ${\cal O}(\alpha^2)$, once the contributions of the $|F_2|^2$ term are added, the distribution has a singularity and scales  as  $1/\sqrt{q^2 -4 m^2_{\ell}}$ at the threshold. However, this singularity is integrable and does not require any regularization. The narrow region, where the   corrections proportional to $|F_2|^2$ start to be relevant, is comprised between $4m_{\mu}^2 < q^2 <4m_{\mu}(1+\delta) $, with $0<\delta < 10^{-4}$. This  gives a tiny but finite contribution to the total BR, which is of the order of $10^{-15}$.

In Fig.\ref{fig2S} we show the contributions of the pure Sommerfeld corrections to $\frac{d\BR^{\Som}}{d\hat{s}}$. Since the Sommerfeld corrections are not additive, accordingly to Eq.(\ref{corr}) we define the corresponding Sommerfeld correction on the differential width distribution as
\bea
d\Gamma^{\Som}&=& (\Omega-1)d\Gamma^{\LO}\, 
\eea
and analogously for the $d{\rm BR}^{\Som}$.

Now, we analyze the impact of the magnetic-dipole and Sommerfeld corrections to the ratio $R_{K^*}$ as defined in Eq.(\ref{RK}). We parametrize these corrections $\Delta R^{M,S}_{K^*}$ as
\bea
R_{K^*}&=&R_{K^*}^{\rm LO}\left(1+\Delta {R^{\M}_{K^*}} +\Delta {R^{\Som}_{K^*}} \right)\, ,
\label{deltaRK}
\eea
where, as before, the suffix LO stands for the SM contribution without magnetic-dipole corrections, while $M$ and $S$ stand for the corrections induced by the magnetic-dipole and Sommerfeld  factor contributions respectively.

We also consider the ratios of distributions $\rho_{K^*}(\hs)$ defined as
\bea
\rho_{K^*}(\hs)&=&\frac{\frac{d\Gamma(B\to K^* \mu^+\mu^-)}{d\hs}}{
  \frac{d\Gamma(B\to K^* e^+e^-)}{d\hs}}\, ,
\label{rho}
\eea
as well as the relative deviation $\delta \rho^{\M,\Som}_{\Ks}(\hs)$ defined as
\bea 
\rho_{K^*}&=&\rho_{\Ks}^{\LO}\left(1+\delta \rho_{\Ks}^{\M}  + \delta \rho_{\Ks}^{\Som}\right)\, ,
\label{deltarho}
\eea
with same notation as above for the quantities with symbols LO,M,S at the top.

\begin{figure}
\begin{center}
\hspace{0.cm}
\includegraphics{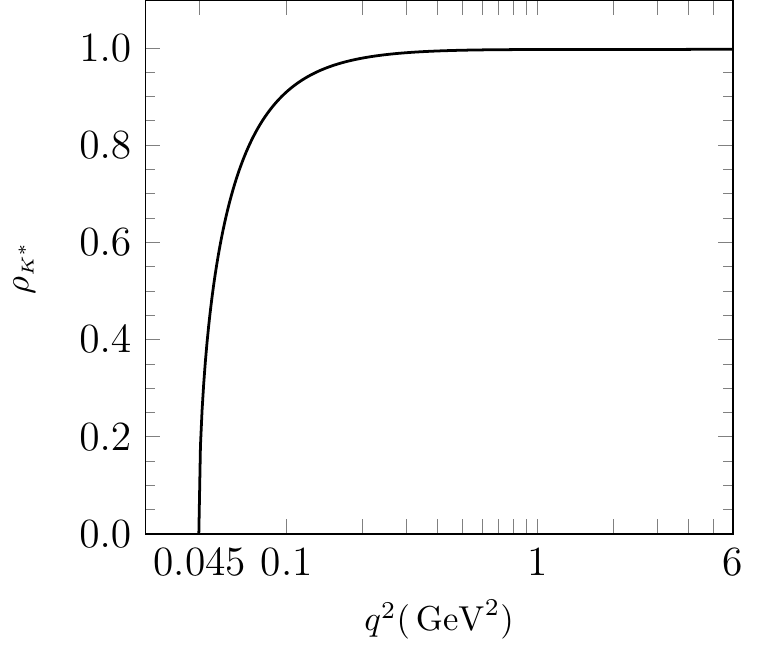}
\hspace{0.7cm}
\includegraphics{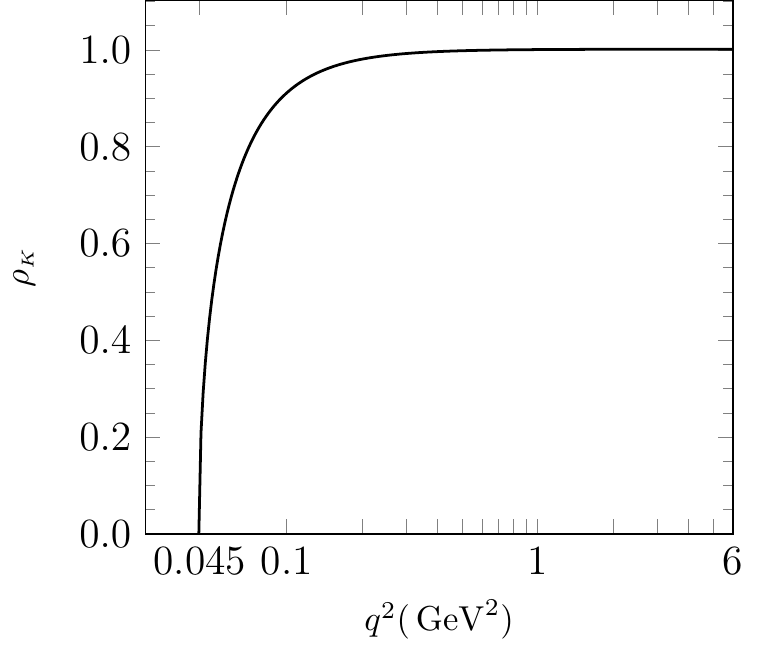}
\caption{The $\rho_{K,K^*}$ functions at the LO in the SM as defined in the text, as a function of $q^2$, for the $B\to K^*\ell^+\ell^-$ (left plot) and
  $B\to K\ell^+\ell^-$ (right plot) decays. }
\label{fig3}
\end{center}
\end{figure}

\begin{figure}
\begin{center}
\hspace{0.cm}
\includegraphics{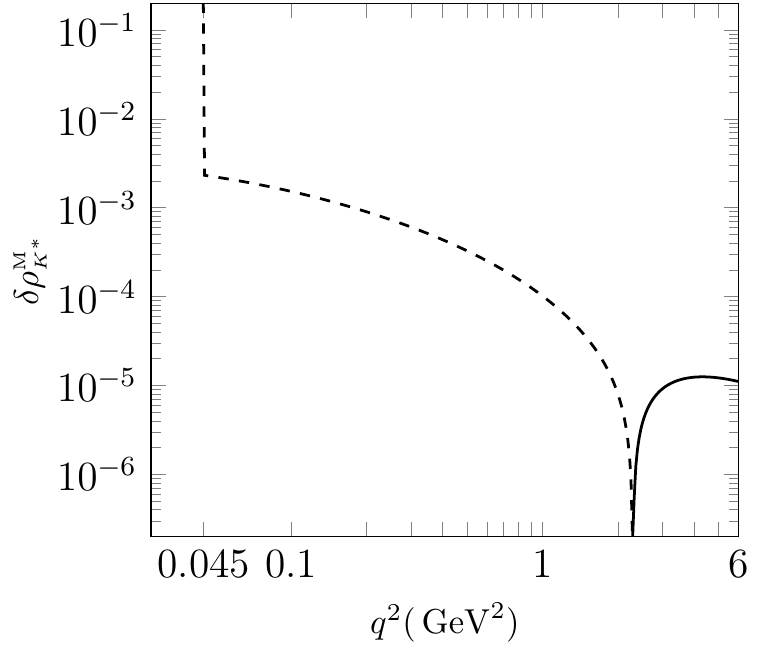}
\hspace{0.7cm}
\includegraphics{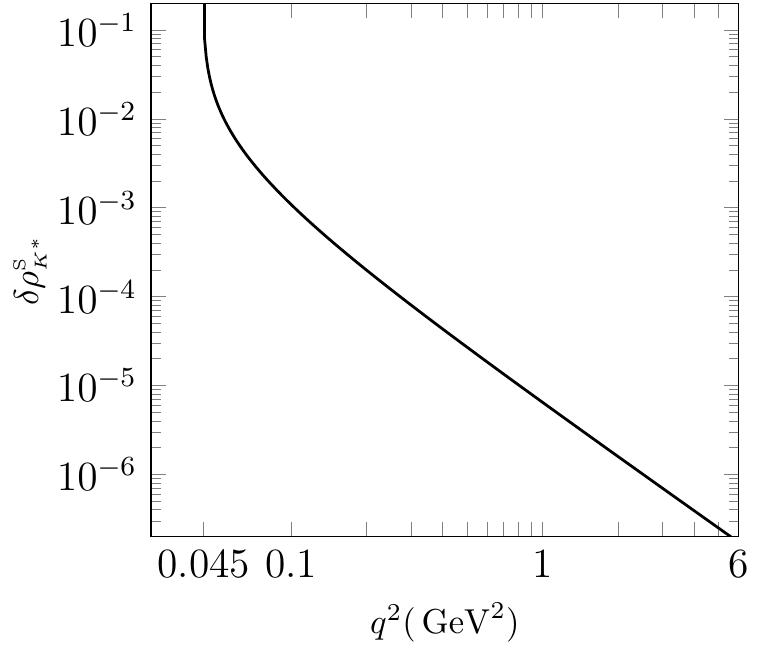}

\caption{The $\delta \rho^{\M}_{\Ks}$ (magnetic-dipole) and  $\delta \rho^{\Som}_{\Ks}$ (Sommerfeld) corrections, as defined in the text, as a function of $q^2$, in the left and right plots respectively.}
\label{fig4}
\end{center}
\end{figure}

In the left plot of  Fig.\ref{fig3}, we show the results for the $\rho^{\LO}_{\Ks}(\hs)$ as a function of $q^2$, for $4m_{\mu}^2 < q^2 < 6{\rm GeV}^2$, while the magnetic-dipole corrections parametrized by  $\delta \rho^{\M}_{K^*}$, are reported in the left plot of Fig.\ref{fig4}. The LO SM contribution to the $\rho$ is almost the same for the $B\to K^*$ and $B\to K$ transitions, the largest difference is of the order of 1-2\% for $q^2<0.5 {\rm GeV}^2$.
Concerning the magnetic-dipole corrections, as we can see from these results, the $\delta \rho^{\M}_{K^*}$ reaches a maximum of the order of $2\times 10^{-3}$ for $q^2$ regions very close to the dimuon  mass threshold,  and drops below $10^{-4}$ for $q^2> 1 {\rm GeV}$. The fact that the distribution does not vanish at the threshold is due to the interference term of SM amplitude at the zero order proportional to the ${\rm Im}[F_2]$ term, which removes the phase space suppression factor.

Concerning the Sommerfeld corrections $\delta \rho^{\Som}_{\Ks}$ as a function of $q^2$, these are reported in the right plot of 
Fig.\ref{fig4}. As we can see from these results, the $\delta \rho^{\Som}_{\Ks}$ function is very peaked in the region close to the threshold, and become smaller than $10^{-5}$ for $q^2> 1 {\rm GeV}$. We stress that, while the magnetic dipole corrections are manifestly LF non-universal, due to the chiral suppression term in the magnetic form factor $F_2$, the Sommerfeld corrections are non-universal only in very narrow regions close to the dimuon threshold $q^2 \simeq 4 m_{\mu}^2$.

In Table \ref{tab3}, we report the values for the magnetic-dipole corrections $\Delta R^{\M}_{K^*}$ integrated in some representative set of bins.  These results have a statistical error of a few percent due to the Montecarlo integration.
As we can see from these results, the largest impact of these corrections affect the regions close to the threshold, in particular on the integrated bins
$[0.0447,0.3]$ and $[0.0447,0.5]{\rm GeV^2}$, we get $\Delta R^{\M}_{K^*}$ negative and  approx  of order of $\Delta R^{\M}_{K^*}\sim {\cal O}(0.1\%)$. Moreover, in the integrated bin region used in the experimental setup, corresponding to $[0.0447,1.1]{\rm GeV^2}$, we get $\Delta R^{\M}_{K^*}\sim 0.6 \times 10^{-3}$. The impact of the magnetic dipole corrections on $R_{K^*}$ becomes totally negligible above the bin regions larger than $1 {\rm GeV}^2$, being smaller than $10^{-5}$ effect.

Concerning the Sommerfeld contribution $\Delta R^{\Som}_{K^*}$, this is positive and relevant only in the narrow region close to the threshold, in particular we get
\bea
\Delta R^{\Som}_{K^*}[4m_{\mu}^2,0.3 {\rm GeV}^2]\sim 2\times 10^{-3}\, ,~~~~~~
\Delta R^{\Som}_{K^*}[4m_{\mu}^2,0.5 {\rm GeV}^2]\sim 1\times 10^{-3} \, .
\eea
Smaller values below $10^{-4}$ are expected for $\Delta R^{\Som}_{K^*}$ when integrated on $q^2_{\rm max}$ bin regions above $1 {\rm GeV}^2$. Contrary to the behavior of the magnetic-dipole corrections, the Sommerfeld contributions for $q^2\gg 4m_{\mu}^2$ are almost LF universal, and cancel out in the ratios $R_{K^*}$. Due to a strong fine-tuning cancellations among the Sommerfeld contributions to the corresponding widths of muon and electron channels, we do not report the results for $\Delta R^{\Som}_{K^*}$ for larger $q^2$ bins, since each integration is affected by a large statistical error, larger than the required precision for the fine-tuning cancellation.

In conclusion, we found that the largest contribution to the $R_{K^*}$ induced by the magnetic-dipole corrections arises in the regions close to the threshold and it is maximum relative effect is of order of a few per mille.  This is well below the present level of experimental precision on $R_{K^*}$ (at least in the $q^2$ bin ranges explored by present experiments)  and it is one order of magnitude smaller that the expected leading QED contributions from the soft photon emissions. Same conclusions for the long-distance contributions induced by the Sommerfeld corrections, which is of the same order as the magnetic-dipole ones in regions very close to the threshold, and completely negligible for $q^2_{\rm max}>1 {\rm GeV^2}$.

\begin{table} \begin{center}    
\begin{tabular}{|c||c|}
\hline 
$\left[ q^2_{\rm min}, q^2_{\rm max}\right]({\rm GeV}^2)$ 
& $\Delta R^{\M}_{K^*}$ 
\\ \hline
 $[0.0447,0.3]$
& $-1.3\times 10^{-3}$ 
\\ \hline
 $[0.0447,0.5]$
& $-1.0\times 10^{-3}$
\\ \hline
 $[0.0447,1.1]$
& $-7.4\times 10^{-4}$
\\ \hline
 $[1.1,6]$
& $2.5\times 10^{-6}$
\\ \hline \hline
 $[0.5,0.8]$
& $-2.2\times 10^{-4}$
\\ \hline
 $[0.8,1]$
& $-1.2\times 10^{-4}$
\\ \hline
 $[1,3]$
& $-1.8\times 10^{-5}$
\\ \hline
 $[3,6]$
& $1.2\times 10^{-5} $
\\ \hline
\end{tabular} 
\caption[]{
  Results for $\Delta R^{\M}_{K^*}$, the relative QED radiative correction of $R_{K^*}$ for $B\to K^* \ell^+\ell^-$ induced by the
  magnetic-dipole corrections as defined in the text, for a representative set of integrated $q^2$ bins in the range $q^2_{\rm min}< q^2 <q^2_{\rm max}$.
}
\label{tab3}
\end{center} \end{table}

Finally, we  report below for completeness the corresponding results for
the ratio  $R^{\M}_{K^*}$ and its corresponding deviation $\Delta R^{\M}_{K^*}$ induced by magnetic-dipole corrections, in the case of dilepton $\tau^+\tau^-$ final states, normalized with respect to the dimuon and electron final states. In particular, we define
\bea
R^{\tau}_{K^*}(\ell)&=&\ddfrac{\int_{q^2_{\rm min}}^{q^2_{\rm max}}
  \ddfrac{d\Gamma(B\to K^*\tau^+\tau^-)}{dq^2}dq^2}
{\int_{q^2_{\rm min}}^{q^2_{\rm max}} 
  \ddfrac{d\Gamma(B\to K^*\ell^+\ell^-)}{dq^2}dq^2}\, ,
\label{RKtau}
\eea
where $\ell=\mu,e$.
For the $B\to K^* \tau^+\tau^-$ decay, the allowed range of $q^2$ is pretty narrow, where $q^2_{\rm min}=4m_{\tau}^2$ and $q^2_{\rm max}=12.6{\rm GeV}^2$. Integrating the $R^{\tau}_{K^*}(\ell)$ on this range of $q^2$ we get
\bea
R^{\tau}_{K^*}(\mu)\simeq R^{\tau}_{K^*}(e)=0.39\, ,~~~~~ \Delta R^{\tau}_{K^*}(\mu)\,=\, 7.5\times 10^{-5}\, ,~~~\Delta R^{\tau}_{K^*}(\mu)\,=\, 7.9\times 10^{-5}\, .
\eea
where $\Delta R^{\tau}_{K^*}(\ell)$ is defined as in Eq.(\ref{deltaRK}). As we can see from this results, despite the fact that the magnetic dipole contribution is chirally enhanced by the tau mass, the magnetic-dipole correction $\Delta R^{\tau}_{K^*}(\ell)$ is quite small and of the order of $10^{-5}$. The reason is that in the $\tau$ case, the further suppression comes from kinematic due to the reduced allowed range of $q^2$.

\subsection{The $B\to K$ transition}
We extend here the same analysis presented above, concerning the magnetic-dipole and Sommerfeld corrections, to the $B\to K \ell^{+}\ell^{-}$ decays.
Concerning our numerical results, these are obtained by using the central values for the $b_1^i$ and  $f_i(0)$ parameters entering in the parametrization of the form factors in Eq.(\ref{K_formF}), as provided in \cite{Bobeth:2011nj}. The corresponding input values are reported in table \ref{tabinput3}.
\begin{table}\centering
\begin{tabular}{CS[table-format=1.2]S[table-format=+1.1]C}
\toprule
{B \to K \text{ form factors}} &       ${F}$ &   ${b_1}$ & m_\textrm{pole}[\GeV]\\
\midrule
       f_+ &       0.34 &       ~~-2.1 &    5.412 \\

       f_0 &       0.34 &       ~~-4.3 &          \\

f_\mathrm{T} &       0.39 &       ~~-2.2 &    5.412 \\
\bottomrule
\end{tabular}
\caption[]{Central values of the $F$ and $b_1$ parameters and resonance masses $m_{\rm pole}$, entering in the evaluation of the $B\to K$ form factors $f_{+,0,T}$ for the $B\to K$ transitions, as provided in \cite{Bobeth:2011nj}. For the $f_0$ form factor no resonance mass is associated.}
\label{tabinput3}
\end{table}
In the right plot of Fig.\ref{fig1} we present the results for LO SM contribution to the  $d{\rm BR}/d\hs$ distribution versus $q^2$, while the corresponding results for magnetic-dipole corrections $d{\rm BR}^{\M}/d\hs$ are shown in the right-plot of Fig.\ref{fig2}. As for the $B\to K^*$ transitions, the magnetic-dipole corrections contain only the contribution of the interference between magnetic-dipole amplitude with the corresponding LO SM one. As we can see from these results, the $d{\rm BR}^{\M}/d\hs$ contribution is always positive for $q^2>4m_{\mu}^2$. The LF non-universality of the contribution is manifest. Its relative effect, with respect to the LO SM contribution, is more suppressed than in the $B\to K^*$ transitions and it is roughly one order of magnitude smaller than in $B\to K^*$. This behavior is well in agreement with the naive expectations based on the enhancement of the magnetic-dipole contributions in the  $B\to K^*\ell^+\ell^-$ decays. Indeed, as mentioned in the introduction, this enhancement is mainly due to the fact that the FC magnetic-dipole contribution, proportional to $C_7^{\rm eff}$, receives an enhancement factor in the $B\to K^*$ transitions proportional to the $m^2_B/m_{K^*}^2$ factor. This factor arises due to the contributions of the longitudinal polarizations of $K^*$, while it is absent in the $B\to K$ transitions. 
Since the QED magnetic-dipole corrections are proportional to the $C_7^{\rm eff}$ coefficient, a corresponding enhancement in the ${\rm BR}(B\to K^*\ell^+\ell^-)$ decays is therefore expected, with respect to ${\rm BR}(B\to K\ell^+\ell^-)$.

\begin{figure}
\begin{center}
\includegraphics[width=0.48\textwidth]{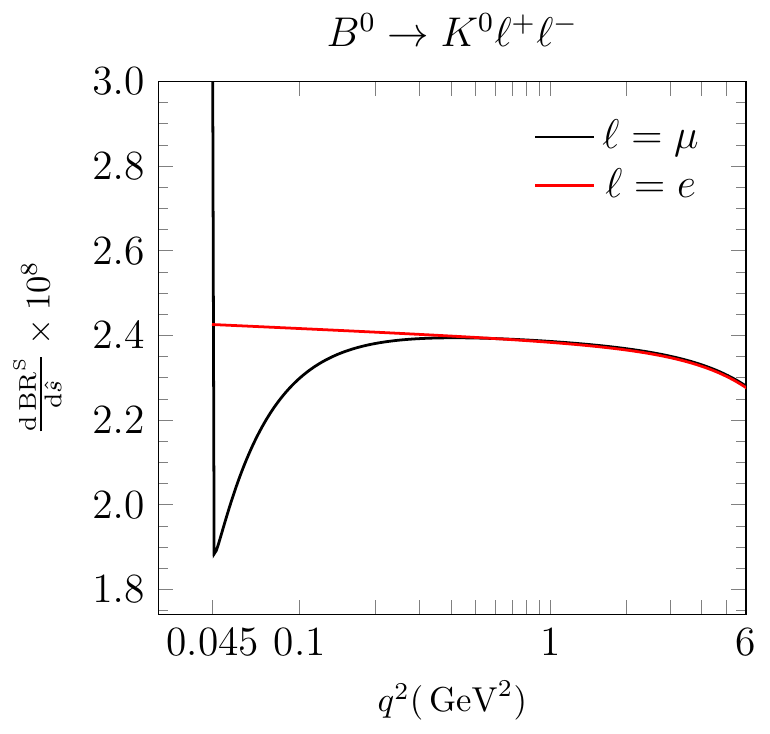}
\hspace{0cm}
\includegraphics[width=0.48\textwidth]{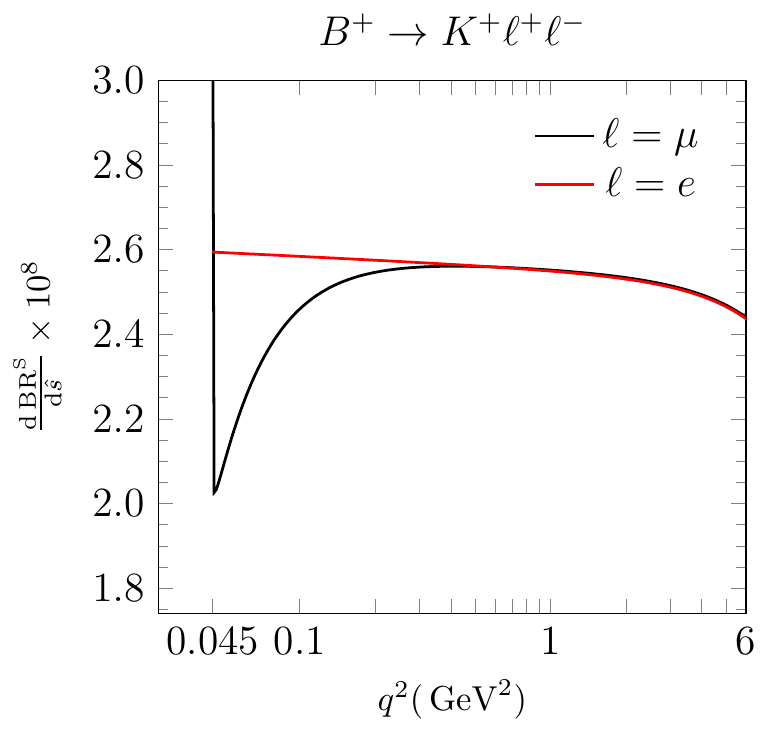}
\caption{Differential distributions of branching ratios for the pure Sommerfeld correction, as a function of $q^2$.  Left and right plots correspond to the
neutral $B^0\to K^0 $ and charged  $B^+\to K^+ $ transitions respectively. }
\label{fig4B}
\end{center}
\end{figure}

The Sommerfeld corrections to the BR distribution $d{\rm BR}/d\hs$ are presented in Fig.\ref{fig4B}, for the neutral (left plots) and charged (right plots) $B\to K$ transitions respectively. For the $B^+\to K^+$ channel we have retained all the Coulomb potential corrections according to the formula in Eq.(\ref{Omega}). In computing these long-distance contributions to the $B^+\to K^+$ channel, we had to numerically integrate over $d\hu$ the convolution of
$d^2{\rm BR}/(d\hs d\hu)$ distribution with the corresponding $\Omega(\hs,\hu)$ Sommerfeld function. As we can see from these results, the effect of the Sommerfeld enhancement becomes quite large and non-universal in regions of $q^2$ quite close to the dimuon threshold. No substantial numerical difference appears between the Sommerfeld corrections to the neutral and charged channels. Indeed, the extra Coulomb corrections in $K^+ \ell^+$ and $K^+ \ell^-$, which are absent in $K^0\ell^+\ell^-$ mode, exactly canceled out in Eq.(\ref{Omega}) at ${\cal O}(\alpha)$ because of the sign difference of $q_i q_j$ products of charges and symmetric behavior in the momentum exchange of final-state leptons $p^+ \leftrightarrow p^-$.

\begin{figure}
\begin{center}
\hspace{0.cm}
\includegraphics[width=0.44\textwidth]{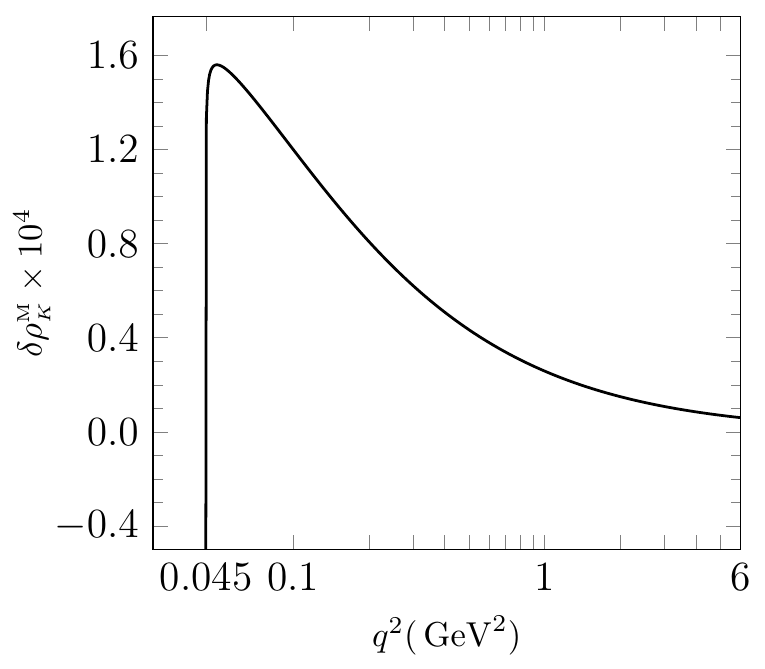}
\hspace{0.7cm}
\includegraphics[width=0.44\textwidth]{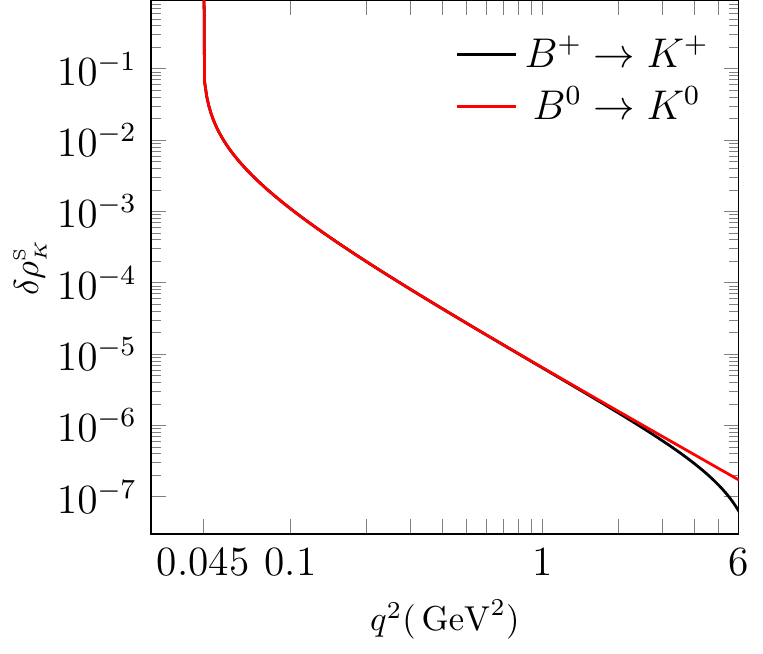}

\caption{The $\delta \rho^{\M}_{\K}(\hs)$ (magnetic-dipole) and  $\delta \rho^{\Som}_{\K}(\hs)$ (Sommerfeld) deviations as defined in the text, as a function of $q^2$, in the left and right plots respectively. The red and black curves in the right plot stand for the Sommerfeld corrections to the neutral and charged $B\to K$ transitions respectively.}
\label{fig5}
\end{center} 
\end{figure}

In the right plot of Fig.\ref{fig3} we report the results for $\rho_K(\hs)$ for the  $B\to K \ell^{+}\ell^{-}$ decays, as a function of $q^2$ in the range
$4m_{\mu}^2 <q^2 < 6 {\rm GeV}^2$, while in Fig.\ref{fig5} we show the corresponding magnetic-dipole and Sommerfeld corrections, respectively $\delta \rho^{\M}_K(\hs)$ (left plot) and $\delta \rho^{\Som}_K(\hs)$ (right plot), as defined in Eq.(\ref{deltarho}) for the $B\to K^*$ transitions and generalized here for the $B\to K$ ones, for both charged and neutral $B$ mesons.

Results for  $\Delta R^{\M}_{K}$, the relative QED radiative correction in $R_K$, induced by the  magnetic-dipole corrections are reported in table \ref{tab4} for a representative set of integrated $q^2$ bins in the range $q^2_{\rm min}< q^2 <q^2_{\rm max}$. Also these results, as in the $\Delta R^{\M}_{K^*}$ case, are affected by a few percent error due to numerical integration. As we can see from the values in table \ref{tab4}, the impact of these corrections is approximately one order of magnitude smaller than in the $R_{K^*}$ case (cfr. table \ref{tab3}), whose larger effect is of order ${\cal O}(10^{-4})$ for $q^2$ bins close to the dimuon mass threshold.

\begin{table} \begin{center}    
\begin{tabular}{|c||c|}
\hline 
$\left[ q^2_{\rm min}, q^2_{\rm max}\right]({\rm GeV}^2)$ 
& $\Delta R^{\M}_{K}$ 
\\ \hline
 $[0.0447,0.3]$
& $9.4\times 10^{-5}$
\\ \hline
 $[0.0447,0.5]$
& $7.4\times 10^{-5}$ 
\\ \hline
 $[0.0447,1.1]$
& $5.0\times 10^{-5}$
\\ \hline
 $[1.1,6]$
& $1.1\times 10^{-5}$
\\ \hline \hline
 $[0.5,0.8]$
& $3.6\times 10^{-5}$
\\ \hline
 $[0.8,1]$
& $2.8\times 10^{-5}$
\\ \hline
 $[1,3]$
& $1.6\times 10^{-5}$
\\ \hline
 $[3,6]$
& $8.0\times 10^{-6} $
\\ \hline
\end{tabular} 
\caption[]{
  Results for  $\Delta R^{\M}_{K}$, the relative QED radiative correction in $R_K$  for $B\to K \ell^+\ell^-$  induced by the  magnetic-dipole corrections as defined in the text, for a representative set of integrated $q^2$ bins in the range
  $q^2_{\rm min}< q^2 <q^2_{\rm max}$.
}
\label{tab4}
\end{center}
\end{table}

Finally, concerning the effect induced by the Sommerfeld corrections on the ratios $R_{K}$ for the neutral $B\to K$ transition, this is positive and relevant only in the narrow region close to the threshold. In particular, by integrating on the $q^2$ bin regions close to the dimuon threshold, we get for the neutral channel
\bea
\Delta R^{\Som}_{K^0}[4m_{\mu}^2,0.3 {\rm GeV}^2]\sim 2\times 10^{-3}\, ,~~~~~~
\Delta R^{\Som}_{K^0}[4m_{\mu}^2,0.5 {\rm GeV}^2]\sim 1\times 10^{-3} \, ,
\eea
and same order results for the Sommerfeld contributions to the charged $B^+\to K^+$ channel.

As discussed above for the $B\to K^*$ transitions, these results have a large statistical error (approx 30\%), due to the lack of precision in the numerical integration. Smaller values below $10^{-4}$ are expected in both charged and neutral $B\to K$ transitions for larger $q^2$ bin regions.
Although, it is quite difficult to experimentally probe $q^2$ regions too close to the dimuon threshold, larger $\Delta R^{\Som}$ corrections of the order of percent could be obtained for both charged and neutral channel, namely $\Delta R^{\Som}_{K^0,K^+}[4m_{\mu}^2,0.08 {\rm GeV}^2]\sim 1\times 10^{-2}$.


\section{The Dark Photon contribution}
Many NP scenarios have been proposed so far to explain the observed discrepancy in $R_{K,K^*}$ with respect the SM predictions, that, if confirmed, should signal the breaking of LFU in weak interactions \cite{Descotes-Genon:2015uva,Altmannshofer:2017fio,Capdevila:2017bsm,DAmico:2017mtc,Altmannshofer:2017yso,Geng:2017svp,Ciuchini:2017mik,Hiller:2017bzc,Alok:2017sui,Hurth:2017hxg,Alguero:2019aa,Alok:2019ufo,Ciuchini:2019usw,Aebischer:2019mlg,Kowalska:2019ley,Alguero:2018nvb,Alguero:2019pjc,Datta:2019zca,Greljo:2015mma,Barbieri:2015yvd,Barbieri:2016las,Bordone:2017anc,Bordone:2017lsy,Buttazzo:2017ixm,Barbieri:2017tuq}.
However, in the majority of these proposals the NP contribution is affecting only the Wilson coefficients of four-fermions contact operators that manifestly break the LFU.
Here we consider a NP scenario that can provide a LF non-universal contribution to the $b\to s \ell^+\ell^-$ transitions via long-distance interactions, mediated by magnetic-dipole operators. In particular, we explore the possibility of a new s-channel contribution to the $b\to s \ell^+\ell^-$ amplitude, mediated by the virtual exchange of a spin-1 field, namely the dark photon.

We restrict our choice to the case of a {\it  massless} dark photon $(\bar{\gamma})$, associated to an unbroken  $U(1)_D$ gauge interaction in the dark sector \cite{Fabbrichesi:2020wbt}.
Indeed, in contrast to the massive case, the {\it massless} dark photon does not have tree-level interactions with ordinary SM fields, even in the presence of a kinetic mixing with ordinary photons \cite{Holdom:1985ag}. However, in this case the dark photon could have interactions with observable SM sector mediated by high-dimensional operators. This can be understood by noticing that, unlike for the massive case, for a {\it massless} dark photon the tree-level couplings to ordinary matter can always be rotated away by matter field re-definitions \cite{Holdom:1985ag}.
On the other hand, ordinary SM photon couples to both the SM and the dark sector, the latter having milli-charged photon couplings strength to prevent macroscopic effects.

However, dark photons can acquire effective SM couplings at one-loop, with heavy scalar messenger fields and/or other particles in the dark sector running in the loops. In this respect, the lowest dimensional operators for couplings to quarks and leptons are provided by the (dimension 5) {\it dark} magnetic-dipole operators. Therefore, unlike the case of the massive dark photon, potentially large $U(1)_D$ couplings in the dark sector would be allowed thanks to the built-in suppression associated to the higher dimensional operators.
We recall here that the dark photon tree-level couplings to SM fields, for light scenarios with masses below the MeV scale, is severely constrained by astrophysics on its milli-charged coupling with ordinary matters \cite{Davidson:1991si,Raffelt:1996wa,Kadota:2016tqq}. Therefore, the only viable way to explain the $R_{K,K^*}$ anomalies by dark photons is to assume a {\it massless} dark photon scenario, thanks also to its possible un-suppressed $U(1)_D$ couplings.

The dark photons scenario (mainly the massive ones) has been extensively considered in the literature, in both theoretical and phenomenological aspects, and it is also the subject of many experimental searches (see \cite{Essig:2013lka,Alexander:2016aln} for recent reviews). Massless dark photon scenarios received particular attentions in the framework of dark sector origin of Dark Matter and also in Cosmology. The role of a {\it massless} dark photon scenario in galaxy formation and dynamics has been explored in \cite{Ackerman:mha,Fan:2013tia,Agrawal:2016quu,Foot:2014uba,Heikinheimo:2015kra},  while it could also help to generate the required long-range forces among dark matter constituents that could predict dark discs of galaxies \cite{Fan:2013tia,Agrawal:2016quu}.

Recently, in this framework, a new paradigm has been proposed that could support the existence of a {\it massless} dark photon interacting with a $U(1)_D$ charged dark sector. This scenario predicts exponential spread SM Yukawa couplings from an unbroken  $U(1)_D$ in the dark sector \cite{Gabrielli:2013jka,Gabrielli:2016vbb}, thus providing a natural explanation for the SM flavor hierarchy, as well as a solution for the missing dark matter constituents. Interesting phenomenological implications of this scenario are predicted \cite{Gabrielli:2014oya,Gabrielli:2016cut,Biswas:2016jsh,Fabbrichesi:2017vma,Fabbrichesi:2017zsc,Barducci:2018rlx,Biswas:2015sha,Biswas:2017lyg,Fabbrichesi:2019bmo}. In particular, at the LHC a real massless dark-photon production in the Higgs boson decay $H\to \gamma \bar{\gamma}$ has been analyzed \cite{Gabrielli:2016cut,Biswas:2016jsh}, including its corresponding signature at the future  $e^+e^-$ colliders \cite{Biswas:2015sha,Biswas:2017lyg}. Since the dark-photon behaves as missing energy in the detector, a resonant monocromatic photon emission at high transverse momentum, plus missing energy is predicted. Recent measurements of this signal have been carried out for the first time by the CMS collaboration at the LHC \cite{Sirunyan:2019xst}, providing a an upper bound on the observed branching ratio of the Higgs boson ${\rm BR}(H\to \gamma \bar{\gamma})< 4.6\%$ at the 95\% confidence level.
Finally, we would like to stress that this scenario can also forecast viable large rates for the decays $K_L\to \pi^0 Q\bar{Q}$ and $K^+ \to \pi^+ Q\bar{Q}$ \cite{Fabbrichesi:2019bmo}, where $Q$ is a light dark fermion of the dark sector, that are  in the sensitivity range of the KOTO experiment \cite{Ahn:2018mvc}, as well as the intriguing possibility of invisible neutral hadrons decays in the dark sector that could either explain the present neutron lifetime puzzle \cite{Barducci:2018rlx}.

In this framework, the magnetic-dipole interactions of SM fermions with dark photons, included the flavor-changing neutral current transitions, have been analyzed in \cite{Gabrielli:2016cut}. This scenario can provide a well defined theoretical framework where to address the origin of such effective couplings.

Inspired by this scenario, we will adopt here a more model independent approach to analyze the impact of such FC neutral current (FCNC) couplings on the $b\to s \ell^+\ell^-$ transitions and $R_{K,K^*}$ anomalies.
In particular, we will assume the existence of effective magnetic-dipole interactions of SM fermions with dark photons, that can affect the $b\to s \bar{\gamma}$ interactions and anomalous magnetic moments $g-2$ of leptons. At this purpose, we introduce the following effective Lagrangian
\bea
    {\cal L}_{\rm eff}=\sum_{q,q'}
    \frac{1}{2\Lambda^L_{q q^{\prime}}}[\bar{q}_R\sigma_{\mu\nu}q^{\prime}_L]F^{\mu\nu}_D +\frac{1}{2\Lambda^R_{q q^{\prime}}}[\bar{q}_L\sigma_{\mu\nu}q^{\prime}_R]F^{\mu\nu}_D+
    \sum_{\ell \ell^{\prime}} \frac{1}{2\Lambda_{\ell\ell^{\prime}}}[\bar{\ell}\sigma_{\mu\nu}\ell] F^{\mu\nu}_D\, ,
    \label{LeffDP}
\eea
where the indices $(q,q^{\prime})$ and $(\ell,\ell^{\prime})$ run over all the quarks and leptons species respectively, $\Lambda_{qq^{\prime}}$ and $\Lambda_{\ell\ell^{\prime}}$  the associated effective scales, and $F^{\mu\nu}_D=\partial^{\mu}A^{\nu}_D-\partial^{\nu}A^{\mu}_D$ is the corresponding $U(1)_D$ field strength associated to the dark photon field $A^{\mu}_D$.

Then, due to the contribution of the Lagrangian in Eq.(\ref{LeffDP}), the dark-photon mediated amplitude for the $b\to s \ell^+\ell^-$ is given by
\bea
M^{\rm DP}&=&-\frac{\eta_R}{\Lambda^R_{bs}\Lambda_{\ell\ell}}
\left[\bar{s}_L\sigma_{\mu\alpha}\hat{q}^{\alpha}b_R\right]\left[\bar{\ell}\sigma^{\mu\beta} \hat{q}_{\beta}\ell\right] \frac{i}{\hs} -
\frac{\eta_L}{\Lambda^L_{bs}\Lambda_{\ell\ell}}
\left[\bar{s}_R\sigma_{\mu\alpha}\hat{q}^{\alpha}b_L\right]\left[\bar{\ell}\sigma^{\mu\beta} \hat{q}_{\beta}\ell\right] \frac{i}{\hs}\, ,
\eea
where  $\eta_{L,R}=\pm 1$ absorbs the overall sign, and we assume the effective scales  $\Lambda^{L,R}_{bs}$ and $\Lambda_{\ell\ell}$ to be positive. In order to simplify the analysis, we will also assume a universal scale $\Lambda_{\ell\ell}$ in the {\it dark} magnetic-dipole contribution to leptons.
For notation of momenta and other symbols we refer to the previous sections.
In the following we will neglect the contribution of the $\Lambda^L_{bs}$ since in the interference term with SM amplitude it vanishes in the strange mass $m_s\to 0$ limit.

Now we can use the results in the previous sections to compute the dark-photon mediated contributions to the BR of $b\to s \ell^+\ell^-$ and on the $B\to (K,K^*)\ell^+\ell^-$. This can be simply obtained by performing a global replacement of $C_7^{\rm eff} F_2$ terms in all previous formulas of BRs. By retaining only the interference terms in the magnetic-dipole corrections, this consists in the following substitution 
\bea
C_7^{\rm eff} F_2\to \frac{\pi\eta}{G_F\sqrt{2}\alpha V^*_{ts} V_{tb}
  \Lambda^2_{\rm eff} \hat{m}_b}\, ,
\eea
where $\Lambda_{\rm eff}\equiv \sqrt{\Lambda^R_{bs}\Lambda_{\ell\ell}}$.
In the following, we will adopt a minimal approach in our analysis by assuming a universal $\Lambda_{\ell\ell}$ scale for both $\ell=e,\mu$ in the lepton sector.\footnote{We will see that conclusions will not change in the case of $\Lambda_{ee} \gg \Lambda_{\mu\mu}$, while the opposite case $\Lambda_{ee} \ll \Lambda_{\mu\mu}$ would be unable to explain the $R_{K^*}$ anomalies, due to the constraints on the $g-2$ of the electron.}

\begin{figure}
\begin{center}
\hspace{0.cm}
\includegraphics{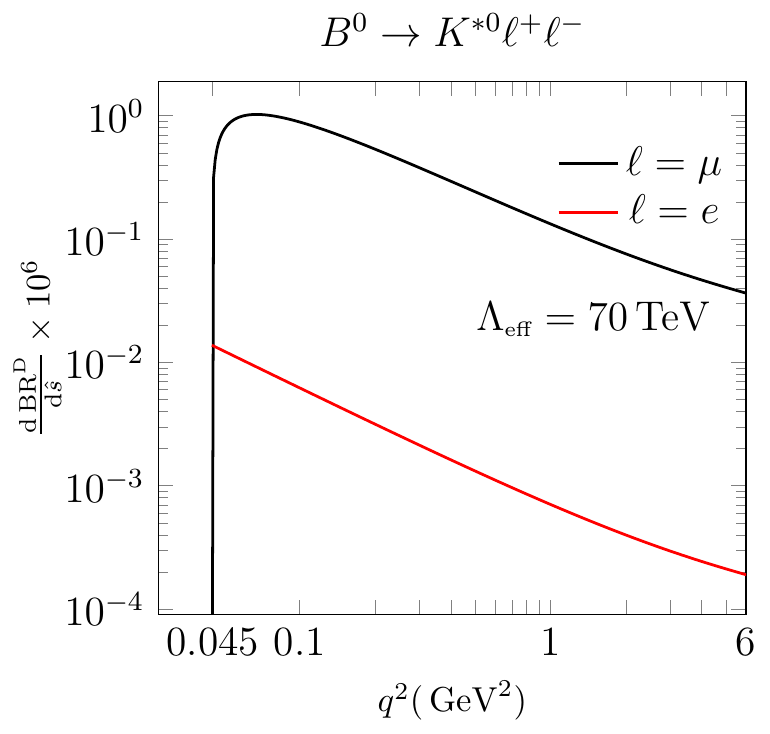}
\hspace{0.7cm}
\includegraphics{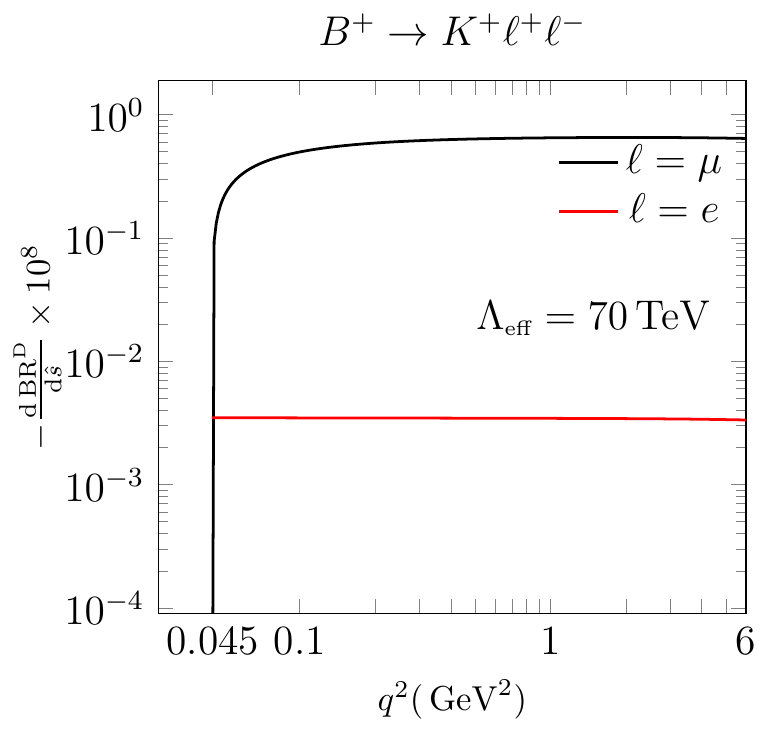}
\caption{The contribution of the magnetic-dipole interaction to the
$d{\rm BR}^{\rm DP}/d\hs$ as a function of $q^2$ for a representative value of $\Lambda_{\eff}=70{\rm TeV}$.}
\label{fig6}
\end{center}
\end{figure}

In Fig.\ref{fig6} we present the results for the differential $d{\rm BR}^{\rm DP}/d\hs$ versus $q^2$ induced by the dark-photon corrections for $B\to K^*$ (left plot) and $B\to K$ (right plot) transitions, for a representative scale of $\Lambda_{\eff}=70{\rm TeV}$. 
These plots should be compared with the corresponding ones of QED magnetic-dipole corrections in Fig.\ref{fig2}. Corresponding results for different values of $\Lambda_{\eff}$ can be simply rescaled from  Fig.\ref{fig2}, by using the relation ${\rm BR}^{\rm DP}\sim 1/\Lambda_{\eff}^2$.
As we can see from these results, the dark-photon induced corrections to the BRs are manifestly LFU-violating. These are of the order of ${\cal O}(10\%)$ for the $B\to K^* \mu^+\mu^-$ channel and roughly two order of magnitude smaller in the $B\to K \mu^+\mu^-$ case. As for the analogous QED corrections discussed in the previous sections, this is due to the fact that the contribution mediated by the magnetic-dipole interactions is enhanced in the $B\to K^*$ transitions with respect to the $B\to K$ ones. Much smaller effects are obtained for the electron-positron final state, due to the chiral suppression proportional to the lepton mass, induced by the interference term with SM amplitude.

Finally, we report below  the predictions for the  $\Delta R^{\rm DP}_{K^*}$ and $\Delta R^{\rm DP}_K$ corrections induced by the dark-photon exchange,  for a particular set of $q^2$ bins (in $\GeV^2$), corresponding to $\eta=1$ and $\Lambda_{\eff}=70 {\rm TeV}$ 
\bea
\Delta R^{\rm DP}_{K^*}[0.045,0.5]&=& 0.13~,~
\Delta R^{\rm DP}_{K^*}[0.045,1.1]= 0.12~,~
\Delta R^{\rm DP}_{K^*}[1.1,6]= 0.046\, ,
\label{RKsDP}
\eea
\bea
\Delta R^{\rm DP}_{K}[0.045,0.5]&=& 0.0054~,~
\Delta R^{\rm DP}_{K}[0.045,1.1]= 0.0058~,~
\Delta R^{\rm DP}_{K}[1.1,6]= 0.0060\, .
\label{RKDP}
\eea
As we can see from the above results, an effective scale around $70{\rm TeV}$ can easily generate corrections of order of $10\%$ on $R_{K^*}$ in the bin [0.045,1.1], while these become of order of $5\%$ in the larger bin [1.1,6]. Clearly, by an appropriate choice of $\Lambda_{\eff}$ in the range $\Lambda_{\eff}\sim 50-70{\rm TeV}$ and $\eta=-1$, the dark-photon correction could easily match the required gap to explain the present SM anomalies on $R_{K^*}$. However, this NP cannot simultaneously account for an analogous explanation of the anomaly in $R_K$ (which would also require a contribution of order of 10\%),  since its effect is of the order of $0.5\%$. To generate a 10\% effect, a smaller scale of the order of $\Lambda\sim 15 {\rm TeV}$ would be required, but this would give a too large contribution to $R_{K^*}$.

In order to see the phenomenological viability of such scenario, we analyze the experimental constraints on the $\Lambda_{\rm eff}$.
For this purpose, we consider two different approaches:
\begin{itemize}
\item[i)] {\bf model independent:} we do not make any assumption on the specific dynamics of NP which generates the effective scales $\Lambda_{bs}$ and $\Lambda_{ll}$ appearing in Eq.(\ref{LeffDP}). In this case, these scales should be considered independent from the corresponding NP contribution to the corresponding effective scales associated to the SM magnetic-dipole operators as in Eq.(\ref{LeffDP}) (with dark photon replaced by the ordinary photon).
\item[ii)] {\bf model dependent:} we assume the effective scales in Eq.(\ref{LeffDP}) to be generated by radiative corrections in specific  renormalizable models for the dark sector. As an example, we take first as benchmark model the one in \cite{Gabrielli:2013jka,Gabrielli:2016vbb}, that predicts these effective scales at 1-loop \cite{Gabrielli:2016cut}.
In this case, the NP provides correlated contributions to both scales associated to the magnetic-dipole operators with ordinary photon and dark photon \cite{Gabrielli:2016cut}. Then, experimental constraints from the $b\to s \gamma$ decay and anomalous magnetic moment of the muon $(g-2)_{\mu}$ can be used to directly constrain the $\Lambda_{bs}^{R}$ and $\Lambda_{\ell\ell}$ scales. We consider then a generalized extension of this model that allows to generate independent NP contributions to these two effective scales, by a suitable choice of the free parameters, that would theoretically support the main assumption of the model independent analysis.
\end{itemize}

\subsection{Model independent analysis} 
We start by analyzing the first hypothesis i). Since dark-photons behave as missing energy at colliders, a direct bound on the scales $\Lambda^{L,R}_{bs}$ can arise from the upper bounds on the BRs of the decay $b\to s X_{\rm inv}$ (where $X_{\rm inv}$ stands for the inclusive invisible channel) or from the $B \to K^* \nu \bar{\nu}$ decay, where the invisible system $\nu \bar{\nu}$ is replaced by a massless dark-photon.\footnote{The decay process $B \to K^* \nu \bar{\nu}$ could also set constraints on the effective scale associated to the dark magnetic-dipole interactions for the $b\to s \bar{\gamma}^*$ transitions, with $\bar{\gamma}$ an off-shell dark-photon. However, the bounds in this case would not enter here directly since they would depend also by the choice of the effective scale of dark-magnetic dipole operator associated to neutrinos.} Indirect bounds could come instead from the $B_s\bar{B}_s$ mixing.

We first consider the constraints from the inclusive decay  $b\to s X_{\rm inv}$ since these do not depend on the model predictions for the hadronic-form factors. In this case, the ${\rm BR} (b\to s \bar{\gamma})$ can be conventionally expressed through the experimental BR of semileptonic
B decay $B\to X_c \bar{\nu}e$ \cite{Gabrielli:2016cut}
\bea
    {\rm BR} (b\to s \bar{\gamma})&=&\frac{12 {\rm BR}^{\rm exp}(B\to X_c \bar{\nu}e)}{G_F^2 |V_{cb}|^2 m_b^2 f_1(z_{cb})}\left(\frac{1}{
        (\Lambda_{bs}^L)^2}+\frac{1}{(\Lambda_{bs}^R)^2}\right)\, ,
\eea
where the function $f_1(x)=1-8x+8x^3-x^4-12x^2\log{x}$. An experimental bound on 
the ${\rm BR} (b\to s X_{\rm inv})<{\cal O}(10\%)$ \cite{Ciuchini:1996vw}  might set some constraints on the scale $\Lambda_{bs}^R$, in particular we get
\bea
\Lambda_{bs}\gsim 3\times 10^3 {\rm TeV}\,
\label{limLambdabs}
\eea
where we assumed for simplicity $\Lambda_{bs}\equiv \Lambda_{bs}^L=\Lambda_{bs}^R$.

Recently, the Belle collaboration has measured the decay processes
$B\to h \nu \bar{\nu}$ with $h$ corresponding to various mesons, including $h$ as  kaons $K$ and $K^*$, pions- and rho-mesons \cite{Grygier:2017tzo}. Negative searches for these signals, have set upper bounds on the corresponding BRs, all compatible with SM expectations. In the particular case of the $B\to K^* \nu \bar{\nu}$ decay, the following upper bound on the corresponding BR has been reported ${\rm BR}(B\to K^* \nu \bar{\nu}) < 2.7\times 10^{-5}$ at 90\% C.L. \cite{Grygier:2017tzo}. The invisible $\nu\bar{\nu}$ system can be detected as missing energy $E_{\rm miss}$, with a continuous invariant mass. A kinematic cut  on the missing energy in the center of mass of B meson decay, corresponding to $E_{\rm miss} > 2.5 {\rm GeV}$ has been implemented in the analysis.
In principle, one can think that these upper bound could also be valid for
the $B\to K^* \bar{\gamma}$ decay, with $\bar{\gamma}$ a massless on-shell dark photon, since the latter behaves as missing energy in the detector, with a massless invariant mass. Indeed, the dark-photon energy in the same frame is fixed and corresponds to $E_{\bar{\gamma}}=E_{\rm miss}=2.56 {\rm GeV}$. However, it is not actually correct to apply these limits to the $B\to K^* \bar{\gamma}$ decay, since a dedicated analysis for this case is missing \cite{Belle-pcomm}.
Nevertheless, in the following we will assume the Belle upper bound on the ${\rm BR}(B\to K^* \nu \bar{\nu})$ to be valid also for the $B\to K^* \bar{\gamma}$ decay process and analyze its phenomenological implications for the dark-photon scenario.

Concerning the analogous upper bound on the $B\to K \nu \bar{\nu}$ process \cite{Grygier:2017tzo}, notice that this cannot be applied to the $B\to K \bar{\gamma}$ decay process, since the BR for this process exactly vanishes in the massless dark-photon case, due to angular momentum conservation \cite{Fabbrichesi:2017vma}.

Finally, using the effective Lagrangian in Eq.(\ref{LeffDP}), and the hadronic matrix elements in  Eq.(\ref{KstarME}), we obtain for the corresponding decay width
\bea
\Gamma(B\to K^* \bar{\gamma}) &=& \frac{m_B^3\left[|T_1(0)|^2+|T_2(0)|^2\right]\left(1-r_{K^*}\right)^3}  {\Lambda^2_{bs} 32\pi}\, ,
\eea
where the $T_{1}(0)=T_2(0)\simeq 0.28$ are the hadronic form factors evaluated at $\hat{s}=0$ and defined in Eq.(\ref{KstarME}). For the BR we get
\bea
{\rm BR}(B\to K^* \bar{\gamma})=4.94 \times 10^5 \left(\frac{\rm TeV}{\Lambda_{bs}}\right)^2\, .
\eea
Then, by using the upper limits of  ${\rm BR}(B\to K^* \nu \bar{\nu}) < 2.7\times 10^{-5}$, we get
\bea
\Lambda_{bs} > 1.35 \times 10^5 {\rm TeV}\, .
\label{Bellebounds}
\eea
As we can see, the above upper bound on $\Lambda_{bs}$ is about 45 times stronger than the corresponding one from the inclusive decay $b\to s X_{\rm inv}$.

Concerning the limits on the effective scale $\Lambda_{\mu\mu}$, the corresponding magnetic-dipole vertex with dark photon could give a contribution to the anomalous magnetic moment of the muon $a^{\rm DP}_\mu\equiv (g-2)^{\rm DP}_{\mu}$ at 1-loop, with a virtual muon and  dark-photon fields running inside. The result is divergent in the effective theory (due to the double insertion of the magnetic-dipole operator in the 1-loop contribution) and so the magnetic-dipole operator needs to be renormalized. Notice that, in the full  UV completion of the theory, responsible to generate the low-energy effective magnetic-dipole interaction of dark-photon with SM fermions \cite{Gabrielli:2016cut}, this contribution corresponds to a 3-loop effect and it turns out to be finite, although it will be dependent from the model structure of the UV theory. However, one can estimate this contribution at low energy by subtracting the divergence thus renormalizing the operator (we used the $\overline{{\rm DR}}$ scheme). In particular, we have computed it in dimensional reduction (DRED) scheme, and, after subtracting the $1/{\bar{\epsilon}}=1/\epsilon-\gamma_E+\log{(4\pi)}$ divergence, coming from the momentum-loop integration in $D=4-\epsilon$ dimensions, we get
\bea
a^{\rm DP}_{\mu} =-\frac{3 m^2_{\mu}}{32 \pi^2 \Lambda_{\mu\mu}^2}\left(5+4
  \log{\left(\frac{\mu^2}{m_{\mu}^2}\right)}\right) \, .
\eea
Then, by evaluating the above contribution at the muon-mass renormalization scale $\mu=m_{\mu}$ scale, we obtain \cite{Fabbrichesi:2020wbt}
\bea
a^{\rm DP}_{\mu} =-\frac{15 m^2_{\mu}}{32 \pi^2 \Lambda_{\mu\mu}^2} \, ,
\label{amudark}
\eea
that is in agreement (by order of magnitudes) with the n\"aive estimation based on dimensional analysis and chiral structures of the operators involved. The result in Eq.(\ref{amudark}) should be understood as a rough estimation of the exact (finite) result that can be obtained in a whole UV completion of the theory. Notice that, also the minus sign in front of Eq.(\ref{amudark}) should not be taken as granted, since the correct sign prediction can only be obtained in the UV completion of the theory and it will be then model dependent.

At the moment there is a $3.7\sigma$ deviation from the experimental measurement and SM prediction, in particular the  discrepancy $\Delta a_{\mu}$ is at the level \cite{Bennett:2006fi,Blum:2018mom}
\bea
\Delta a_{\mu}=a_{\mu}^{\rm exp}-a_{\mu}^{\rm SM}=(2.74\pm 0.73)\times 10^{-9}\, .
\eea
where the term in parenthesis summarizes the $1\sigma$ error. If we impose that the above NP contribution to $a_{\mu}$ lies within the $2\sigma$ error band of $\delta a_{\mu}$, taking (conservatively) the largest effect, we get
\bea
 |a^{\rm DP}_{\mu}| < 4.2 \times 10^{-9}\,
\label{boundsg2mu}
\eea
that from Eq.(\ref{amudark}) implies $\Lambda_{\mu\mu}> 355 {\rm GeV}$. If instead of the $4.2 \times 10^{-9}$ upper limit we would require the less conservative and more stringent constraint $|a^{\rm DP}_{\mu}|< 2\sigma\sim 1.5\times 10^{-9}$, that would also make sense in the case of a negative contribution to $a_{\mu}$, we get $\Lambda_{\mu\mu} > 594 {\rm GeV}$.

Now, by combining the above bounds on $\Lambda_{\mu\mu}$ from Eq.(\ref{boundsg2mu}) with the bounds on the
$\Lambda_{bs}$ scale derived from the $b\to s X_{\rm inv}$ in Eq.(\ref{limLambdabs}), we get 
\bea
\Lambda_{\rm eff}\gsim 34 {\rm TeV}\, ,
\eea
that would be pushed up to $\Lambda_{\rm eff}\gsim 42 {\rm TeV}$ if $|a^{\rm DP}_{\mu}|<2\sigma$, 
where the effective  $\Lambda_{\rm eff}$ appearing above is defined as $\Lambda_{\rm eff}=\sqrt{\Lambda_{\mu\mu}\Lambda_{bs}}$.
From these results we conclude that, the scale required to explain the $R_{K^*}$ anomalies
(which is of the order of  $70~ {\rm TeV}$) is well allowed by both the $(g-2)_{\mu}$ and $b\to s X_{\rm inv}$ constraints.

Now, assuming the Belle bounds on $B\to K^*\nu \bar{\nu}$ could be applied to $B\to K^* \bar{\gamma}$, from Eq.(\ref{Bellebounds}) we get
\bea
\Lambda_{\rm eff}\gsim 220 {\rm TeV}\, .  
\eea
Then, by rescaling the results in Eq.(\ref{RKsDP}) by a factor $(70/220)^2\sim 0.1$, we can see that these constraints would reduce the allowed dark-photon contribution to $R_K^*$ to a maximum 1-2\% effect, ruling out the possibility to fully explain the $R_K^*$ anomaly in terms of a NP dark-photon contribution.

Regarding the other potential constraint from the $B_s \bar{B}_s$ mixing, induced by the magnetic-dipole interactions in Eq.(\ref{LeffDP}), this computation requires in principle the evaluation of a non-perturbative long-distance effect. However, we can make use of perturbation theory for the magnetic-dipole operator. In this case the tree-level contribution induced by exchange of the virtual dark-photon between $B$ and $\bar{B}^0$ is zero
\bea
\Delta M_{B_s}&\sim & \langle B^0_s(q) |[\bar{s}\sigma_{\mu\nu}q^{\nu}b]|0\rangle
\langle 0 |[\bar{s}\sigma_{\mu\nu}q^{\nu}b]| \bar{B}^0_s(q)\rangle\frac{1}{m_B^2} =0\, , \eea
since the corresponding $B$ matrix elements vanish due to the energy and angular momentum conservation. Then, the next (non-vanishing) contribution is expected to appear  at higher loops, and therefore to be suppressed.\footnote{However, considering the non-perturbative aspect of such computation, that involves the evaluation of non-perturbative long-distance contributions induced by the magnetic-dipole interactions among external hadron states, it is difficult to correctly estimate the magnitude of this effect. A more careful analysis would be required in this case, that goes beyond the purposes of the present paper.}

\subsection{Model dependent analysis}

Here we consider the model dependent analysis for the predictions
of the effective scales in Eq.(\ref{LeffDP})   
in the {\it massless} dark photon scenario. This is based on a benchmark model
for the dark sector, inspired by \cite{Gabrielli:2013jka,Gabrielli:2016vbb}. This scenario was proposed as a solution of the Flavor hierarchy problem in the SM, where the SM Yukawa couplings are predicted to arise radiatively from a dark sector and exponentially spread.
This model contains dark fermions of up-  ($Q^{U_i}$) and down-type ($Q^{D_i}$, which are singlet under the full SM gauge interactions, and a set of heavy (above TeV scale) scalar messenger fields $S^{U_i,D_i}_{L,R}$, the latter carrying the same SM internal quantum numbers of quarks and leptons. Below we restrict our discussion to the quark sector, but it can be straightforwardly generalized to the lepton sector.

The dark fermions $Q^{U,D}$ couple to  the SM fermions by means of Yukawa-like interactions  given by 
\bea
{\cal L}  &\supset& \ g_R \Big\{ 
S^{\U_i \dag }_R \left[ \bar{Q}^{\U_i}_L (\rho^{\U}_R)_{ij} q^j_R \right]  +
S^{\D_i \dag }_R \left[ \bar{Q}^{\D_i}_L (\rho^{\D}_R)_{ij} q^j_R\right]   \Big \} 
\nonumber \\
 &+&
 g_L \Big\{
S^{\U_i\dag }_L \left[\bar{Q}^{\U_i}_R (\rho^{\U}_L)_{ij}q^j_L\right]  
+S^{\D_i\dag }_L\left[\bar{Q}^{\D_i}_R (\rho^{\D}_L)_{ij} q^j_L\right]  \Big\}
 + \mbox{H.c.}\, ,
\label{L}
 \eea 
where the index $i$ run over the family generations, and $q^i_{L,R}$ are the usual SM $SU(2)_L$ doublet and singlet quark fields respectively.
In Eq.(\ref{L}), the fields $S^{\U_i,\D_i}_{L}$ and $S^{\U_i,\D_i}_{R}$ are the messenger scalar particles, respectively doublets and singlets of the SM $SU(2)_L$ gauge group as well as  $SU(3)$ color triplets (color indices are implicit in Eq.(\ref{L}). The various symmetric matrices $(\rho)_{ij}=(\rho)_{ji}$ are the result of  the diagonalization of the mass matrices in the mass eigenstates of both the  SM and dark fermions, and provide the required generation mixing to contribute to the flavor physics. The messenger  fields are also charged under the $U(1)_{\D}$ gauge interaction, and carry the same quantum $U(1)_D$ charges as the dark fermions they are coupled to. For more details of the model we refer to
\cite{Gabrielli:2016vbb,Gabrielli:2013jka}.

\begin{figure}
\begin{center}
\hspace{0.3cm}
\includegraphics[width=0.8\textwidth]{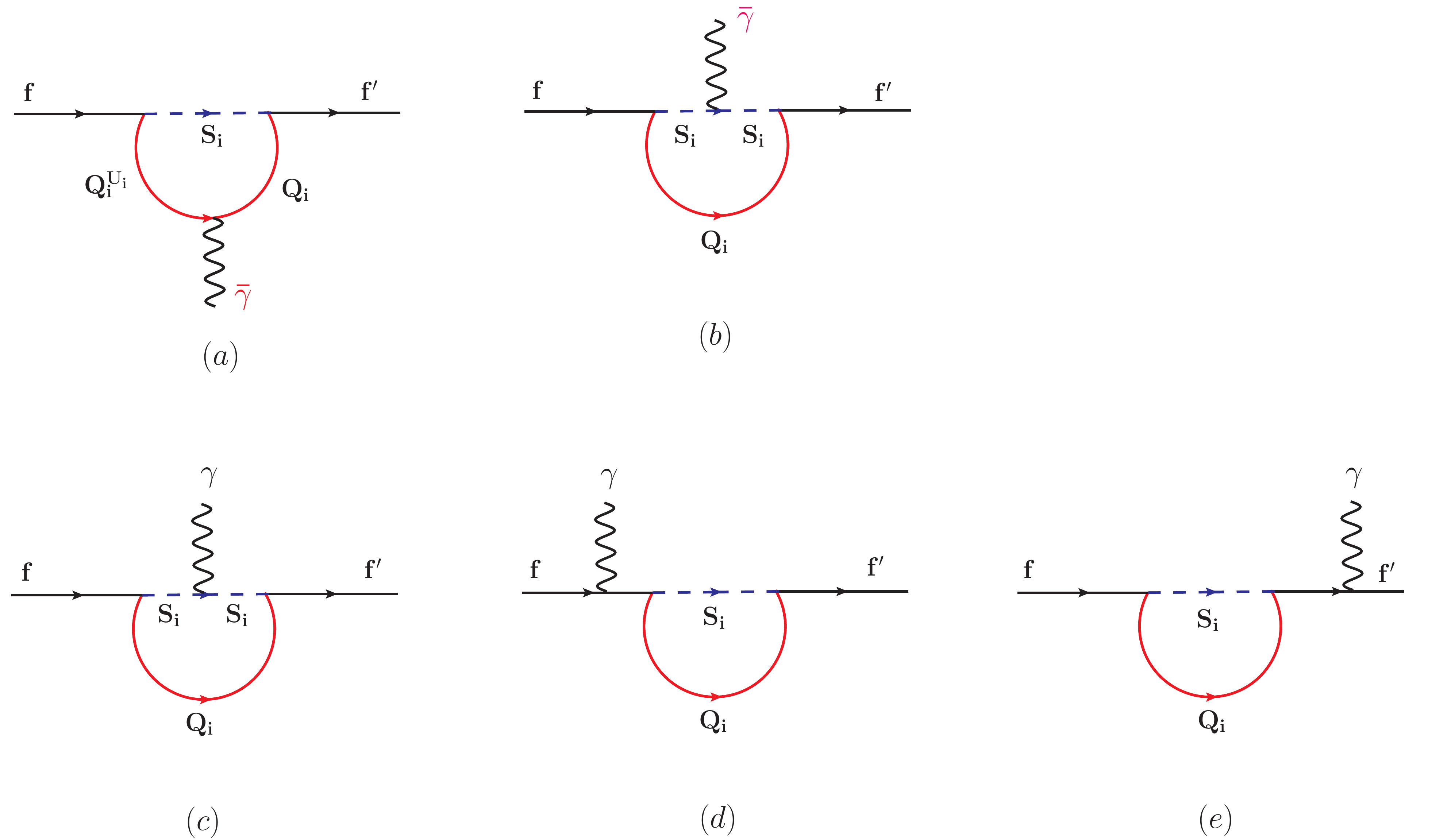}
\caption{Feynman diagrams for the contributions to the dark-photon $\bar{\gamma}$ (a-b) and  photon $\gamma$  (c-e) magnetic-dipole interactions \cite{Gabrielli:2016cut}. Dark continuous lines correspond to external SM fermions $f,f^{\prime}$, (red continuous lines) and (blue dashed lines) refer to generic dark-fermion ($Q_i$) and scalar messenger ($S_i$) propagators respectively.}
\label{figDP}
\end{center}
\end{figure}

The Lagrangian in Eq.(\ref{L}) can induce at 1-loop contributions to both the usual magnetic-dipole interactions with ordinary photon and the effective interactions in Eq.(\ref{LeffDP}) \cite{Gabrielli:2016cut}
\bea
    {\cal L}_{\rm eff}&\supset& \sum_{q,q'}
    \frac{1}{2\bar{\Lambda}^L_{q q^{\prime}}}[\bar{q}_R\sigma_{\mu\nu}q^{\prime}_L]F^{\mu\nu} +\frac{1}{2\bar{\Lambda}^R_{q q^{\prime}}}[\bar{q}_L\sigma_{\mu\nu}q^{\prime}_R]F^{\mu\nu}+
    \sum_{\ell \ell^{\prime}} \frac{1}{2\bar{\Lambda}_{\ell\ell^{\prime}}}[\bar{\ell}\sigma_{\mu\nu}\ell] F^{\mu\nu}\, ,
    \label{Leffphoton}
\eea
that add to the corresponding SM contributions. For more details, see
\cite{Gabrielli:2016cut}. The corresponding Feynman diagrams are given in Fig.\ref{figDP}.

The predictions for the above scales in the $b\to s \gamma\, (\bar{\gamma})$ transitions are 
\bea
\frac{1}{\Lambda_{bs}}&\sim& \frac{e_Dg_Lg_R \rho_R\rho_L\xi m_Q}{16 \pi^2 m_S^2} f_2(x,\xi)\,, ~~~~
\frac{1}{\bar{\Lambda}_{bs}} \sim  \frac{e g_Lg_R \rho_R\rho_L\xi m_Q}{16 \pi^2 m_S^2} \bar{f}_2(x,\xi)\, ,
\label{scalesmodel}
\eea
where  $e_D$ and $e$ are the unity of $U(1)_D$ and EM charges respectively, and $0<\xi<1$ parametrizes the mixing in the left-right sector of the messenger mass matrix. For the definitions of the dark-fermion-messenger-quark couplings $g_{L,R}$ and $\rho_{L,R}$ see Ref.\cite{Gabrielli:2016cut} for more details. Here $f_2$ and $\bar{f}_2$ are the corresponding loop functions, with $x=m_Q^2/\bar{m}_S^2$, where $m_Q$ and $\bar{m}_S$ are the mass of the heaviest dark fermion and the average mass of messenger fields running in the loop respectively. The analytical expressions for the functions $f_2(x,y)$ and $\bar{f}_2(x,y)$ are \cite{Gabrielli:2016cut}\footnote{Notice that the functions $f_2(x,y)$ and $\bar{f}_2(x,y)$ appearing here correspond to the $\xi f_2(x,y)$ and $\xi \bar{f}_2(x,y)$ ones respectively in the notations of \cite{Gabrielli:2016cut}.}
\bea
f_2(x,y)&=&\left\{\frac{1-x+y+(1+y)\log{\left(\frac{x}{1+y}\right)}}{
  2\left(1-x+y\right)^2}\right\}\, -\left\{y \to -y\right\}
\nonumber\\
\bar{f}_2(x,y)&=&\left\{\frac{(1+y)^2-x^2+2x\left(1+y\right)
   \log{\left(\frac{x}{1+y}\right)}}{4\left(x-1-y\right)^3}\right\}\, -\left\{y \to -y\right\}\, .
\nonumber
\eea
Formally we have the same expressions as in Eq.(\ref{scalesmodel}) for the effective scales $\Lambda_{\ell\ell}$  and $\bar{\Lambda}_{\ell\ell}$, with the same loop functions $f_2$ and $\bar{f}_2$, where one should replace the couplings, corresponding messenger mass, dark-fermion mass with the ones entering in the loop contribution. The only difference, is that in the flavor-diagonal transitions the $\rho_{L,R}$ parameters should be set to 1.

\begin{figure}
\begin{center}
\hspace{0.3cm}
\includegraphics[width=0.5\textwidth]{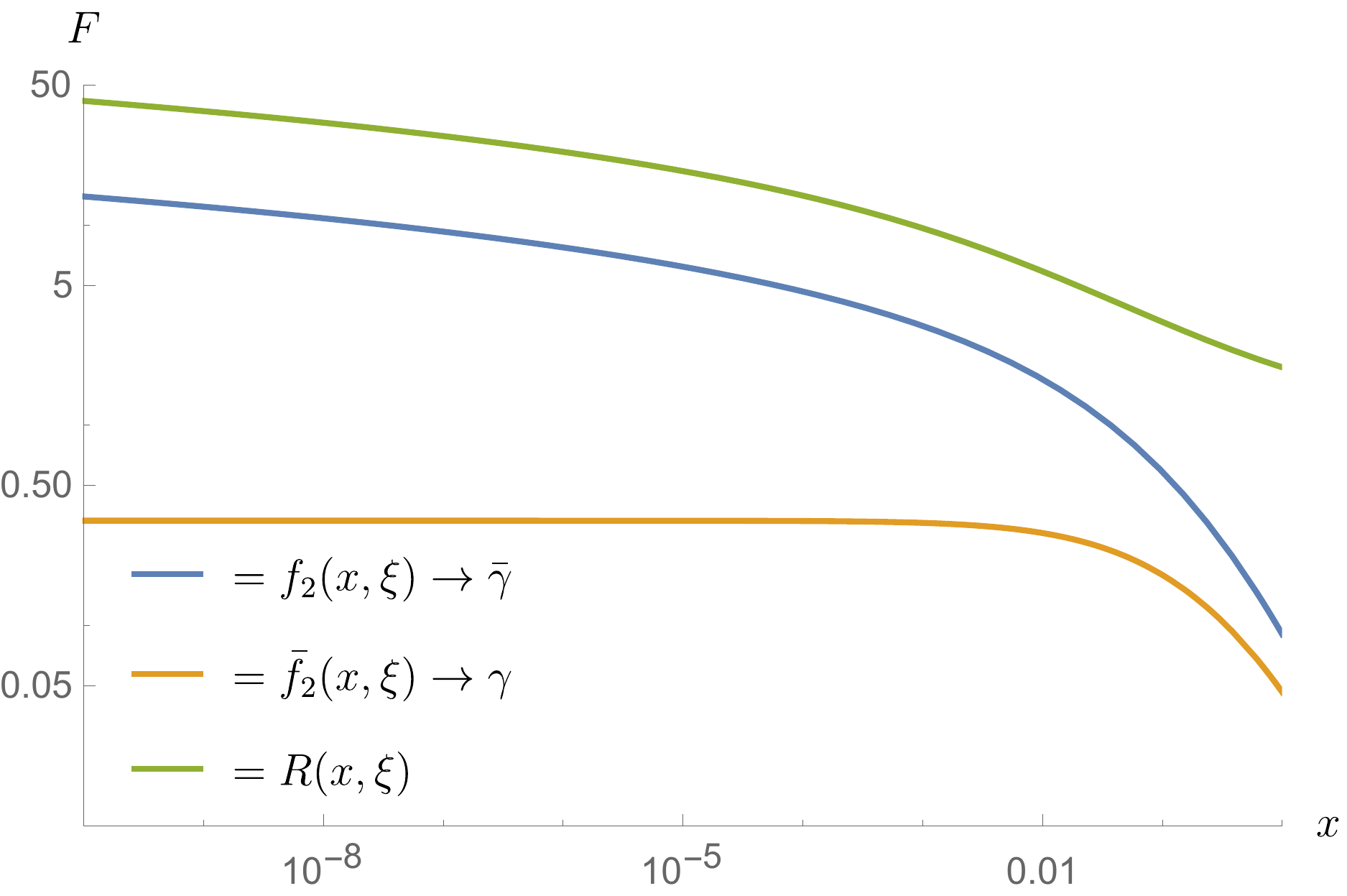}
\caption{Plots of the function $F$ corresponding to the ratio $R(x,\xi)=\frac{f_2(x,\xi)}{\bar{f}_2(x\,\xi)}$, and the loop functions $f_2$  and $\bar{f}_2$ associated to the magnetic-dipole scales for dark-photon ($\bar{\gamma}$)  and photon ($\gamma$) respectively, versus $x=m^2_Q/m^2_S$, with $m_Q$ and $m_S$ the dark fermion and  average messenger mass respectively.  Plots correspond to the representative value of the mixing parameter (in the messenger mass sector) $\xi=\frac{1}{2}$.}
\label{Rfunct}
\end{center}
\end{figure}

A more predictive result of this model is the ratio of the two scales, that depends only by the $U(1)_D$ strength in the dark sector $\alpha_D$, and the ratio of the heaviest dark-fermion mass running in the loop over the average messenger mass, namely
\bea
\bar{\Lambda}^{L,R}_{q q^{\prime}}&\simeq&\Lambda^{L,R}_{q q^{\prime}}
\sqrt{\frac{\alpha_D}{\alpha}} R(x,\xi)\, ,
\label{Lambdaratio}
\eea
where $R(x,\xi)=f_{2}(x,\xi)/\bar{f}_{2}(x,\xi)$. In Fig.\ref{Rfunct} we plot the ratio $R(x,\xi)$ and the functions $f_{2}(x,\xi)$,$\bar{f}_2(x,\xi)$ versus $x$ for the representative value of $\xi=0.5$. Other values of $0<\xi<1$ does not change the whole picture. As we can see a small ratio is reached for $x\ll 1$. This is due to a $\log(x)$ enhancement in the limit $x\ll 1$, corresponding to a small dark-fermion mass. This originates from the diagram  in Fig.\ref{figDP}(a) where the dark-photon is coupled to internal dark-fermion lines, which gives rise to the pure $\log{x}$ term. The latter is absent in the photon contribution in Fig.\ref{figDP}[(c)-(e)], being the dark-fermions electrically neutral. Therefore, potential large enhancement can be achieved, up to 2 order of magnitude, in the ratio $\frac{\Lambda^{L,R}_{q q^{\prime}}}{\bar{\Lambda}^{L,R}_{q q^{\prime}}}$ for large values of $\alpha_D$ couplings, i.e.
$\alpha_D\sim 0.1$ and small $x$.
In particular, for a realistic benchmark point of $m_Q=1\GeV$, $m_S=5\TeV$, and $\alpha_D=0.1$ we get  the following relation between the two scales
\bea
\bar{\Lambda}^{L,R}_{q q^{\prime}}\sim 110\, \Lambda^{L,R}_{q q^{\prime}}\, ,
\label{rescaling}
\eea
that shows a large enhancement.

Now, we provide the lower bound on the effective scale $\Lambda_{\rm eff}$ coming from the constraints on the $B\to X_s\gamma$ and $g-2$ of the muon.
By using the $B\to X_s\gamma$ constraints at 95 C.L., with the corresponding  $\BR(B\to X_s\gamma)$ evaluated at the next-to-leading (NLO) in QCD, we can derive a (conservative) lower bound on the effective scale $\bar{\Lambda}^{{\scriptscriptstyle R}}_{bs}$, which is given by \cite{Gabrielli:2016cut}
\bea
\bar{\Lambda}^{{\scriptscriptstyle R}}_{bs} > \pi \left(\sqrt{2} G_Fm_b |V_{ts}^*V_{tb}| C_7(M_W) R_7^{\rm min}\right)^{-1} &\simeq& 3.8 \times 10^4 {\rm TeV}\, ,
\label{bound1}
\eea
where $R_7^{\rm min} =0.139$ at $2\sigma$ (see \cite{Gabrielli:2016cut} for derivation), and other SM inputs can be found in table \ref{tabinput1}.

By taking into account the Lagrangian in
Eq.(\ref{Leffphoton}), we have for its contribution to  $(g-2)_{\mu}$
\bea
\Delta a_{\mu}&=&\frac{2m_{\mu}}{e\bar{\Lambda}_{\ell\ell}}\, .
\nonumber
\eea
Then, by applying the constraint in Eq.(\ref{boundsg2mu}) to $\Delta a_{\mu}$ above we get
\bea
\bar{\Lambda}_{\ell\ell}> 1.8\times 10^5{\rm TeV}\, .
\label{bound2}
\eea
Using the predicted values of the effective scales for the photon couplings $\bar{\Lambda}_{bs}$ in Eq.(\ref{scalesmodel}), we find for the colored messengers mass
\bea
m_S >\sqrt{\frac{e\Lambda_{\rm bound} m_Q\xi \bar{f}_2(x,\xi)}{16\pi^2}}\, ,
\label{boundmS}
\eea 
where $\Lambda_{\rm bound}=3.8\times 10^4 {\rm TeV}$,  with $x=m_Q^2/m_S^2$, where $m_Q$ and $m_S$ stand for the associated dark-fermion and messenger mass respectively entering in the effective scale, and $e$ is units of electric charge. Above, in order to maximize the effect, we assumed the other parameters to be of order one, namely the couplings $g_{L,R},\rho_{L,R}=1$.
As we can see for the plot in Fig. \ref{Rfunct} the loop function $\bar{f}_2$ for small $x\ll 1$ tends to a constant value of the order $\bar{f}_2\sim 0.33$ for $\xi=1/2$. By replacing $\bar{f}_2\to 0.33$ and $\xi=1/2$ inside Eq.(\ref{boundmS}), we get for the lower bound on the colored messenger mass
\bea
m_S \gsim 156 \GeV \sqrt{\frac{m_Q}{\GeV}}\, .
\label{mSboundcol}
\eea
The result in Eq.(\ref{mSboundcol}) provides the minimum value of the dark fermion mass $m_Q$, as a function of $m_S$, that is sitting on the lower bound scale $\Lambda_{\rm bound}$ required by $b\to s \gamma$ namely $\bar{\Lambda}_{bs} \ge \Lambda_{\rm bound}$. Analogously, for the lower bound on electroweak messengers mass from $(g-2)_{\mu}$ constraint we find
\bea
m_S \gsim 337 \GeV \sqrt{\frac{m_Q}{\GeV}}\, .
\label{mSboundEW}
\eea
For instance, by imposing the (conservative) experimental lower bounds on colored and electroweak messenger masses sectors from direct searches at colliders, respectively $m_S>1$TeV and $m_S> 300$ GeV, we get from Eq.(\ref{mSboundcol}) and Eq.(\ref{mSboundEW})  $m_Q \gsim 40 \GeV$ and $m_Q\gsim 1\GeV$ respectively.

Then, by replacing $x\to m_Q^{\rm min}/m^{\rm min}_S$  inside the $R(x,\xi)$ function, with the minimum values $m^{\rm min}_S$ and $m^{\rm min}_Q$ given above for colored and EW sectors, we get from  Eq.(\ref{Lambdaratio}) the lower bounds $\Lambda_{\rm bs}> 1200 {\rm TeV}$ and $\Lambda_{\rm \mu\mu}> 2500 {\rm TeV}$ respectively  for $\alpha_D=0.1$, that corresponds to
\bea
\Lambda_{\rm eff}\gsim 1700\, {\rm TeV}\sqrt{\frac{0.1}{\alpha_D}}\, .
\label{boundLambda}
\eea
This is the minimum allowed value of the effective scale $\Lambda_{\rm eff}$ entering in the dark magnetic-dipole corrections to $R_{K,K^*}$ required to satisfy the bounds from $b\to s \gamma$ and $(g-2)_{\mu}$ constraints.

As we can see from Eq.(\ref{boundLambda}), for $\alpha_D=0.1$ this scale is approximately 20 times larger than the required one $(70 \TeV)$ to generate large deviation of order 10\% on the $R_{K^*}$ via {\it massless} dark-photon exchanges. Even taking a large $U(1)_D$ coupling in the dark sector, bordeline with perturbation theory ($\alpha_D\sim 1$), the lower bound on $\Lambda_{\rm eff}$ would be still 6 times larger than the required one.
These conclusions will not be affected by a different choice for $g_{L,R},\rho_{L,R}$ and $\xi$ in the perturbative regime, since different values of these couplings could be reabsorbed in a rescaling of the $m_S$ and $m_Q$ masses.
Therefore, we conclude that in the framework of the particular dark sector scenario \cite{Gabrielli:2013jka,Gabrielli:2016vbb}, aimed to solve the flavor hierarchy problem, the $b\to s \gamma$ and $(g-2)_{\mu}$ constraints can fully rule out the possibility to explain the present $R_{K^*}$ anomaly.

\subsection{A viable dark sector model}
    
Here we will show the existence of a dark-sector model that would allow to circumvent the $b\to s \gamma$ and $(g-2)_{\mu}$ constraints, while providing potential sizeable contributions to the dark magnetic-dipole operators. This will require a straightforward extension of the dark-sector model analyzed in the previous section. Then, the phenomenological implication of this scenario for the $R_{K^*}$ anomaly will just fit into the model-independent analysis already discussed in section 6.1, and  we will not be repeated here.

The main idea consists in adding new contributions to the diagrams (c),(d),(e) in Fig.\ref{figDP}, that exactly cancel out or strongly suppress the contributions to the  magnetic-dipole operators, while providing a non-vanishing contribution to the dark magnetic-dipole operators induced by the diagrams (a),(b) in Fig.\ref{figDP}. The minimal model can be realized by requiring a degenerate $SU(2)$ mass replica of dark-fermions and messenger fields and suitable $U(1)_D$ charges and couplings of the portal sector.

In order to simplify both the discussion and notations, let us restrict the analysis to one flavor of dark-fermion $Q$ and corresponding messenger fields $S_L$, $S_R$, that as usual stands for SM $SU(2)_L$ gauge doublets and singlets respectively. Then, only diagonal transitions in the portal sector of Eq.(\ref{L}) are involved. The generalization of this mechanism to the full Lagrangian involving all flavors, including the off-diagonal flavor portal interactions in Eq.(\ref{L}), will be straightforward. The same conclusions derived for the one flavor analysis will hold in the general case too.

This extension consists in replacing the dark fermion and messenger fields of a specific flavor, with a  degenerate replica of an internal global $SU(2)$, that in vectorial notation reads
\bea
Q\to \hat{Q}=\left(
\begin{array}{c}
Q^{1}\\
Q^{2}\\
\end{array}
\right)\, , ~~~  \hat{S}_{L,R}\to\left(
\begin{array}{c}
S^{1}_{L,R}\\
S^{2}_{L,R}\\
\end{array}
\right)\, ,
\eea
where  $\hat{Q}$ and $\hat{S}_{L,R}$ can be interpreted as $SU(2)$ doublets. Each $S^{\alpha}_{L}$ and $S^{\alpha}_{R}$ fields, with $\alpha=1,2$, are also understood to $SU(2)_L$ doublets and singlets of the SM gauge group respectively. In the following we will omit the SM gauge indices for simplicity.

We assume the non-interacting free Lagrangian of this model to possess an exact {\it global} $SU(2)$ symmetry.
However, this symmetry will be broken by the interactions, in particular the dark $U(1)_D$ gauge interaction and the portal interaction.
The $U(1)_D$ gauge invariance requires the $U(1)_D$ charges $q$ of each species $\alpha$ to be equal, namely
\bea
q(Q^{\alpha})=q(\hat{S}^{\alpha}_{L,R})\,. 
\eea
Now we choose different  $U(1)_D$ charges within the same $SU2)$  multiplet, namely $q_1\equiv q(Q^{1})=q(\hat{S}^{1}_{L,R})$ and $q_2\equiv (Q^{2})=q(\hat{S}^{2}_{L,R})$, with the additional requirement $q_1\neq q_2$. This last condition explicitly breaks $SU(2)$ and it is crucial for the realization of the mechanism that we will explain below.

Concerning the portal interaction, this is a simple generalization of the one in Eq.(\ref{L}).  In particular, if we restrict the discussion to one flavor, the corresponding Lagrangian is
\bea
    {\cal L}_{\rm portal}  &=&g_R S^{\dag}_R \left[ \bar{\hat{Q}}^{\alpha}_L q_R \right] + \bar{g}_L\hat{S}^{\dag }_L T_3 \left[\bar{\hat{Q}}_R q _L\right]  
 + \mbox{H.c.}\, ,
\label{L2}
 \eea 
 where $q_L$ and $q_R$ indicate a generic SM fermions $SU(2)_L$ doublet  and singlet respectively, and the sum over the spin, $SU(2)$, colors and  $SU(2)_L$ gauge indices is understood. As in section 6.2, the terms in parenthesis $[\cdots]$ indicate the bi-spinorial products. Here $T_3$ in the second term proportional to $g_L$  stands for the third (diagonal) generator of the $SU(2)$ group, and its role will be understood in the following. The symmetries of the theory do not forbid the existence of the interaction term  $g_L\hat{S}^{\dag }_L\left[\bar{\hat{Q}}_R q _L\right]$ in the portal Lagrangian. Let us assume for the moment vanishing the contribution of this interaction term, by setting its overall constant $g_L\to 0$ or choosing it much smaller than all other constants $g_R$ and $g_L^{\prime}$.

 Now we go to the computation of the magnetic-dipole transitions. Assuming for the moment the contribution to a diagonal flavor transition of a SM fermion of charge $e$, the conclusions will be valid also for the general off-diagonal flavor case. The mass term of the dark fermion sector is $SU(2)$ degenerate as well as the masses of the $\hat{S}_L$ and $\hat{S}_R$, including the off-diagonal mixing mass term $m^2_{LR}$. By requiring a left-right symmetry we can impose the masses  of the $\hat{S}_L$ and $\hat{S}_R$ to be identical, and set them equal to $m_S$.

Consider first the contribution to the SM magnetic-dipole transitions. In this case, from the analogous diagrams (c),(d),(e) in Fig. \ref{figDP}, we get for the following result for the effective coupling $\bar{\Lambda}$ associated to the magnetic-dipole operator $[\bar{q}\sigma_{\mu\nu}q]F^{\mu\nu}$
 \bea
 \frac{1}{\bar{\Lambda}}\,\sim\, e{\rm Tr}[T_3]\frac{\bar{g}_Lg_R\xi m_Q }{m_S^2 16\pi^2}f_2(x,\xi)=0\, .
 \label{condition}
 \eea
 where the symbols $x$ and the mixing parameter $\xi$ are defined in the same way as in section 6.2. This contribution exactly vanishes due to the ${\rm Tr}[T_3]=0$ condition. Notice that the EM charge, factorizes since by gauge invariance inside the loop must circulate the same charge of the external SM fermion.

If we go now to compute the contribution to the dark-magnetic dipole operator $[\bar{q}\sigma_{\mu\nu}q]\bar{F}^{\mu\nu}$, and defining $\Lambda$ the corresponding associated effective scale, we obtain by means of analogous diagrams (a) and (b) of Fig.\ref{figDP}
\bea
\frac{1}{\Lambda}\,\sim\, e_D{\rm Tr}[T_3 \hat{q}_D]\frac{\bar{g}_Lg_R\xi m_Q }{m_S^2 16\pi^2}f_2(x,\xi) \,=\, e_D(q_1-q_2)\frac{\bar{g}_Lg_R\xi m_Q }{m_S^2 16\pi^2}f_2(x,\xi)  \neq 0\, ,
\label{condition1}
 \eea
provided $q_1\neq q_2$, where $e_D$ is the unit charge of $U(1)_D$ and
 $\hat{q}=
  \left( {\begin{array}{cc}
   q_1 & 0 \\
   0 & q_2 \\    \end{array} } \right)$
the corresponding charge eigenvalues matrix. Notice, this non-vanishing contribution is proportional to the $SU(2)$ breaking term connected to the $U(1)_D$ charge operator $\hat{q}$. The same argument can be easily generalized to include the other flavors and off-diagonal terms.
 Same results would be obtained if the $T_3$ matrix would have been inserted in the first term of the right-hand-side of Eq.({\ref{L2}) of the portal interaction.

   A last comment regarding the stability of these results under radiative corrections. First, notice that the cancellation among the diagrams is realized by assuming the exact degeneracy among the masses of dark fermions and messenger fields (as well as the mixing term) belonging to the $SU(2)$ multiplet. Now, this is a tree-level condition and radiative $U(1)_D$ corrections for instance are expected to break this degeneracy since $q_1\neq q_2$. However this splitting should be finite and small effect, being proportional to radiative ${\cal O}(\alpha_D/4\pi)$ correction. Then, we expect that the residual effect of this partial cancellation among diagrams, due to radiative corrections, is small being loop suppressed, as long as we keep the theory in the perturbative regime.
Second, we have choose the coupling $g_L$ to be vanishing, or much smaller than the other couplings in the portal interaction, namely $g_L\ll (g_R, \bar{g}_L)$. This condition is understood to be valid at some energy scale. However, we expect radiative corrections to regenerate the $g_L$ coupling at a different scales. Since the running of the couplings is governed by a perturbative renormalization group flow, we do not expect the hierarchy among the $g_L$ coupling and the other couplings, to be dramatically changed by (perturbative) radiative corrections. In other words, being an independent parameter, one can always choose the $g_L$ value in such a way to arbitrarily suppress the new physics contribution to the SM magnetic-dipole operators discussed above, leaving basically unchanged the conclusions of this analysis under the effect of radiative corrections.
   
In conclusion, we proved the existence of at least one viable model of the dark sector that allows to circumvent the $(g-2)_{\mu}$ and $b\to s \gamma$ constraints, while providing large contributions to the dark magnetic dipole operator. This can be realized by a suitable of choice of the free parameters of the theory. Then this model could provide a well motivated theoretical support for the main conclusions of the model independent analysis exposed above.

It is worth mentioning here that the massless dark-photon exchange does not affect the other $R_{D}$ ($R_{D^*}$) observables, the ratio of the
branching fraction of $\bar{B}\to D \tau \bar{\nu}_{\tau}$ ($\bar{B}\to D^* \tau \bar{\nu}_{\tau}$) to that of $\bar{B}\to D \ell \bar{\nu}_{\ell}$ ($\bar{B}\to D^* \ell \bar{\nu}_{\ell}$), where significant discrepancies have been also observed with respect to the SM  predictions. Indeed, these observables are connected to the tree-level charge currents, while the massless dark-photon can mediate only neutral  magnetic-dipole currents.

\section{Conclusions}
We evaluated the impact of the QED magnetic-dipole corrections to the final lepton states, for the widths and branching ratios of the $B\to (K,K^*)\ell^+\ell^-$ decays. We also included the Sommerfeld correction factor in the corresponding widths, which reabsorbs the re-summation of the leading logs terms induced by the long-distance contributions in the virtual Coulomb corrections.

Using the current cuts on $q^2$  adopted by the LHCb collaborations, we found that these corrections do not exceed a few per mille effect on $R_{K^*}$, depending on the integrated $q^2$ bin regions, while these are one order of magnitude smaller in $R_{K}$. The largest contribution is achieved in $q^2$ regions close to the dimuon mass threshold and it is one order of magnitude smaller than the typical corrections induced by the QED soft and collinear photon emissions. In particular, corrections on $R_{K^*}$ are of the order of $0.1\%$ and $0.07\%$ for $4m_{\mu}^2 < q^2 < 0.5 \GeV^2$ and $4m_{\mu}^2 < q^2 < 1.1 \GeV^2$ ranges respectively, while they drop down to less than $10^{-4}$ for $q^2>1\GeV$. Concerning the $R_K$, we found that the corresponding deviations are approximately one order of magnitude smaller than in the $R_{K^*}$ case, in almost all integrated regions of $q^2$. The enhanced effect of magnetic-dipole corrections in the $B\to K^*$ transitions, against the $B\to K$ ones, is due to the contribution of the spin-1 longitudinal polarization of $K^*$.

Regarding the Sommerfeld corrections to $R_{K,K^*}$, these dominate only in the $q^2$ region close to the dilepton mass threshold. In particular, we found them to be of the order of 0.1\% for both the $B\to K^*$, $B^0\to K^0$ and $B^+\to K^+$ transitions, in the integrated range of $4m_{\mu}^2 < q^2 < 0.5 \GeV^2$. However, unlike the magnetic-dipole correction, the Sommerfeld correction almost entirely satisfies LFU for  $q^2 > 0.5 \GeV^2$, and its impact on $R_{K,K^*}$ falls rapidly to zero above this $q^2$ threshold. Moreover, we found that, the contributions of both Sommerfeld and magnetic-dipole corrections are opposite in sign and tend to cancel in the $B\to K^*\ell^+\ell^-$ decays, while they add coherently with same sign in the  $B^{(0,+)}\to K^{(0,+)}\ell^+\ell^-$ decays.
In conclusion, a high experimental precision on the $R_{K,K^*}$ measurements of the order of per mille would be required in order to explore the sensitivity of the $R_{K,K^*}$ to magnetic-dipole corrections.

Finally, we analyzed the role of a potential NP, mediated by magnetic-dipole interactions, in explaining or softening the discrepancies in the present measurements of the $R_{K,K^*}$ observables.
In particular, we considered the role a {\it massless} dark-photon scenario. The {\it massless} dark photon, unlike the massive one, has not any tree-level milli-charged interactions with SM matter fields, and mainly couple via magnetic-dipole interactions with SM fermions. In this respect, we considered two possible approaches. The first one, in which we assume the NP to contribute mainly to the effective magnetic-dipoles of SM fermions with dark-photon. The second one, based on a renormalizable models for the dark sector, in which the NP predicts correlated contributions to both the usual magnetic-dipole interactions of SM fermions with ordinary photons and the corresponding ones with a dark-photon.

By using a model independent approach, we found that a {\it massless} dark-photon exchange could give a ${\cal O}(10\%)$ deviation on $R_{K^*}$ compatible with all present constraints from dark sector searches, flavor physics and $(g-2)_{\mu}$, providing a viable interpretation of the present discrepancies on the $R_{K^*}$ anomaly. On the other hand, a modest effect on  $R_K$ is found which could be at the best of a few percent effect, and cannot account for the observed $R_K$ anomalies. The dark-photon contributions to $R_{K^*}$ could be further constrained if we assume that the recent $B\to K^* \nu\bar{\nu}$ constraints could be applied to the massless dark-photon production via $B\to K^* \bar{\gamma}$ decay, with $\bar{\gamma}$ detected as missing energy. In particular, these constraints would reduce the maximum available contribution to $R_{K^*}$ within a 1-2\% effect, ruling out the possibility to fully explain the $R_{K^*}$ anomaly in terms of a massless dark-photon contribution. However, at present no dedicated analysis for the search of $B\to K^* \bar{\gamma}$ decay process is available that could support the validity of this assumption.

We found that the results of the  model independent analysis could be well motivated and supported by a renormalizable dark sector scenario that is a suitable extension of the one proposed in  \cite{Gabrielli:2013jka,Gabrielli:2016vbb} and predicts the existence of a {\it massless} dark photon. This model allows to bypass the  $b\to s \gamma$ constraints at the NLO and $(g-2)_{\mu}$ constraints,  while providing large contributions to the relevant dark magnetic-dipole operators involved in the $b\to s \ell^+\ell^-$ transitions.
This required to double the dark-fermions and messengers fields of the model in \cite{Gabrielli:2013jka,Gabrielli:2016vbb} by promoting them to degenerate $SU(2)$ doublets.

In conclusion, if future measurements should reduce the gap with SM predictions in $R_K$, while increasing the discrepancy in $R_{K^*}$, this might be interpreted as a smoking gun signature of a LFU-violating long-distance interactions mediated by a {\it massless} dark photon exchange.
However, new analyses would be required in this case to disentangle the  effect of a {\it massless} dark photon exchange from other potential new physics sources, including a dedicated experimental search for the $B\to K^* \bar{\gamma}$ decay process.

\section*{Acknowledgments}
We thank M. Fabbrichesi, P. Goldenzweig, D. Marzocca, B. Mele, M. Nardecchia, L. Silvestrini, and D. Tonelli for useful comments and discussions. E.G. would like to thank the CERN Theory Department for the kind of hospitality during the preparation of this work. 



\begin{thebibliography}{99}
  

\bibitem{Misiak:1992bc} 
  M.~Misiak,
  Nucl.\ Phys.\ B {\bf 393}, 23 (1993)
  Erratum: [Nucl.\ Phys.\ B {\bf 439}, 461 (1995)].
  
\bibitem{Buras:1994dj} 
  A.~J.~Buras and M.~Munz,
  Phys.\ Rev.\ D {\bf 52}, 186 (1995)
   [hep-ph/9501281].
  
\bibitem{Buchalla:1995vs} 
  G.~Buchalla, A.~J.~Buras and M.~E.~Lautenbacher,
  Rev.\ Mod.\ Phys.\  {\bf 68}, 1125 (1996)
  [hep-ph/9512380].
\bibitem{Chetyrkin:1996vx} 
  K.~G.~Chetyrkin, M.~Misiak and M.~Munz,
  Phys.\ Lett.\ B {\bf 400}, 206 (1997), 
  Erratum: [Phys.\ Lett.\ B {\bf 425}, 414 (1998)]
  [hep-ph/9612313].

\bibitem{Bobeth:1999mk} 
  C.~Bobeth, M.~Misiak and J.~Urban,
  Nucl.\ Phys.\ B {\bf 574}, 291 (2000)
  [hep-ph/9910220].

\bibitem{Bobeth:2001jm} 
  C.~Bobeth, A.~J.~Buras, F.~Kruger and J.~Urban,
  Nucl.\ Phys.\ B {\bf 630}, 87 (2002)
  [hep-ph/0112305].
 
\bibitem{Gambino:2003zm} 
  P.~Gambino, M.~Gorbahn and U.~Haisch,
  Nucl.\ Phys.\ B {\bf 673}, 238 (2003)
  [hep-ph/0306079].
  
\bibitem{Gorbahn:2004my} 
  M.~Gorbahn and U.~Haisch,
  Nucl.\ Phys.\ B {\bf 713}, 291 (2005)
  [hep-ph/0411071].

\bibitem{Gorbahn:2005sa} 
  M.~Gorbahn, U.~Haisch and M.~Misiak,
  Phys.\ Rev.\ Lett.\  {\bf 95}, 102004 (2005)
  [hep-ph/0504194].

\bibitem{Altmannshofer:2008dz} 
  W.~Altmannshofer, P.~Ball, A.~Bharucha, A.~J.~Buras, D.~M.~Straub and M.~Wick,
  JHEP {\bf 0901}, 019 (2009)
  [arXiv:0811.1214 [hep-ph]].

\bibitem{Cho:1996we} 
  P.~L.~Cho, M.~Misiak and D.~Wyler,
  Phys.\ Rev.\ D {\bf 54}, 3329 (1996)
  [hep-ph/9601360].

\bibitem{Ali:1999mm} 
  A.~Ali, P.~Ball, L.~T.~Handoko and G.~Hiller,
  Phys.\ Rev.\ D {\bf 61}, 074024 (2000)
  [hep-ph/9910221]. 



  
\bibitem{Khodjamirian:2010vf} 
  A.~Khodjamirian, T.~Mannel, A.~A.~Pivovarov and Y.-M.~Wang,
  JHEP {\bf 1009}, 089 (2010)
  [arXiv:1006.4945 [hep-ph]].
  
\bibitem{Beylich:2011aq} 
  M.~Beylich, G.~Buchalla and T.~Feldmann,
  Eur.\ Phys.\ J.\ C {\bf 71}, 1635 (2011)
  [arXiv:1101.5118 [hep-ph]].


\bibitem{Bobeth:2011nj} 
  C.~Bobeth, G.~Hiller, D.~van Dyk and C.~Wacker,
  JHEP {\bf 1201}, 107 (2012)
  [arXiv:1111.2558 [hep-ph]].

  
\bibitem{DescotesGenon:2012zf} 
  S.~Descotes-Genon, J.~Matias, M.~Ramon and J.~Virto,
  JHEP {\bf 1301}, 048 (2013)
  [arXiv:1207.2753 [hep-ph]].

\bibitem{Khodjamirian:2012rm} 
  A.~Khodjamirian, T.~Mannel and Y.~M.~Wang,
  JHEP {\bf 1302}, 010 (2013)
  [arXiv:1211.0234 [hep-ph]].

\bibitem{Jager:2012uw} 
  S.~Jäger and J.~Martin Camalich,
  JHEP {\bf 1305}, 043 (2013)
  [arXiv:1212.2263 [hep-ph]].

\bibitem{Descotes-Genon:2014uoa} 
  S.~Descotes-Genon, L.~Hofer, J.~Matias and J.~Virto,
  JHEP {\bf 1412}, 125 (2014)
  [arXiv:1407.8526 [hep-ph]].


\bibitem{Jager:2014rwa} 
  S.~Jäger and J.~Martin Camalich,
  Phys.\ Rev.\ D {\bf 93}, no. 1, 014028 (2016)
  [arXiv:1412.3183 [hep-ph]].

\bibitem{Straub:2015ica} 
  A.~Bharucha, D.~M.~Straub and R.~Zwicky,
  JHEP {\bf 1608}, 098 (2016)
  [arXiv:1503.05534 [hep-ph]].
  
  
\bibitem{Ciuchini:2015qxb} 
  M.~Ciuchini, M.~Fedele, E.~Franco, S.~Mishima, A.~Paul, L.~Silvestrini and M.~Valli,
  JHEP {\bf 1606}, 116 (2016)
  [arXiv:1512.07157 [hep-ph]].


\bibitem{Capdevila:2017ert} 
  B.~Capdevila, S.~Descotes-Genon, L.~Hofer and J.~Matias,
  JHEP {\bf 1704}, 016 (2017)
  [arXiv:1701.08672 [hep-ph]].

\bibitem{Chobanova:2017ghn} 
  V.~G.~Chobanova, T.~Hurth, F.~Mahmoudi, D.~Martinez Santos and S.~Neshatpour,
  JHEP {\bf 1707}, 025 (2017)
  [arXiv:1702.02234 [hep-ph]].
   

  
\bibitem{Aaij:2014ora} 
  R.~Aaij {\it et al.} [LHCb Collaboration],
  Phys.\ Rev.\ Lett.\  {\bf 113}, 151601 (2014)
  [arXiv:1406.6482 [hep-ex]].

\bibitem{Aaij:2019wad} 
  R.~Aaij {\it et al.} [LHCb Collaboration],
  Phys.\ Rev.\ Lett.\  {\bf 122}, no. 19, 191801 (2019)
  [arXiv:1903.09252 [hep-ex]].

\bibitem{Aaij:2017vbb} 
  R.~Aaij {\it et al.} [LHCb Collaboration],
  JHEP {\bf 1708}, 055 (2017)
  [arXiv:1705.05802 [hep-ex]].
  
\bibitem{BelleMoriond}
 {\bf  Belle}  Collaboration, {\it Search for $B\to \ell \nu \gamma$ and $B\to \mu \nu_{\mu}$ and Test of Lepton Universality with ${\rm R}(K^*)$ at Belle}, Seminar of M. Prim at Rencontres de Moriond EW, 2019.

\bibitem{Abdesselam:2019wac} 
  A.~Abdesselam {\it et al.} [Belle Collaboration],
  arXiv:1904.02440 [hep-ex].

\bibitem{Hiller:2003js} 
  G.~Hiller and F.~Kruger,
  Phys.\ Rev.\ D {\bf 69}, 074020 (2004)
  [hep-ph/0310219].


\bibitem{Hiller:2014yaa} 
  G.~Hiller and M.~Schmaltz,
  Phys.\ Rev.\ D {\bf 90}, 054014 (2014)
  [arXiv:1408.1627 [hep-ph]].

\bibitem{Bouchard:2013mia} 
  C.~Bouchard {\it et al.} [HPQCD Collaboration],
  Phys.\ Rev.\ Lett.\  {\bf 111}, no. 16, 162002 (2013)
  Erratum: [Phys.\ Rev.\ Lett.\  {\bf 112}, no. 14, 149902 (2014)]
  [arXiv:1306.0434 [hep-ph]].
  
\bibitem{Bordone:2016gaq} 
  M.~Bordone, G.~Isidori and A.~Pattori,
  Eur.\ Phys.\ J.\ C {\bf 76}, no. 8, 440 (2016)
  [arXiv:1605.07633 [hep-ph]].

 


  

\bibitem{Descotes-Genon:2015uva}
S.~Descotes-Genon, L.~Hofer, J.~Matias, and J.~Virto {\em JHEP} {\bf 06} (2016)
  092 [arXiv:1510.04239 [hep-ph]].

\bibitem{Altmannshofer:2017fio}
W.~Altmannshofer, C.~Niehoff, P.~Stangl, and D.~M. Straub {\em Eur. Phys. J.}
  {\bf C77} (2017), no.~6 377 [arXiv:1703.09189 [hep-ph]].

\cite{Capdevila:2017bsm}
\bibitem{Capdevila:2017bsm}
B.~Capdevila, A.~Crivellin, S.~Descotes-Genon, J.~Matias, and J.~Virto {\em
  JHEP} {\bf 01} (2018) 093 [arXiv:1704.05340 [hep-ph]].

\bibitem{DAmico:2017mtc}
G.~D'Amico, M.~Nardecchia, P.~Panci, F.~Sannino, A.~Strumia, R.~Torre, and
  A.~Urbano {\em JHEP} {\bf 09} (2017) 010 [arXiv:1704.05438 [hep-ph]].

\bibitem{Altmannshofer:2017yso}
W.~Altmannshofer, P.~Stangl, and D.~M. Straub {\em Phys. Rev.} {\bf D96}
  (2017), no.~5 055008 [arXiv:1704.05435 [hep-ph]].

\bibitem{Geng:2017svp}
L.-S. Geng, B.~Grinstein, S.~J{\"a}ger, J.~Martin~Camalich, X.-L. Ren, and
  R.-X. Shi {\em Phys. Rev.} {\bf D96} (2017), no.~9 093006 [arXiv:1704.05446 [hep-ph]].

\bibitem{Ciuchini:2017mik}
M.~Ciuchini, A.~M. Coutinho, M.~Fedele, E.~Franco, A.~Paul, L.~Silvestrini, and
  M.~Valli {\em Eur. Phys. J.} {\bf C77} (2017), no.~10 688 [arXiv:1704.05447 [hep-ph]].

\bibitem{Hiller:2017bzc}
G.~Hiller and I.~Nisandzic {\em Phys. Rev.} {\bf D96} (2017), no.~3 035003 [arXiv:1704.05444 [hep-ph]].

  
\bibitem{Alok:2017sui}
A.~K. Alok, B.~Bhattacharya, A.~Datta, D.~Kumar, J.~Kumar, and D.~London {\em
  Phys. Rev.} {\bf D96} (2017), no.~9 095009  [arXiv:1704.07397 [hep-ph]].

\bibitem{Hurth:2017hxg}
T.~Hurth, F.~Mahmoudi, D.~Martinez~Santos, and S.~Neshatpour {\em Phys. Rev.}
  {\bf D96} (2017), no.~9 095034 [arXiv:1705.06274].

\bibitem{Alguero:2019aa}
M.~Alguer{\'o}, B.~Capdevila, A.~Crivellin, S.~Descotes-Genon, P.~Masjuan,
  J.~Matias, and J.~Virto [arXiv:1903.09578 [hep-ph]].

\bibitem{Alok:2019ufo}
A.~K. Alok, A.~Dighe, S.~Gangal, and D.~Kumar [arXiv:1903.09617 [hep-ph]].

\bibitem{Ciuchini:2019usw}
M.~Ciuchini, A.~M. Coutinho, M.~Fedele, E.~Franco, A.~Paul, L.~Silvestrini, and
  M.~Valli [arXiv:1903.09632 [hep-ph]].

\bibitem{Aebischer:2019mlg}
J.~Aebischer, W.~Altmannshofer, D.~Guadagnoli, M.~Reboud, P.~Stangl, and D.~M.
  Straub [arXiv:1903.10434 [hep-ph]].

\bibitem{Kowalska:2019ley}
  K.~Kowalska, D.~Kumar and E.~M.~Sessolo,
  Eur.\ Phys.\ J.\ C {\bf 79} (2019) no.10,  840
  [arXiv:1903.10932 [hep-ph]].
 
\bibitem{Alguero:2018nvb}
M.~Alguer{\'o}, B.~Capdevila, S.~Descotes-Genon, P.~Masjuan, and J.~Matias [arXiv:1809.08447 [hep-ph]].

\bibitem{Alguero:2019pjc}
M.~Alguer{\'o}, B.~Capdevila, S.~Descotes-Genon, P.~Masjuan, and J.~Matias
  [arXiv:1902.04900 [hep-ph]].

\bibitem{Datta:2019zca}
A.~Datta, J.~Kumar, and D.~London [arXiv:1903.10086 [hep-ph]].



\bibitem{Greljo:2015mma}
A.~Greljo, G.~Isidori, and D.~Marzocca {\em JHEP} {\bf 07} (2015) 142 [arXiv:1506.01705 [hep-ph]].

\bibitem{Barbieri:2015yvd}
R.~Barbieri, G.~Isidori, A.~Pattori, and F.~Senia {\em Eur. Phys. J.} {\bf C76}
  (2016), no.~2 67 [arXiv:1512.01560 [hep-ph]].

\bibitem{Barbieri:2016las}
R.~Barbieri, C.~W. Murphy, and F.~Senia {\em Eur. Phys. J.} {\bf C77} (2017),
  no.~1 8 [arXiv:1611.04930 [hep-ph]].

\bibitem{Bordone:2017anc}
M.~Bordone, G.~Isidori, and S.~Trifinopoulos {\em Phys. Rev.} {\bf D96} (2017),
  no.~1 015038 [arXiv:1702.07238 [hep-ph]].

\bibitem{Bordone:2017lsy}
M.~Bordone, D.~Buttazzo, G.~Isidori, and J.~Monnard {\em Eur. Phys. J.} {\bf
  C77} (2017), no.~9 618 [arXiv:1705.10729 [hep-ph]].

\bibitem{Buttazzo:2017ixm}
D.~Buttazzo, A.~Greljo, G.~Isidori, and D.~Marzocca {\em JHEP} {\bf 11} (2017)
  044 [arXiv:1706.07808 [hep-ph]].

\bibitem{Barbieri:2017tuq}
R.~Barbieri and A.~Tesi {\em Eur. Phys. J.} {\bf C78} (2018), no.~3 193 [arXiv:1712.06844 [hep-ph]].


\bibitem{Sommerfeld}
  A.~Sommerfeld, Ann. Phys. {\bf 403} (1931) 257.
\bibitem{Fermi}
  E.~Fermi,
  ``An attempt of a theory of beta radiation. 1.,''
  Z.\ Phys.\  {\bf 88}, 161 (1934).

  
\bibitem{Isidori:2007zt} 
  G.~Isidori,
  Eur.\ Phys.\ J.\ C {\bf 53}, 567 (2008)
  doi:10.1140/epjc/s10052-007-0487-0
  [arXiv:0709.2439 [hep-ph]].

  
\bibitem{Fabbrichesi:2020wbt}
M.~Fabbrichesi, E.~Gabrielli and G.~Lanfranchi,
``The Dark Photon'',
[arXiv:2005.01515 [hep-ph]].
  
\bibitem{Holdom:1985ag} 
  B.~Holdom,
  Phys.\ Lett.\  {\bf 166B}, 196 (1986).

\bibitem{Tanabashi:2018oca} 
  M.~Tanabashi {\it et al.} [Particle Data Group],
  Phys.\ Rev.\ D {\bf 98}, no. 3, 030001 (2018).
  doi:10.1103/PhysRevD.98.030001

\bibitem{Gabrielli:1998sw} 
  E.~Gabrielli and U.~Sarid,
  Phys.\ Rev.\ D {\bf 58}, 115003 (1998)
  doi:10.1103/PhysRevD.58.115003
  [hep-ph/9803451].


\bibitem{Aubert:2009qda} 
  B.~Aubert {\it et al.} [BaBar Collaboration],
  Phys.\ Rev.\ D {\bf 81}, 032003 (2010)
  [arXiv:0908.0415 [hep-ex]].



\bibitem{Appelquist:2004mn}
T.~Appelquist, M.~Piai and R.~Shrock,
Phys. Lett. B \textbf{593}, 175-180 (2004)
[arXiv:hep-ph/0401114 [hep-ph]];
T.~Appelquist, M.~Piai and R.~Shrock,
Phys. Lett. B \textbf{595}, 442-452 (2004)
[arXiv:hep-ph/0406032 [hep-ph]].

  
\bibitem{Weinberg:1965nx} 
  S.~Weinberg,
  Phys.\ Rev.\  {\bf 140}, B516 (1965).



\bibitem{Davidson:1991si} 
  S.~Davidson, B.~Campbell and D.~C.~Bailey,
  Phys.\ Rev.\ D {\bf 43}, 2314 (1991).

\bibitem{Raffelt:1996wa} 
  G.~G.~Raffelt,
  ``Stars as laboratories for fundamental physics : The astrophysics of neutrinos, axions, and other weakly interacting particles,''
  Chicago, USA: Univ. Pr. (1996) 664 p

\bibitem{Kadota:2016tqq} 
  K.~Kadota, T.~Sekiguchi and H.~Tashiro,
  ``A new constraint on millicharged dark matter from galaxy clusters,''
  [arXiv:1602.04009 [astro-ph.CO]].
  
\bibitem{Essig:2013lka} 
  R.~Essig {\it et al.},
  ``Working Group Report: New Light Weakly Coupled Particles,''
  [arXiv:1311.0029 [hep-ph]].

\bibitem{Alexander:2016aln} 
  J.~Alexander {\it et al.},
  ``Dark Sectors 2016 Workshop: Community Report,''
  [arXiv:1608.08632 [hep-ph]].

\bibitem{Ackerman:mha} 
  L.~Ackerman, M.~R.~Buckley, S.~M.~Carroll and M.~Kamionkowski,
  Phys.\ Rev.\ D {\bf 79}, 023519 (2009)
  [arXiv:0810.5126 [hep-ph]].

\bibitem{Fan:2013tia} 
  J.~Fan, A.~Katz, L.~Randall and M.~Reece,
  Phys.\ Rev.\ Lett.\  {\bf 110}, no. 21, 211302 (2013)
  [arXiv:1303.3271 [hep-ph]].


\bibitem{Agrawal:2016quu} 
  P.~Agrawal, F.~Y.~Cyr-Racine, L.~Randall and J.~Scholtz,
  JCAP {\bf 1705}, 022 (2017)
  [arXiv:1610.04611 [hep-ph]].

\bibitem{Foot:2014uba} 
  R.~Foot and S.~Vagnozzi,
  Phys.\ Rev.\ D {\bf 91}, 023512 (2015)
  [arXiv:1409.7174 [hep-ph]].

\bibitem{Heikinheimo:2015kra} 
  M.~Heikinheimo, M.~Raidal, C.~Spethmann and H.~Veermäe,
  Phys.\ Lett.\ B {\bf 749}, 236 (2015)
  [arXiv:1504.04371 [hep-ph]].

 
\bibitem{Gabrielli:2013jka} 
  E.~Gabrielli and M.~Raidal,
  Phys.\ Rev.\ D {\bf 89}, no. 1, 015008 (2014)
  [arXiv:1310.1090 [hep-ph]].
  
\bibitem{Gabrielli:2016vbb} 
  E.~Gabrielli, L.~Marzola and M.~Raidal,
  Phys.\ Rev.\ D {\bf 95}, no. 3, 035005 (2017)
  [arXiv:1611.00009 [hep-ph]].

  
\bibitem{Gabrielli:2014oya} 
  E.~Gabrielli, M.~Heikinheimo, B.~Mele and M.~Raidal,
  Phys.\ Rev.\ D {\bf 90}, no. 5, 055032 (2014)
  [arXiv:1405.5196 [hep-ph]].


\bibitem{Gabrielli:2016cut} 
  E.~Gabrielli, B.~Mele, M.~Raidal and E.~Venturini,
  Phys.\ Rev.\ D {\bf 94}, no. 11, 115013 (2016)
  [arXiv:1607.05928 [hep-ph]].

 
\bibitem{Fabbrichesi:2017vma} 
  M.~Fabbrichesi, E.~Gabrielli and B.~Mele,
  Phys.\ Rev.\ Lett.\  {\bf 119}, no. 3, 031801 (2017)
  [arXiv:1705.03470 [hep-ph]].

\bibitem{Fabbrichesi:2017zsc} 
  M.~Fabbrichesi, E.~Gabrielli and B.~Mele,
  Phys.\ Rev.\ Lett.\  {\bf 120}, no. 17, 171803 (2018)
  [arXiv:1712.05412 [hep-ph]].

\bibitem{Barducci:2018rlx} 
  D.~Barducci, M.~Fabbrichesi and E.~Gabrielli,
  Phys.\ Rev.\ D {\bf 98}, no. 3, 035049 (2018)
  [arXiv:1806.05678 [hep-ph]].

\bibitem{Biswas:2016jsh}
S.~Biswas, E.~Gabrielli, M.~Heikinheimo and B.~Mele,
Phys. Rev. D \textbf{93}, no.9, 093011 (2016)
[arXiv:1603.01377 [hep-ph]].
  
\bibitem{Biswas:2015sha}
S.~Biswas, E.~Gabrielli, M.~Heikinheimo and B.~Mele,
JHEP \textbf{06}, 102 (2015)
[arXiv:1503.05836 [hep-ph]].

\bibitem{Biswas:2017lyg}
S.~Biswas, E.~Gabrielli, M.~Heikinheimo and B.~Mele,
Phys. Rev. D \textbf{96}, no.5, 055012 (2017)
[arXiv:1703.00402 [hep-ph]].

\bibitem{Fabbrichesi:2019bmo}
M.~Fabbrichesi and E.~Gabrielli,
Eur. Phys. J. C \textbf{80}, no.6, 532 (2020)
[arXiv:1911.03755 [hep-ph]].

\bibitem{Sirunyan:2019xst}
A.~M.~Sirunyan \textit{et al.} [CMS],
JHEP \textbf{10}, 139 (2019)
[arXiv:1908.02699 [hep-ex]].


\bibitem{Ahn:2018mvc} 
  J.~K.~Ahn {\it et al.} [KOTO Collaboration],
  Phys.\ Rev.\ Lett.\  {\bf 122}, no. 2, 021802 (2019)
  [arXiv:1810.09655 [hep-ex]].



\bibitem{Ciuchini:1996vw} 
  M.~Ciuchini, E.~Gabrielli and G.~F.~Giudice,
  Phys.\ Lett.\ B {\bf 388}, 353 (1996)
  Erratum: [Phys.\ Lett.\ B {\bf 393}, 489 (1997)]
  [hep-ph/9604438].

  
\bibitem{Grygier:2017tzo}
J.~Grygier \textit{et al.} [Belle],
Phys. Rev. D \textbf{96}, no.9, 091101 (2017)
[arXiv:1702.03224 [hep-ex]].

\bibitem{Belle-pcomm}
Private communications with the Belle Collaboration.

\bibitem{Bennett:2006fi} 
  G.~W.~Bennett {\it et al.} [Muon g-2 Collaboration],
  Phys.\ Rev.\ D {\bf 73}, 072003 (2006)
  [hep-ex/0602035].


\bibitem{Blum:2018mom} 
  T.~Blum {\it et al.} [RBC and UKQCD Collaborations],
  Phys.\ Rev.\ Lett.\  {\bf 121}, no. 2, 022003 (2018)
  [arXiv:1801.07224 [hep-lat]].

\end{thebibliography}
\end{document}